\documentclass[11pt,a4paper]{article}              

\usepackage[margin=25mm]{geometry}
\usepackage{graphicx}
\usepackage{colortbl}
\usepackage{algorithm}
\usepackage{dsfont}
\usepackage{subcaption}
\usepackage{algpseudocode}
\usepackage{comment}
\usepackage{wrapfig}

\usepackage{fancyhdr}
\usepackage{amssymb,amsfonts,amsmath,amsthm}
\usepackage{natbib}
\usepackage{hyperref}
\usepackage{cleveref}
\Crefformat{equation}{Eq.~(#2#1#3)}
\crefmultiformat{equation}
  {Eqs.~(#2#1#3)}          
  { and~(#2#1#3)}         
  {, (#2#1#3)}            
  { and~(#2#1#3)}         
  
\usepackage{booktabs}
\usepackage[table,xcdraw]{xcolor}
\usepackage[font=small, width=\linewidth, justification=centering, skip=10pt]{caption}

\newcommand{\cm}{{\mathcal M}}

\newcommand{\cx}{{\mathcal X}}
\newcommand{\cy}{{\mathcal Y}}

\newcommand{\ve}[1]{\boldsymbol{#1}}			

\newcommand{\mat}[1]{\boldsymbol{#1}}			

\newcommand{\tr}{^{\textsf T}}				

\newcommand{\enu}{ , \, \dots \,,}

\newcommand{\norm}[1]{\lVert #1\rVert}

\newcommand{\prt}[1]{\left(#1\right)}			
\newcommand{\acc}[1]{\left\{#1\right\}}			
\newcommand{\abs}[1]{\left| #1 \right|}			
\newcommand{\eqdef}{\stackrel{\text{def}}{=}}

\newcommand{\Prob}[1]{{\mathbb P}\left( #1 \right)}	

\newcommand{\ned}{N_{\text{ED}}}

\newlength\myindent
\usepackage{authblk}
	
\renewcommand{\ned}{n_{\textrm{ED}}}
\newcommand{\cna}{\hat{C}_{n,\alpha}}
\newcommand{\cnasplit}{\hat{C}_{\ncal,\alpha}^S}
\newcommand{\nval}{n_{\textrm{val}}}
\newcommand{\ncal}{n_{\textrm{cal}}}
\newcommand{\Sval}{\mathcal{S}^{\textrm{val}}}
\newcommand{\Scal}{\mathcal{S}^{\textrm{cal}}}
\newcommand{\nr}{n_{\textrm{R}}}
\newcommand{\errval}{\varepsilon_{\textrm{val}}}
\newcommand{\errgen}{\varepsilon_{\textrm{gen}}}

\newcommand{\riloo}{\mathcal{R}_i^{\textrm{LOO}}}
\newcommand{\riloosgn}{\mathcal{R}_i^{\textrm{sgn,LOO}}}
\newcommand{\yed}{\ve{y}_{\textrm{ED}}}
\newcommand{\yval}{\ve{y}_{\textrm{val}}}

\graphicspath{{./Figures}}

\begin{document}

\title{Conformal prediction for full and sparse polynomial chaos expansions} 

\author[1]{Arthur Hatstatt} 
\author[2]{Xujia Zhu}
\author[1]{Bruno Sudret}

\affil[1]{ETH Zürich, 

    \vspace{-0.5em}

    Chair of Risk, Safety and Uncertainty Quantification,

 \vspace{-0.5em}
  
 Stefano-Franscini-Platz 5, 8093 Zurich, Switzerland}

\affil[2]{Université Paris-Saclay, CNRS,

  \vspace{-0.5em}

    CentraleSupélec, Laboratoire des Signaux et Systèmes,

    \vspace{-0.5em}

  3 rue Joliot Curie , 91190 Gif-sur-Yvette, France}

 \date{January 15, 2026}

\maketitle

\abstract{
Polynomial Chaos Expansions (PCEs) are widely recognized for their efficient computational performance in surrogate modeling. Yet, a robust framework to quantify local model errors is still lacking. While the local uncertainty of PCE prediction can be captured using bootstrap resampling, other methods offering more rigorous statistical guarantees are needed, especially in the context of small training datasets. Recently, conformal predictions have demonstrated strong potential in machine learning, providing statistically robust and model-agnostic prediction intervals. Due to its generality and versatility, conformal prediction is especially valuable, as it can be adapted to suit a variety of problems, making it a compelling choice for PCE-based surrogate models. In this contribution, we explore its application to PCE-based surrogate models. More precisely, we present the integration of two conformal prediction methods, namely the full conformal and the Jackknife+ approaches, into both full and sparse PCEs. For full PCEs, we introduce computational shortcuts inspired by the inherent structure of regression methods to optimize the implementation of both conformal methods. For sparse PCEs, we incorporate the two approaches with appropriate modifications to the inference strategy, thereby circumventing the non-symmetrical nature of the regression algorithm and ensuring valid prediction intervals. Our developments yield better-calibrated prediction intervals for both full and sparse PCEs, achieving superior coverage over existing approaches, such as the bootstrap, while maintaining a moderate computational cost. 
}

\vspace{20pt}
  {\bf Keywords}: Conformal predictions -- Surrogate modeling -- Sparse polynomial chaos expansions -- Local error measure -- Conformalized PCE

\maketitle


\newpage
\section{Introduction}

Modern scientific and engineering research on complex systems increasingly relies on numerical simulations. These simulations are typically performed using computational models, also known as simulators. High-fidelity models often require substantial computational time, making each single model evaluation costly. This poses significant challenges for uncertainty quantification \citep{SmithUQBook2014}, which aims to characterize how input uncertainties propagate and affect the model output quantities of interest. Because such an analysis requires numerous model evaluations, it often faces bottlenecks due to limited computational resources and time constraints. This issue can be mitigated by employing surrogate models, which are usually built by applying statistical techniques to a limited number of evaluations of the original complex model. Once fitted onto the simulation data, surrogate models are computationally inexpensive, making them suitable for many-query analyses such as global sensitivity analysis, optimization, and rare event estimation.


Polynomial chaos expansions (PCEs) \citep{Ghanem1991,Xiu2002,Soize2004} are nowadays one of the most mature surrogate models widely applied in uncertainty quantification and optimization. Such a surrogate model represents a deterministic function as a series of orthogonal polynomials that form a basis for the space of square-integrable functions with respect to the probability measure of the input parameters. Common approaches to construct PCE include stochastic Galerkin methods \citep{Babuska01}, projection methods \citep{LeMaitreBook2010,XiuBook2010}, and regression methods \citep{Berveiller2006}. Among these, regression methods have gained widespread popularity due to their nonintrusive nature, flexibility, and their link with advanced statistical techniques, especially when dealing with complex models. In data-rich scenarios, ordinary least-squares could be employed with a large set of orthogonal polynomials to fit a so-called \emph{full PCE}, where all terms up to a specified degree are included in the expansion. However, limitations arise in data-scarce scenarios, where the number of potential regressors is limited by the available data. Interestingly, most physical systems are governed primarily by main effects and low-order interactions, a feature known as the \textit{sparsity-of-effects principle} \citep{Montgomery:2004,BlatmanThesis}. This principle has inspired the development of compressive sensing \citep{Candes_sparsity} and, more specifically, sparse PCEs \citep{Blatman2010a,BlatmanJCP2011,Doostan2011,Jakeman2015,LuethenIJUQ2022,LuethenCMAME2023}. Moreover, sparse PCE has demonstrated comparable performance to state-of-the-art machine learning techniques in purely data-driven contexts, as shown by \citet{Torre2019a}, making it a particularly versatile tool.

The performance of polynomial chaos expansions is typically assessed using \textit{global} error estimates, such as the mean-squared error, which is defined over the whole input space. However, in some applications, such as anomaly detection, it is crucial to quantify the model error at the individual prediction level, taking into account the uncertainty in model construction. 

Local prediction intervals provide measures of the possible range of true model outputs at a given confidence level, reflecting the epistemic uncertainty inherent to the model construction. Methods for constructing these intervals can be categorized according to the assumptions upon which they are built. Empirical prediction intervals can be constructed by assuming a distribution for the residuals \citep{Schmoyer_pi,Olive_empirical_pi}. However, these methods typically yield intervals with \textit{asymptotic} validity, limiting their applicability in data-scarce contexts. A prominent alternative is \textit{bootstrap resampling}, introduced by \citet{Efron1979}, with specific adaptations for regression \citep{Stine1985}. Nevertheless, bootstrap methods also rely on asymptotic correctness, and \citet{Efron1983} demonstrated that bootstrap error estimates often underestimate true errors, leading to poor coverage in small-sample scenarios. Despite these limitations, bootstrap techniques can still provide valuable insights into local model errors, as shown by their integration with PCE for active learning \citep{MarelliSS2018}. However, more robust methods with stronger statistical guarantees are desirable.

Recently, \textit{conformal prediction} has gained significant attention for its ability to construct prediction intervals without assuming specific data size or model distributions. Conformal predictors offer inherent flexibility for diverse applications, including classification and regression (as remarkably summarized by \citet{Angelopoulos2023}). First introduced by \citet{Vovk1998} and further developed in \citet{Vovk1999}, conformal prediction found application in classification-related machine learning problems. Despite being more challenging, \citet{Vovk1999b} attempted to apply the approach to regression problems, offering a conformal prediction framework for ridge regression problems. Continuous improvements to regression problems have been presented by \citet{Lei2013}, \citet{Lei2018}, and \citet{Barber2021}, among others, allowing this framework to provide statistical guarantees of valid coverage for local predictions for linear models. Recently, conformal prediction has been successfully applied to Gaussian process regression, providing coverage guarantees \citep{Papadopoulos2024, Jaber2025}. Conformal prediction encompasses various procedures, such as \textit{full conformal} \citep{Vovk2005}, \textit{split conformal} \citep{Papadopoulos2002}, and \textit{Jackknife+} \citep{Barber2021}, closely related to \textit{cross-conformal} \citep{Vovk2018}, each differing in computational requirements and coverage guarantees. In uncertainty quantification, where data is often limited, methods that require dataset splitting should generally be avoided. Thus, our work focuses on two methods: the full conformal approach and Jackknife+. The split conformal procedure will serve as a reference for comparing the various methods in numerical experiments with toy functions where an independent large-size validation set is available.

The primary goal of this work is to develop conformal predictors for full and sparse polynomial chaos expansions. For full PCE, computational shortcuts leveraging the structure of regression and classical algebra results, inspired by \cite{BlatmanThesis}, are introduced. However, these shortcuts are not directly applicable to sparse PCE because the selected sparse basis depends on the available data. Thus, an alternative approach based on \citet{Lei2019b} is adapted to enable the use of conformal predictors with sparse PCE at a moderate computational cost. 

The remainder of this paper is organized as follows. Sections 2 and 3 introduce the theoretical background on conformal prediction and polynomial chaos expansion, respectively. Section 4 presents the comparison framework between methods, while Section 5 discusses the application of conformal procedures to full PCE. The extension to sparse PCE is addressed in Section 6. Finally, Section 7 provides a discussion of the results and their practical implications.

\section{Conformal predictions}

\subsection{Introduction}
\label{sec:intro_symm}
Conformal predictors aim to infer prediction intervals for new input parameters with theoretically founded statistical coverage, without making assumptions about either the distribution of the data or the model structure. Broadly speaking, the main idea is to ensure that the residuals for new predictions align with the statistical distribution of the residuals of the selected surrogate observed over the available data. According to \citet{Vovk2005}, conformal prediction generally requires two conditions. First, the algorithm used to build the surrogate model is assumed to treat the data \textit{symmetrically}, i.e., the outcome of the algorithm remains invariant under reordering of the data. Second, all the data points are assumed to be \textit{exchangeable}, meaning that their joint distribution is invariant under any permutation of the points. This implies that observing any sequence of the same data points is equiprobable, making the statistical treatment insensitive to the indexing of the data points.

The general framework for conformal prediction and the various methods are presented in the following, which is inspired by the description of \citet{Angelopoulos2023}. Consider a deterministic computational model $\cm$,  defined as the mapping: 
\begin{align}\label{eq:1}
\begin{split}
    \cm:& ~\mathcal{D}_{\ve{X}}\subset \mathbb{R}^M\rightarrow\mathbb{R}^{Q}\\
    &~ \ve{x}\rightarrow \ve{y}=\cm(\ve{x}) .
\end{split}
\end{align}

 Without loss of generality, the present study concentrates on models with scalar outputs ($Q=1$), since multiple outputs involve minor changes to the methodology. Consider now an experimental design $\cx=\acc{\ve{x}_i\in \mathcal{D}_{\ve{X}},\, i=1 \enu n}$ and the corresponding model evaluations $\cy=\acc{y_i=\cm(\ve{x}_i),\, i=1 \enu n}$. This data defines the training set in a machine learning context, from which we want to fit a surrogate model $\hat{\cm}~:\mathcal{D}_{\ve{X}}\rightarrow \mathbb{R}$. For simplicity let us denote $\mathcal{S}=\acc{(\ve{x}_i,y_i),\, i=1 \enu n}$ this training set. 
The goal of conformal prediction is to provide an interval at an unseen data point $\ve{x}_{n+1}$, denoted by $\hat{C}_{n,\alpha}(\ve{x}_{n+1})$, such that, for a prescribed confidence $1-\alpha, ~\alpha\in(0,1)$: 
\begin{equation}
\label{eq:cov_guarantee}
    \mathbb{P}\prt{\cm(\ve{X}_{n+1})\in \hat{C}_{n,\alpha}(\ve{X}_{n+1})}\geq 1-\alpha.    
\end{equation}
Conformal predictions typically rely on a \emph{conformal score}, also referred to as the conformity (or nonconformity) score in the literature, which measures the agreement of the prediction and the data. For regression problems as in the present study, the absolute error $\abs{y_i-\hat{\cm}(\ve{x}_i)}$ is a common choice. However, other types of problems (such as classification) can be addressed within a similar framework using different scores.

Since the construction of confidence intervals requires the analysis of the statistics of the distribution of conformity scores, a specific definition of quantiles is presented by \cite{Barber2021} in order to account for the limited size of the data set. The empirical $(1-\alpha)$-quantile of the empirical distribution of values $\acc{v_i,~i=1 \enu n}$ is taken as: 
\begin{equation}
    \hat{q}_{n,\alpha}^+(v_i) = \textnormal{the}~\lceil (1-\alpha)(n+1)\rceil\textnormal{-th smallest value of }\acc{v_1,...,v_n}.
\end{equation}
Analogously, the $\alpha$-quantile of the empirical values is defined as: 
\begin{equation}
    \hat{q}_{n,\alpha}^-(v_i) = \textnormal{the}~\lfloor \alpha(n+1)\rfloor\textnormal{-th smallest value of }\acc{v_1,...,v_n}.
\end{equation}
In these equations, $\lfloor x \rfloor$ (resp. $\lceil x \rceil$) denotes the floor (resp. ceiling) of $x$, i.e., the greatest integer smaller than or equal to $x$ (resp. the smallest integer greater than or equal to $x$).

\subsection{Full conformal prediction}
In this section, we detail the full conformal prediction framework. The main idea is to train the surrogate model denoted $\hat{\cm}^y$ on an augmented set $\mathcal{S}^{aug}=\mathcal{S}\cup (\acc{\ve{x}_{n+1},y^{\textrm{trial}}})$ for different values of the unknown prediction $y^{\textrm{trial}}$. The trial value is considered conformal based on the accordance between the statistics of conformity scores over the training data and at the considered trial value. More precisely, the prediction interval is given by
\begin{equation}
    C^{FC}_{n,\alpha} = \acc{y: \abs{y-\hat{\cm}^{y}(\ve{x}_{i+1})}\leq q^+_{n,\alpha}\prt{\acc{\abs{y_i-\hat{\cm}^{y}(\ve{x}_{i})}:i=1,\ldots,n}}}.
\end{equation}
A straightforward approach would be grid search, as presented in \Cref{alg:1}. This method requires trying $n_{\textrm{trial}}$ various values of $y^{\textrm{trial}}$, and the resolution of the generated confidence interval depends on the grid refinement.  

\begin{algorithm}[H]
\small
\caption{Full conformal prediction (general formulation)}
\label{alg:1}
\begin{algorithmic}
\State $\cna^{FC} \gets  [~]$
\For{$k=~1:n_{\textrm{trial}}$}

    \State \textbf{Step 1: }\textnormal{Train the surrogate} $\hat{\cm}^{y}$\textnormal{ based on} $\mathcal{S}\cup (\ve{x}_{n+1},y_k^{\textrm{trial}})$

    \State \textbf{Step 2: }\textnormal{Evaluate the quantile of the conformity scores over the data } $\mathcal{S}$\textnormal{ : }\\ \hspace{2.5em}$\hat{q}_{n,\alpha}^{+}(\mathcal{R})$\textnormal{, where }$\mathcal{R}=\acc{\abs{y_i-\hat{\cm}^{y}(\ve{x}_i)},\,i=1\enu n}$

     \If{$\abs{y_k^{\textrm{trial}}-\hat{\cm}^{y}(\ve{x}_{n+1})}\leq \hat{q}_{n,\alpha}^{+}$}

        \State $\cna^{FC} \gets \cna^{FC}\cup y_k^{\textrm{trial}}$
    
    \EndIf
\EndFor
\State $\cna^{FC} \gets [\min(\cna^{FC}),\max(\cna^{FC})]$
\end{algorithmic}
\end{algorithm}

It should be noted that this formulation is adequate if the distribution of residuals is approximately symmetrical, justifying the use of the absolute value of the residuals. However, this may not be the case in practice. This remark leads to the alternative \textit{asymmetrical formulation} presented in \Cref{alg:2} \citep{Linusson2014_signed_residual}.

\begin{algorithm}[H]
\small
\caption{Full conformal prediction (asymmetrical formulation)}
\label{alg:2}
\begin{algorithmic}
\State $\cna^{FC} \gets [~]$
\For{$k=~1:n_{\textrm{trial}}$}

    \State \textbf{Step 1: }\textnormal{Train the surrogate} $\hat{\cm}^{y}$\textnormal{ based on} $\mathcal{S}\cup (\ve{x}_{n+1},y_k^{\textrm{trial}})$

    \State \textbf{Step 2: }\textnormal{Evaluate the quantile of the conformity scores over the data } $\mathcal{S}$\textnormal{ : }\\ \hspace{2.5em}$\hat{q}_{n,\alpha/2}^{+}(\mathcal{R})$\textnormal{ and }$\hat{q}_{n,\alpha/2}^{-}(\mathcal{R})$\textnormal{, where }$\mathcal{R}=\acc{y_i-\hat{\cm}^{y}(\ve{x}_i),\,i=1\enu n}$

     \If{$\hat{q}_{n,\alpha/2}^{-} \leq y_k^{\textrm{trial}}-\hat{\cm}^{y}(\ve{x}_{n+1})\leq \hat{q}_{n,\alpha/2}^{+}$}

        \State $\cna^{FC} \gets \cna^{FC}\cup y_k^{\textrm{trial}}$
    
    \EndIf
\EndFor
\State $\cna^{FC} \gets [\min(\cna^{FC}),\max(\cna^{FC})]$
\end{algorithmic}
\end{algorithm}

The main issue with full conformal prediction is that it requires the training of a new surrogate model for each value of $y_k^{\textrm{trial}}$. This can become intractable in high-dimensional problems when using high-degree polynomials. Because the grid search is rather inefficient for searching the minimum and maximum of the set of conformal values, we propose an alternative approach in this work. Because our interest lies in finding a prediction interval rather than a set of conformal values (that may be disjoint), the computation of the bounds of $\cna$ is reformulated as a root-finding problem. The equations solved for the lower and upper bounds of the prediction interval read: 

\begin{align}
    \textrm{BC}^+(y)\eqdef y-\hat{\cm}^y(\ve{x}_{n+1})-\hat{q}_{n,\alpha/2}^+\prt{\acc{y_i-\hat{\cm}^y(\ve{x}_i),\,i=1\enu n}}=0,\label{eq:4}\\
    \textrm{BC}^-(y)\eqdef y-\hat{\cm}^y(\ve{x}_{n+1})-\hat{q}_{n,\alpha/2}^-\prt{\acc{y_i-\hat{\cm}^y(\ve{x}_i),\,i=1\enu n}}=0~.\label{eq:5}
\end{align}
These two equations are solved using a root-finding algorithm, which may vary depending on the surrogate model formulation, or can be solved analytically in the case of full PCE, as detailed in the following. Each iteration of the search algorithm requires the training of the surrogate $\hat{\cm}^{y}$ and the computation of quantiles $\hat{q}^-_{n,\alpha/2}$ and $\hat{q}^+_{n,\alpha/2}$.

\subsection{Split conformal prediction}
As mentioned in the previous section, full conformal predictions rely on numerous model trainings, which is expected to result in a high computational cost. Split conformal prediction has been proposed to alleviate this burden. The main idea is to split the available data into a training set $\mathcal{S}^{\textrm{train}}$, used to construct the surrogate model $\hat{\cm}$, and a calibration set $\Scal$, in order to evaluate statistics of the surrogate model over unseen data. \Cref{eq:6} defines the prediction interval for the symmetrical split conformal setting:
\begin{align}
\begin{split}
\label{eq:6}
    \cnasplit(\ve{x}_{n+1})&=\hat{\cm}(\ve{x}_{n+1})\pm\hat{q}_{\ncal,\alpha}^+(\mathcal{R}),\\
    \textnormal{with}~\mathcal{R}&=\acc{|y_i-\hat{\cm}(\ve{x}_i)|,\,(\ve{x}_i,y_i)\in \Scal}.
\end{split}
\end{align}
Similarly, the \textit{asymmetrical split conformal} prediction interval can be defined as 
\begin{align}
\label{eq:7}
    \cnasplit(\ve{x}_{n+1})&=[\hat{\cm}(\ve{x}_{n+1})+\hat{q}_{\ncal,\alpha/2}^-(\mathcal{R}^{\textrm{sgn}})~;~\hat{\cm}(\ve{x}_{n+1})+\hat{q}_{\ncal,\alpha/2}^+(\mathcal{R}^{\textrm{sgn}})],  \\
    \textnormal{with }\mathcal{R}^{sgn}&=\acc{y_i-\hat{\cm}(\ve{x}_i),\,(\ve{x}_i,y_i)\in \Scal}.
\end{align}
By construction, split conformal prediction intervals are guaranteed to provide at least $1-\alpha$ coverage, that is, at least $\lfloor(1-\alpha)\ncal\rfloor$ points from the calibration set belong to the split conformal prediction interval. The main issue with this method arises in a context where only few data is available. As a matter of fact, only part of the information would be used to build the surrogate model due to the data split. This can be deleterious to the quality of the model in data-scarce contexts. Thus, this method is of lesser interest for small-data configurations, which are recurrent in surrogate modeling problems, as opposed to standard machine learning setups. Nonetheless it will be used for comparison purposes in the sequel with toy-function models where a large validation set is affordable.

\subsection{Jackknife+}
\label{sec:jackknife+}
An elegant alternative to full conformal prediction has been presented by \citet{Barber2021}. Analogously to the full conformal approach, this method requires fitting multiple surrogate models (in fact, $n+1$ models). First, let us define the leave-one-out surrogates $\hat{\cm}_{-i}$, trained over $\mathcal{S} \backslash (\ve{x}_i,y_i)$, that is the initial dataset but $(\ve{x}_i,y_i)$. The leave-one-out residuals are defined as $\riloo=\abs{y_i-\hat{\cm}_{-i}(\ve{x}_i)}$. In the symmetrical setting, the \textit{Jackknife+} prediction interval is constructed as follows: 

\begin{equation}
\label{eq:7b}
\begin{split}
    \cna^{J+}(\ve{x}_{n+1}) = \left[ \hat{q}_{n,\alpha}^-\left(\acc{\hat{\cm}_{-i}(\ve{x}_{n+1})-\riloo,\, i=1\enu n}\right)\right.~,\\
    ~\left.\hat{q}_{n,\alpha}^+\left(\acc{\hat{\cm}_{-i}(\ve{x}_{n+1})+\riloo,\, i=1\enu n}\right) \right]. 
\end{split}
\end{equation}

According to \cite{Barber2021}, this construction yields a theoretically guaranteed coverage of level $(1-2\alpha)$, that is,  
\begin{equation}
    \Prob{\cm(\ve{X}_{n+1})\in \cna^{J+}(\ve{X}_{n+1})}\geq 1-2\alpha, 
\end{equation}
whereas the empirically observed coverage usually tends to be close to the prescribed level $(1-\alpha)$. 

The asymmetric formulation of the \textit{Jackknife+} confidence interval is obtained considering quantiles of the signed leave-one-out residuals $\riloosgn=y_i-\hat{\cm}_{-i}(\ve{x}_i)$, as described in \Cref{eq:8}. The exact same theoretical guarantee still holds.  
\begin{equation}
\label{eq:8}
\begin{split}
    \cna^{J+}(\ve{x}_{n+1}) = \left[ \hat{q}_{n,\alpha/2}^-\left(\acc{\hat{\cm}_{-i}(\ve{x}_{n+1})+\riloosgn,\, i=1\enu n}\right)\right.~,\\
    \left.\hat{q}_{n,\alpha/2}^+\left(\acc{\hat{\cm}_{-i}(\ve{x}_{n+1})+\riloosgn,\, i=1\enu n}\right) \right].
\end{split}
\end{equation}

It should be noted that in the asymmetrical setting, other values $\alpha_{lb}\neq\alpha/2$ and $\alpha_{ub}\neq\alpha/2$ can be chosen as long as $\alpha_{lb}+\alpha_{ub}=\alpha$. However, this choice is conditional on the distribution of the residuals and requires manual parameter tuning, which is not considered in this study.

\subsection{Coverage of prediction intervals}
\label{sec:coverage}

One key idea about conformal prediction, as detailed by \citet{Angelopoulos2023}, is that the coverage of conformal prediction intervals conditionally on the training set is a random quantity. Therefore, the theoretical guarantee in \Cref{eq:cov_guarantee} holds on average when repeating the conformal prediction procedure over multiple different training sets. This property is referred to as \emph{marginal coverage}. Therefore, the achieved coverage should be represented in terms of distribution in order to compare methods with each others. Moreover, the theoretical distribution of coverage for the split conformal procedure over different calibration sets follows a Beta distribution, as introduced by \citet{Vovk2012}. The distribution of coverage conditionally on the training set $\mathcal{S}$ for an infinite validation set on which the empirical coverage $1-\hat{\alpha}$ is computed, referred to as \emph{training conditional coverage}, is expressed by: 
\begin{equation}
   1-\hat{\alpha}\thicksim \textnormal{Beta}(\ncal+1-l,l),
\end{equation}
where
\begin{equation}
    l=\lfloor(\ncal+1)\alpha \rfloor~.
\end{equation}
Besides, the effect of the finiteness of the validation set can also be captured analytically, leading the observed coverage $1-\hat{\alpha}$ to follow a so-called Beta-Binomial distribution, as noted by \cite{Angelopoulos2023}. More precisely, the observed coverage is the average of Beta-distributed indicator functions, corresponding therefore to a binomial random variable with a Beta-distributed parameter: 
\begin{equation}
    1-\hat{\alpha}\thicksim \frac{1}{\nval}\textnormal{Binom}(\nval,\xi)~\textnormal{where}~\xi \thicksim \textnormal{Beta}(\ncal+1-l,l).
\end{equation}
However, for a large enough validation set with $\nval\geq10^6$, the effect of the binomial part of the distribution is negligible and the distribution of empirical coverage rapidly converges to the Beta distribution \citep{Angelopoulos2023}.

Furthermore, both Jackknife+ and full conformal methods offer marginal coverage guarantees. Training conditional coverage for these procedures can be established under assumptions of algorithmic stability \citep{Liang2025}. Notably, these algorithmic stability conditions are satisfied in our context for both linear regression (full PCE) and Lasso regression \citep{Xu2011}. However, the guaranteed coverage level is reduced by a factor depending on the setting-specific algorithmic behavior, which could be significant in scenarios where the number of training points and regressors are comparable. To assess the effect of weakened guarantees on the specific examples, we directly compare the marginal coverage distribution of full conformal prediction against the theoretical distribution of split conformal prediction with an equivalent calibration set size $\Tilde{n}_{\textrm{cal}}$ in the sequel, for the same size of validation set. This equivalent calibration set size is the one that, with the split conformal method, yields the same distribution of marginal coverage as observed with the full conformal procedure.

\section{Polynomial chaos expansion}

Polynomial chaos expansion (PCE) is a well-established surrogate modeling technique for uncertainty quantification purposes \citep{Xiu2002}. Consider again the function defined in \Cref{eq:1}, where the input vector $\ve{X}$ is an $M$-dimensional random vector with mutually independent components and probability density function $f_{\ve{X}}$. Given the Hilbert space of functions with finite variance under $f_{\ve{X}}$, a basis of orthogonal polynomials can be constructed, say $\{\psi_{\ve{\alpha}}\::\ve{\alpha}~\in~\mathbb{N}^M\}$. The construction of the polynomial basis functions is based on a product of univariate polynomials, leveraging the independence property of the components of $\ve{X}$. In this way, the multi-index $\ve{\alpha}$ denotes the considered degrees of each univariate polynomial (for more details, see \citet{Marelli2021} for an efficient implementation in UQLab \citep{MarelliICVRAM2014}).

Let $Y=\cm(\ve{X})$ be the random variable obtained by propagating the uncertainty represented by $f_{\ve{X}}$ through the computational model $\cm$. Provided that $Y$ has finite variance, we could represent it as: 
\begin{equation}
    Y = \sum_{\ve{\alpha}\in \mathbb{N}^M} c_{\ve{\alpha}}\psi_{\ve{\alpha}}(\ve{X}),
\end{equation}
where the series converges in the mean-square sense. A surrogate model $\hat{\cm}$ is thus obtained by considering a truncated series: 
\begin{equation}
    \hat{\cm}(\ve{x}) = \sum_{\ve{\alpha}\in \mathcal{A}} c_{\ve{\alpha}}\psi_{\ve{\alpha}}(\ve{x}). 
\end{equation}

The associated coefficients $\acc{c_{\ve{\alpha}},\,\ve{\alpha}\in \mathcal{A}}$ may be computed by ordinary least squares \citep{Berveiller2006}. Given the available data $\mathcal{S}=\acc{(\ve{x}_i,y_i),\,i=1\enu n}$, referred to as the experimental design (ED) in the uncertainty quantification literature, and the enumerated multi-indices in $\mathcal{A}$ as $\acc{\ve{\alpha}_j,\,j=0\enu P-1}$, one can build the column vector of model responses $\yed=(y_1\enu y_n)\tr$ and the regression matrix $\ve{\Psi}\in\mathbb{R}^{n\times P}$ with entries $\ve{\Psi}_{ij}=\psi_{\ve{\alpha}_j}(\ve{x}_i)$. Denote by $\ve{c}=(c_{\ve{\alpha}_1}\enu c_{\ve{\alpha}_P})\tr$ the column vector of PCE coefficients. Minimizing the mean-square error between the PCE predictions and the true model responses $\yed$ on the experimental design leads to the minimization problem: 
\begin{equation}
\label{eq:18}
    \hat{\ve{c}} = \min_{\ve{c}\in\mathbb{R}^P} \norm{\ve{\Psi}\ve{c}-\yed}_2^2.
\end{equation}
This is an ordinary least-squares (OLS) problem, whose solution reads: 
\begin{equation}
\label{eq:ols_sol}
    \hat{\ve{c}} = (\ve{\Psi}\tr\ve{\Psi})^{-1}\ve{\Psi}\tr\yed
\end{equation}

Because only a few regressors contribute in a remarkable manner to the model's response, the solution is expected to be sparse. This is a notable characteristic of most physical problems, as theorized by \citet{Montgomery:2004} as the sparsity-of-effects phenomenon. Relying on this postulate, as shown by \citet{Candes_sparsity}, sparse regression methods can find robust solutions to underdetermined systems of linear equations. The problem is reformulated using a penalty term: 
\begin{equation}
\label{eq:19}
    \hat{\ve{c}}=\min_{\ve{c}\in\mathbb{R}^P} \norm{\ve{\Psi}\ve{c}-\yed}_2^2+\lambda \norm{\ve{c}}_1
\end{equation}

This minimization problem, referred to as $\ell_1$-constrained regression can be solved using various algorithms, including least angle regression (LARS) \citep{Efron2004,BlatmanJCP2011}, orthogonal matching pursuit (OMP) \citep{Tropp2007,Doostan2011}, subspace pursuit (SP) \citep{Dai2009,Diaz2018} or Bayesian compressive sensing (BCS) \citep{Tipping2001,Ji2008,Sargsyan2014,Tsilifis2020} to name only a few. Interested readers are referred to the review papers of \citet{LuethenSIAMJUQ2021,LuethenIJUQ2022} for a detailed comparison of their performance. Note that, in the context of PCE, the first sparse solver has been adapted from LARS by \citet{BlatmanThesis,BlatmanJCP2011}. As the LARS solver provides a set of sparse solutions, i.e., with an increasing number of regressors, validation-based model selection is required, which can be achieved using leave-one-out cross-validation. Once the best sparse basis has been formed, the corresponding coefficients may be estimated by OLS: this procedure is referred to as \textit{Hybrid LARS} in the literature.

\section{Comparison framework}

The scope of this paper is to present algorithms to efficiently apply conformal predictors to (sparse) polynomial chaos expansions. Moreover, we aim to evaluate the quality of the local prediction intervals offered by the conformal predictors presented in Section~2, namely the full conformal and the Jackknife+, in a comprehensive way. We also compare our results with the ones obtained using bootstrap \citep{Efron1979}. Albeit lacking non-asymptotical theoretical guarantees of coverage, bootstrap remains a commonly used technique in distribution-free error estimation for uncertainty quantification. In order to allow absolute comparison between prediction interval sizes, a reference case is also defined using split conformal prediction.

\subsection{Bootstrap}
Different variations of bootstrap applied to regression problems are available in the literature. Some of them consider bootstrapping the observed residuals over the training set to build new surrogates and evaluate quantiles of a forecast error \citep{Stine1985}. We consider approach of this paper is trajectory-based, as introduced by \cite{MarelliSS2018}, and relies on the construction of $B$ bootstrap surrogate models $\hat{\cm}^b,b\in[1,B]$ using $B$ specific training sets \emph{resampled with replacement} from the original data set $\mathcal{S}$. As described in \Cref{eq:9}, prediction intervals are associated with quantiles of point predictions from the $B$ bootstrap surrogates, denoted by $\hat{q}_{B,\alpha/2}^{\pm }$. This method is referred to as \textit{percentile bootstrap}. 
\begin{equation}
\label{eq:9}
\begin{split}
    \cna^B(\ve{x}_{n+1})=\left[\hat{q}_{B,\alpha/2}^-\left(\acc{\hat{\cm}^b(\ve{x}_{n+1}),\,b=1\enu B}\right)\right., \\ \left. \hat{q}_{B,\alpha/2}^+\left(\acc{\hat{\cm}^b(\ve{x}_{n+1}),\,b=1\enu B}\right) \right]
\end{split}
\end{equation}

\subsection{Reference case}
The \textit{reference case} is designed so as to get trustworthy values of prediction intervals by leveraging the availability of a large number of data points for the considered analytical models. Here, we use split conformal prediction with an ``overkill'' calibration set of size $\ncal=10^6$, whereas the surrogate model $\hat{\cm}$ is built on $\ned$ data points. The advantage of this approach is that the coverage is only affected by the finiteness of the validation set (see discussion at the end of Section~2.5). The obtained width of the prediction interval provides a reasonable value for comparison with other methods. The formal expression of the reference case reads
\begin{align}
\label{eq:20}
    \cna^{\textrm{ref}}(\ve{x}_{n+1})&=\hat{\cm}(\ve{x}_{n+1})\pm\hat{q}_{\ncal,\alpha}^+(\mathcal{R}_i), \\
    \textnormal{where }\mathcal{R}_i&=\acc{|y_i-\hat{\cm}(\ve{x}_i)|,\,i=1\enu \ncal}.
\end{align}

\subsection{Comparison metrics}
\label{subsec:metrics}
The quality of the established prediction intervals is analyzed in terms of coverage and width. While a high coverage is always desirable, a large interval width may become useless for making informed decisions. The width of the interval should also be proportional to the quality of the regression procedure. First, the performance of the prediction intervals is evaluated on a large validation set $\Sval$ of size $\nval= 1{,}000$. The empirical coverage $1-\hat{\alpha}$ is defined as 
\begin{equation}
\label{eq:11}
    1-\hat{\alpha} = \frac{\sum_{j=1}^{\nval}\mathds{1}(y_j \in \hat{C}_{n,\alpha}(\ve{x}_j))}{\nval}. 
\end{equation}

Moreover, according to \citet{Demay2021}, the comparison between the target coverage and the empirical coverage for a given set of levels $\acc{\alpha_i\in(0,1),\,i=1\enu K}$ indicates the overall quality of the prediction intervals. These associated quantities are often presented in the form of a $(1-\alpha,1-\hat{\alpha})$ plot. For most uncertainty quantification problems and decision-making settings, high coverage levels are desired. Therefore, in the subsequent comparisons, we select the following target coverage levels:
\begin{equation}
    1-\alpha=\acc{0.5~,~0.6~,~0.7~,~0.8~,~0.85~,~0.9~,~0.95}.
\end{equation}

As detailed above, provided that the obtained coverage level is satisfactory, the width of the constructed prediction intervals is a second proxy for their quality. Taking a step back, another notable feature is the variation of interval widths over the input parameter space. Indeed, when building a surrogate model, the amount of available information about the model at a specific location is directly related to the distance from this specific point to the neighboring points of the training set. Therefore, ideally, confidence intervals should be narrower close to training points and wider in regions where less information is available as it is a built-in feature for Gaussian process surrogate models \citep{Santner2003}. To assess this desired property, we present the prediction interval width as a function of the distance to the closest point in the training set in the subsequent analyses.

Quantifying the accuracy and predictive quality of the constructed metamodel $\hat{\cm}$ usually relies on the estimation of the (relative) generalization error of the form: 
\begin{equation}
\label{eq:gen_error}
    \errgen = \frac{\mathbb{E}\left[ (\cm(\ve{X})-\hat{\cm}(\ve{X}))^2    \right]}{\textrm{Var}[\cm(\ve{X})]}. 
\end{equation}

In the case of an available validation set, the most natural generalization error estimate is the relative mean-square validation error, defined as

\begin{equation}
    \errval = \frac{\sum_{i=1}^{\nval}\left(\cm(\ve{x}_i)-\hat{\cm}(\ve{x}_i)\right)^2}{\sum_{i=1}^{\nval}\left(\cm(\ve{x}_i) - \hat{\mu}_{\yval} \right)^2}, 
\end{equation}
where $\hat{\mu}_{\yval}=\frac{1}{\nval}\sum_{i=1}^{\nval}\cm(\ve{x}_i)$ is the sample mean of the validation set. Leave-one-out, or any other cross-validation error, also provides estimates of the generalization error in the absence of an additional validation set. However, generalization error measures are global and do not provide local insight. Conformal prediction intervals, on the contrary, provide such a measure, as they describe the uncertainty in a point prediction. Therefore, and to be consistent with \Cref{eq:gen_error}, we introduce the \emph{normalized interval width} $\norm{\cna}^{*}$ as follows:

\begin{equation}
\label{eq:norm_width}
    \norm{\hat{C}_{n,\alpha}}^{*}=\frac{\norm{\hat{C}_{n,\alpha}}^2}{\textrm{Var}\left[\yval\right]},
\end{equation}
where $\norm{\hat{C}_{n,\alpha}}$ denotes the length of the interval and $\textrm{Var}\left[\yval\right]$ is the empirical variance of the model responses in the validation set supposed to be nonzero. As a consequence, the averaged value of the normalized interval width over the points of the validation set is analogous to the global generalization error estimates discussed above, allowing for direct comparison.  


\section{Conformal prediction for full PCE}

This section presents the specific application of conformal prediction to full PCE and introduces some computational adaptations that allow for drastically reducing the computational requirements. Its numerical performance is subsequently examined.

\subsection{Full conformal predictors}
\label{sec:full_conformal_fpce}
As described in Section 2.3, each new trial value $y_k^{\textrm{trial}}$ at the query point $\ve{x}_{n+1}$ requires to train the surrogate $\hat{\cm}^{y}$ based on the set $\mathcal{S}\cup (\ve{x}_{n+1},y_k^{\textrm{trial}})$. This construction is actually very similar to that of $\hat{\cm}$. In the regression setup, only a rank-1 update of the regression matrix is needed. Let us define the row vector of values of the basis polynomials at $\ve{x}_{n+1}$, denoted $\ve{\phi}(\ve{x}_{n+1})\eqdef \left( \psi_{\ve{\alpha}_0}(\ve{x}_{n+1})~\enu\psi_{\ve{\alpha}_{P-1}}(\ve{x}_{n+1}   \right)$. Therefore, the solution of the OLS problem in \Cref{eq:ols_sol} requires the inversion of $\mat{M}_{n+1}=\mat{\Psi}_{n+1}\tr\mat{\Psi}_{n+1}$, where 
\begin{equation}
    \mat{\Psi}_{n+1} = \begin{pmatrix}
\mat{\Psi} \\
\ve{\phi}(\ve{x}_{n+1})
\end{pmatrix}. 
\end{equation}

Denoting by $\mat{M}=\mat{\Psi}\tr\mat{\Psi}$, this can be simplified by using the Sherman-Morrison formula \citep{Sherman1950}: 
\begin{equation}
        \mat{M}_{n+1}^{-1} = \mat{M}^{-1}-\frac{\mat{M}^{-1}\ve{\phi}(\ve{x}_{n+1})\ve{\phi}\tr(\ve{x}_{n+1}) \mat{M}^{-1}}{1+\ve{\phi}(\ve{x}_{n+1})\tr\mat{M}^{-1}\ve{\phi}(\ve{x}_{n+1})} ~. 
\end{equation}
Finally, the OLS solution can be computed, and the surrogate built on the augmented data can be evaluated at a new point $\ve{x}$ as: 
\begin{equation}
\label{eq:17}
        \hat{\cm}^y(\ve{x}) = \ve{\phi}(\ve{x}) \ve{\hat{c}}^y~, 
\end{equation} 
where $\ve{\hat{c}}^{y} = \mat{M}_{n+1}^{-1}\mat{\Psi}_{n+1}\tr\ve{y}_{\textrm{ED}}^{y}$ with the column vector $\ve{y}_{\textrm{ED}}^y=\left(\ve{y}_{\textrm{ED}}\tr,y_k^{\textrm{trial}}\right)\tr$ being the augmented vector of observations. 

Moreover, following the formulation of \citet{Vovk2005}, we define the hat matrix $\mat{H}_{n+1}$ as: 
\begin{equation}
    \mat{H}_{n+1}=\mat{\Psi}_{n+1} \mat{M}_{n+1}^{-1}\mat{\Psi}_{n+1}\tr~,
\end{equation}
and the column vector of PCE predictions on $\mathcal{S}\cup (\ve{x}_{n+1})$ as $\hat{\cm}(\ve{x}_{\textrm{ED}}^{+1})$. The column vector of non-conformity scores $\ve{R}_{n+1}=\ve{y}_{\textrm{ED}}^y-\hat{\cm}(\ve{x}_{\textrm{ED}}^{+1})$ can be computed as 
\begin{equation}
    \ve{R}_{n+1} = (\mathbb{I}_{n+1}-\mat{H}_{n+1})\ve{y}_{\textrm{ED}}^{y}~, 
\end{equation}
where $\mathbb{I}_{n+1}$ is the identity matrix of size $(n+1)$. Thus, the residuals can be represented as a linear function of the trial value $y_k^{\textrm{trial}}$: 
\begin{equation}
    \ve{R}_{n+1} = \ve{A}_{n+1} + \ve{B}_{n+1}y_k^{\textrm{trial}}~,
\end{equation}
where 
\begin{equation}
    \ve{A}_{n+1} = (\mathbb{I}_{n+1}-\mat{H}_{n+1})\cdot(y_1\enu y_{n},0)\tr
\end{equation}
and 
\begin{equation}
    \ve{B}_{n+1} = (\mathbb{I}_{n+1}-\mat{H}_{n+1})\cdot(0\enu 0,1)\tr~.
\end{equation}

For the full conformal procedure, by denoting $\ve{R}_{n+1,i}$ the $i$-th element of the vector $\ve{R}_{n+1}$, \Cref{eq:4,eq:5} can be expressed as: 
\begin{align}
    \textrm{BC}^+(y_k^{\textrm{trial}})\eqdef \ve{R}_{n+1,n+1}-\hat{q}_{n,\alpha/2}^+\prt{\acc{\ve{R}_{n+1,i}~,\,i=1\enu n}}&=0 \label{eq:4_modif}\\
    \textrm{BC}^-(y_k^{\textrm{trial}})\eqdef \ve{R}_{n+1,n+1}-\hat{q}_{n,\alpha/2}^-\prt{\acc{\ve{R}_{n+1,i}~,\,i=1\enu n}}&=0~.\label{eq:5_modif}
\end{align}
According to the definition of quantiles, which are forced to be values of the set of residuals, there are only $n$ candidate values for the upper and lower bounds of the full conformal prediction intervals given by: 
\begin{equation}
    y_i^{\textrm{trial}} = \frac{\ve{A}_{n+1,i}-\ve{A}_{n+1,n+1}}{\ve{B}_{n+1,n+1}-\ve{B}_{n+1,i}},\,i=1\enu n .
\end{equation}
By computing the residuals for each of those trial values, the one corresponding to the solution of \Cref{eq:4_modif,eq:5_modif} can be easily identified. 

\subsection{Jackknife+}
As defined in \Cref{sec:jackknife+}, the Jackknife+ procedure requires the computation of the $n$ leave-one-out surrogate models $\hat{\cm}_{-i}$. According to \citet{BlatmanThesis}, analytical expressions are available in the context of OLS. Let us denote as before the row vector of the basis polynomials at $\ve{x}_i$ by $\ve{\phi}(\ve{x}_i)$, and define the regression matrix $\ve{\Psi}_{\backslash i}$, being the original regression matrix $\ve{\Psi}$ with the $i$-th row removed. Similarly, the vector of model observations with the $i$-th observation removed is represented by $\ve{y}_{\textrm{ED}\backslash i}$. Using this formalism, 
the computation of the regression coefficients $\hat{\ve{c}}^{i}$ of the $\hat{\cm}_{-i}$ surrogates becomes
\begin{equation}
     \ve{\hat{c}}^{i} = \mat{M}_{\backslash i}^{-1}\mat{\Psi}_{\backslash i} \ve{y}_{\textrm{ED}\backslash i}~,
\end{equation}
where 
\begin{equation}
    \mat{M}_{\backslash i}^{-1} = \mat{M}^{-1} + \frac{\mat{M}^{-1}\ve{\phi}(\ve{x}_i)\ve{\phi}(\ve{x}_i)\tr\mat{M}^{-1}}{1-\ve{\phi}(\ve{x}_i)\tr\mat{M}^{-1}\ve{\phi}(\ve{x}_i)}~.
\end{equation}
Thereby, the leave-one-out residuals can be computed efficiently as
\begin{equation}
     \riloosgn = \frac{\ve{y}_{\textrm{ED},i}-\ve{\phi}(\ve{x}_i)\ve{\hat{c}}}{h_i}~,
\end{equation}
where $h_i$ denotes the $i$-th diagonal element of the matrix $\mat{P}$ given by 
\begin{equation}
    \mat{P} = \mathbb{I}_{n} - \mat{\Psi}    (\mat{\Psi}\tr\mat{\Psi})^{-1}\mat{\Psi}\tr.
\end{equation}
Finally, the $i$-th leave-one-out surrogate evaluated at the query point $\ve{x}_{n+1}$ gives: 
\begin{equation}
     \hat{\cm}_{-i}(\ve{x}_{n+1}) = \ve{\phi}(\ve{x}_{n+1}) \ve{\hat{c}}^{i}~.
\end{equation}
As a result, the prediction intervals given in \Cref{eq:8} can be efficiently evaluated using the equations above.

\subsection{Results}
\label{sec:fpce_results}
This section presents the results of both conformal procedures applied to full PCE. For each case, the amount of training data is selected in such a way that the resulting surrogate models have rather moderate accuracy, a setting where the knowledge of uncertainty on point predictions is of particular interest. This is achieved by choosing experimental designs of moderate size compared to the problem dimensionality, a situation that is common in real case studies. Indeed, for the selected applications, a relative mean-square validation error $\errval = 10^{-6}$ would be easily affordable. However, the width of the conformal prediction intervals is so small in this setting that it is of little interest. 

Moreover, as described in \Cref{sec:coverage}, the observed coverage of conformal prediction is also affected by statistical uncertainty. Therefore, numerical experiments are replicated $\nr$~times in order to study also the distribution of the quantities of interest before drawing conclusions. For each replication, a new experimental design of $\ned$ points is drawn, and all procedures are applied to obtain prediction intervals at each point of the validation set of $\nval$ points. Therefore, the observed coverages $\acc{1-\hat{\alpha}_r,\,r=1\enu \nr}$ are computed for all procedures. Furthermore, for each replication $r$ and a target coverage level $1-\alpha=0.9$, the normalized interval widths as defined in \Cref{eq:norm_width} are computed for all points of the validation set: $\acc{ \norm{\hat{C}_{n,\alpha}(\ve{x}_{\textrm{val},i})}^{*}_r,\, i=1\enu \nval}$. For practical discussion, this set is summarized by its median $q_{50}( \norm{\hat{C}_{n,\alpha}}^*_{r})$, which provides a robust measure of the typical normalized interval width for replication $r$ and by its interquantile range 
$q_{90}( \norm{\hat{C}_{n,\alpha}}^*_{r})- q_{10}( \norm{\hat{C}_{n,\alpha}}^*_{r})$, which provides a measure of the spread of the distribution of normalized interval widths. Note that the latter is equal to zero for the reference case, as for one replication, the split conformal procedure provides only a single interval width for all validation points. Thus it will not be presented in the results.


\subsubsection{Ishigami function}
\label{sec:fpce_results_ishi}
The first example is the Ishigami function, described by an analytical 3-dimensional equation \citep{ishigami_function}: 
\begin{equation}
    f(x_1,x_2,x_3) = \sin(x_{1}) + a\,\sin^2(x_{2})+b\,x_{3}^4\sin(x_{1})
\end{equation}
with parameters $a=7$ and $b=0.07$ respectively. The input variables are uniformly distributed over $[-\pi, \pi]^3$, i.e., $X_i\thicksim \mathcal{U}(-\pi,\pi),\,i=1,2,3$ .

\begin{figure}[H] 
    \centering
    \begin{subfigure}[b]{0.49\textwidth}
        \centering
        \includegraphics[width=\textwidth]{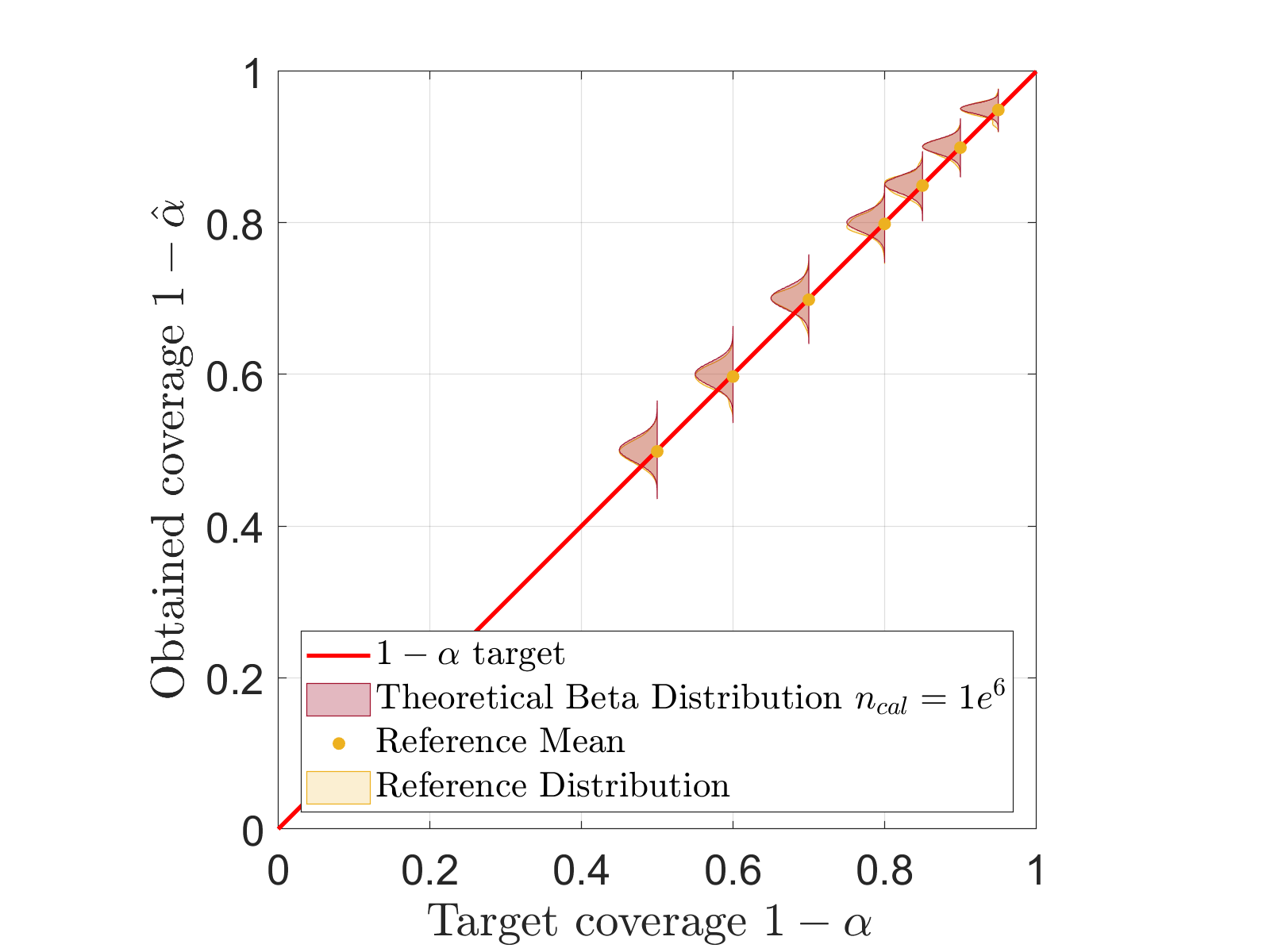}  
        \caption{Reference case}
        \label{fig:fig1_ref}
    \end{subfigure}
    \hfill
    \begin{subfigure}[b]{0.49\textwidth} 
        \centering
        \includegraphics[width=\textwidth]{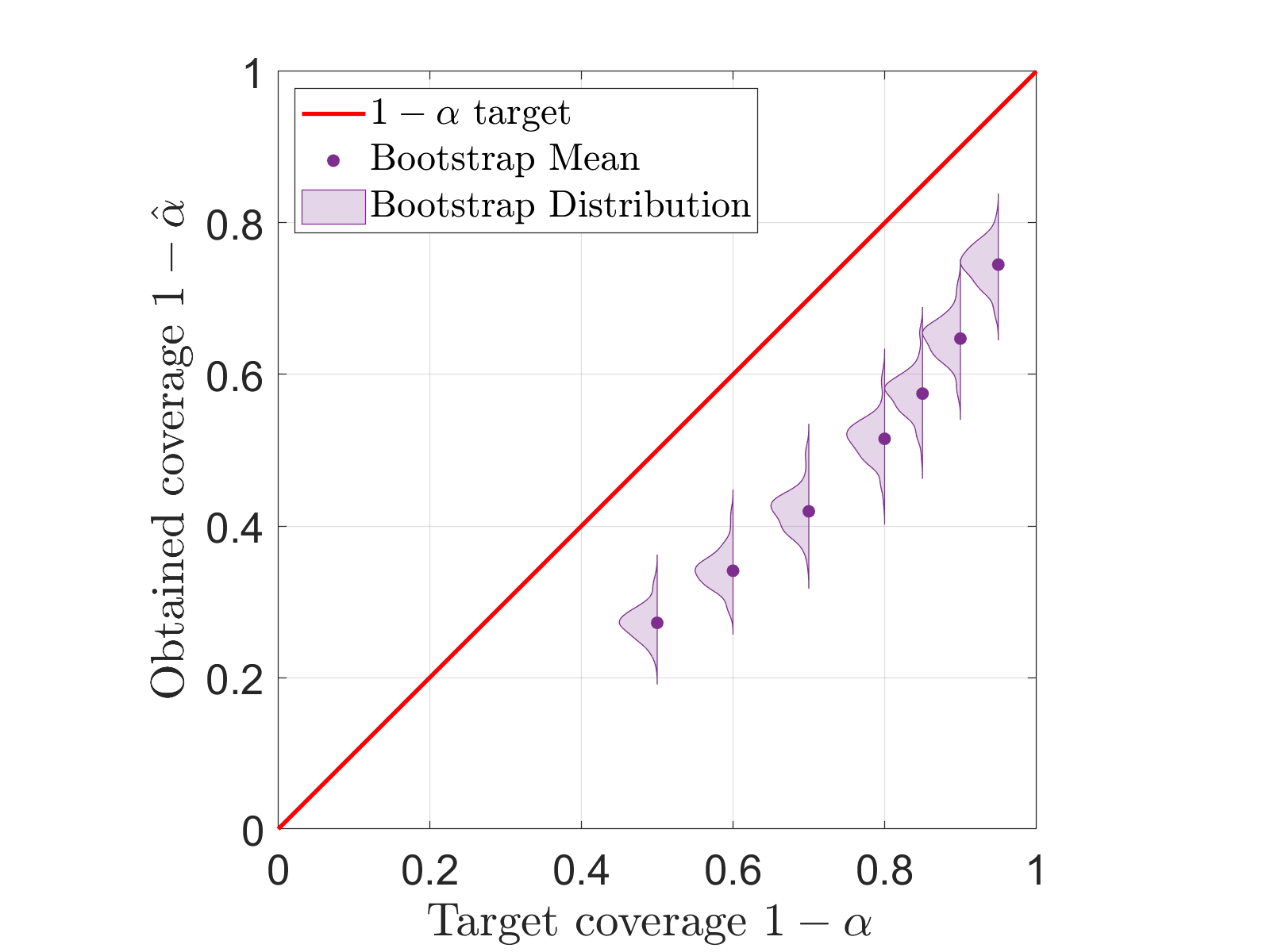}  
        \caption{Bootstrap}
        \label{fig:fig1_boot}
    \end{subfigure}
    \begin{subfigure}[b]{0.49\textwidth}
        \centering
        \includegraphics[width=\textwidth]{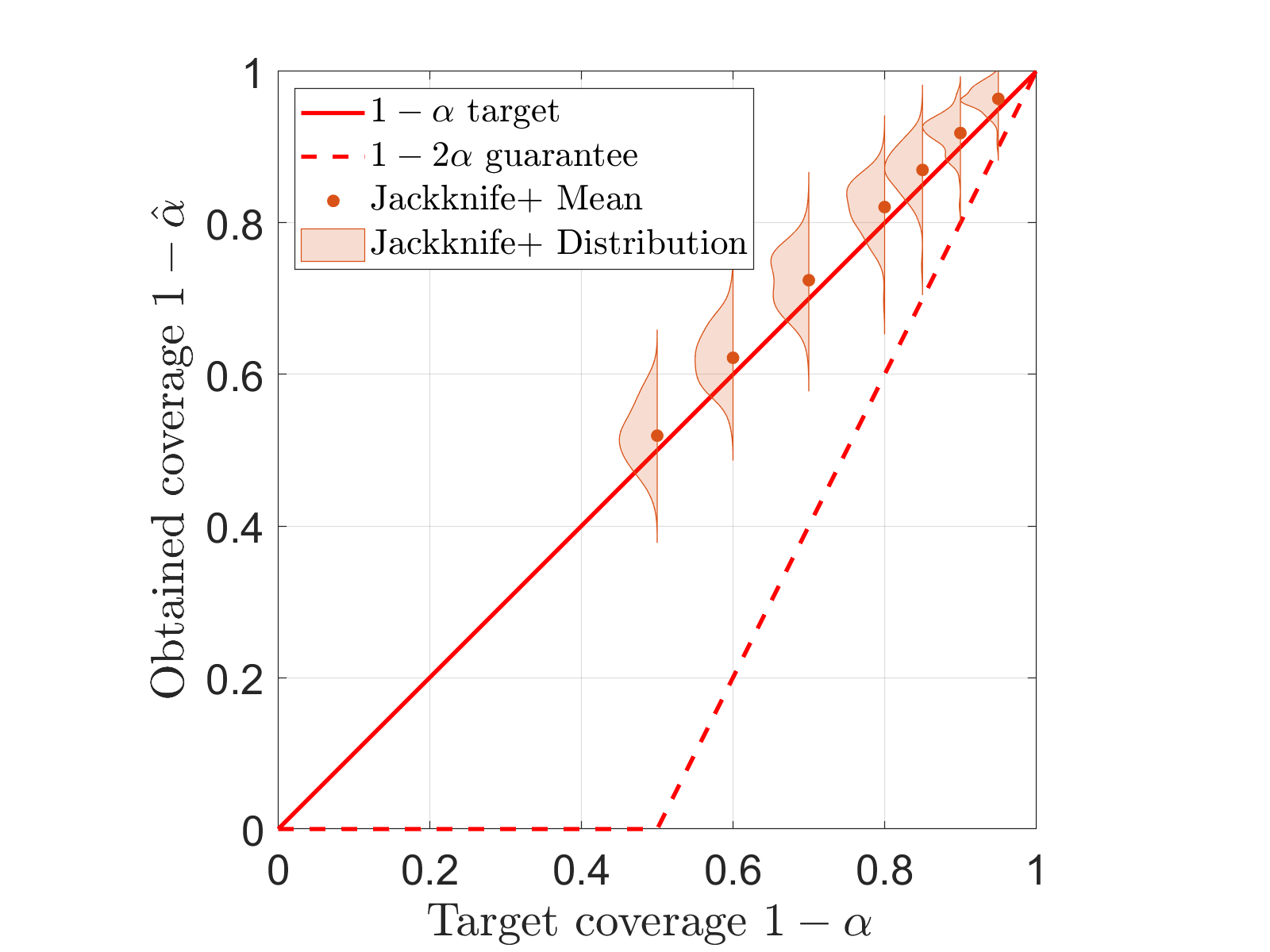} 
        \caption{Jackknife+}
        \label{fig:fig1_j+}
    \end{subfigure}
    \hfill
    \begin{subfigure}[b]{0.49\textwidth}
        \centering
        \includegraphics[width=\textwidth]{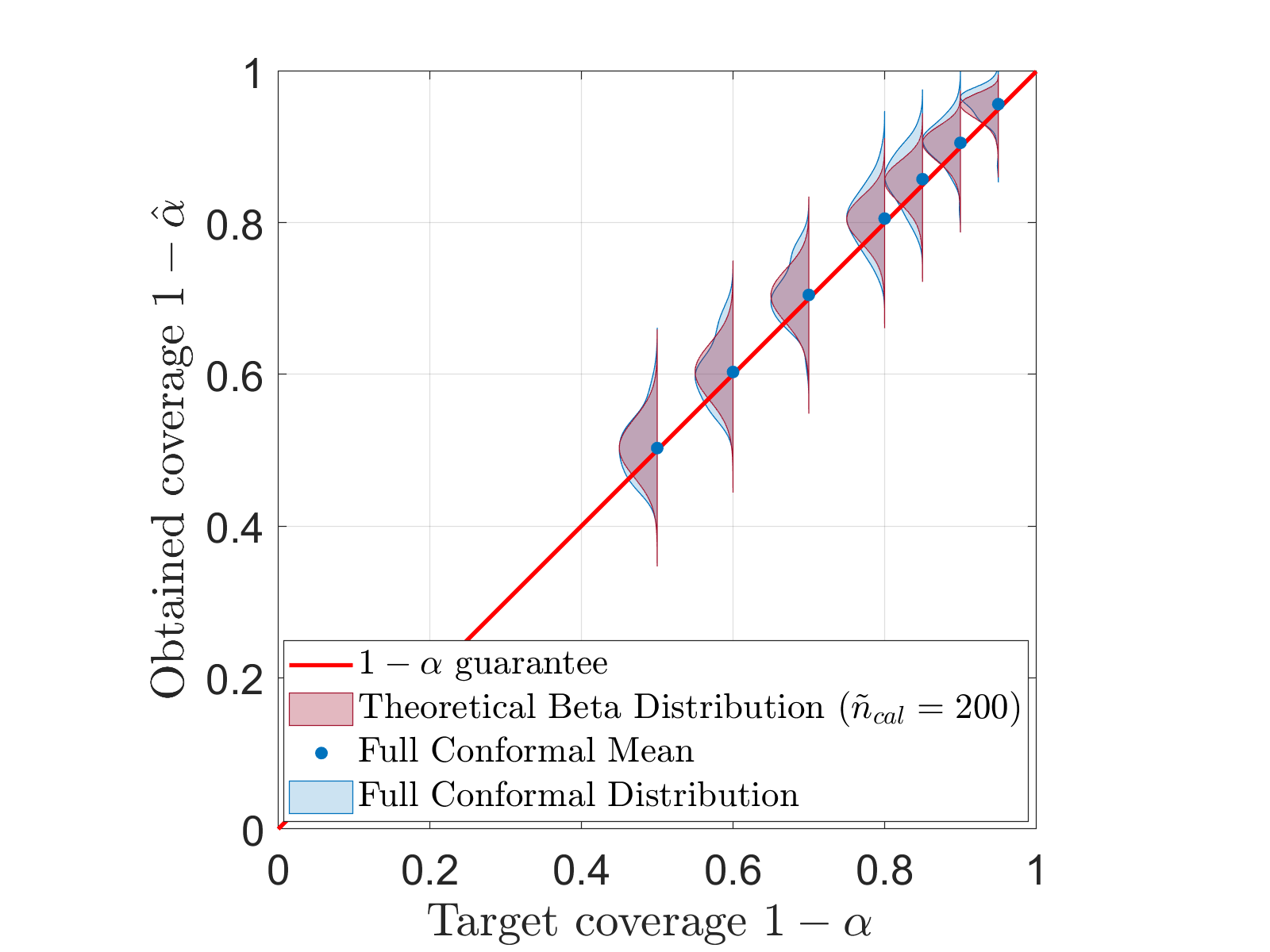}  
        \caption{Full conformal}
        \label{fig:fig1_fc}
    \end{subfigure}
    \caption{Evaluation of prediction intervals built based on the full PCE surrogate of Ishigami function. Settings of the experiment: $\ned=200$, $\nval=1{,}000$, PCE degree $p=5$, number of regressors $P=56$, $\errval\approx 2\cdot 10^{-1}$, $\nr=100$.}
    \label{fig:11}
\end{figure}

For this analysis, we chose experimental designs of $\ned$ points obtained by Latin Hypercube Sampling (LHS) for each replication and a PCE degree of $p=5$. The latter yields $P=56$ regressors. The OLS solution yields a relative mean-square error $\errval \approx 2\times 10^{-1}$ on a validation set of size $\nval=1{,}000$ points. For each replication,
prediction intervals are constructed using the different methods for each of the $\nval$ validation points. This process is repeated $\nr=100$ times, and the results for the coverage levels are presented in terms of both distribution and mean value in \Cref{fig:11}. 

First, \Cref{fig:fig1_ref} confirms that the theoretical (beta) distributions of the coverage (red) are identical to the obtained distributions of coverage for the reference split conformal procedure (yellow). For both full conformal and Jackknife+ prediction methods, the theoretical guarantees in terms of coverage are achieved as shown in \Cref{fig:fig1_fc,fig:fig1_j+}. The distribution of coverage of the full conformal method could be achieved by the split conformal method with an additional calibration set of size $\Tilde{n}_{\textrm{cal}}=200$. In the particular setting of this experiment, bootstrap (\Cref{fig:fig1_boot}) achieves a rather poor coverage, since the observed coverage (distribution of $\nr$ replications and mean value) is significantly lower than the target one. 

\begin{figure}[H] 
    \centering
    \begin{subfigure}[b]{0.49\textwidth}
        \centering
        \includegraphics[width=\textwidth]{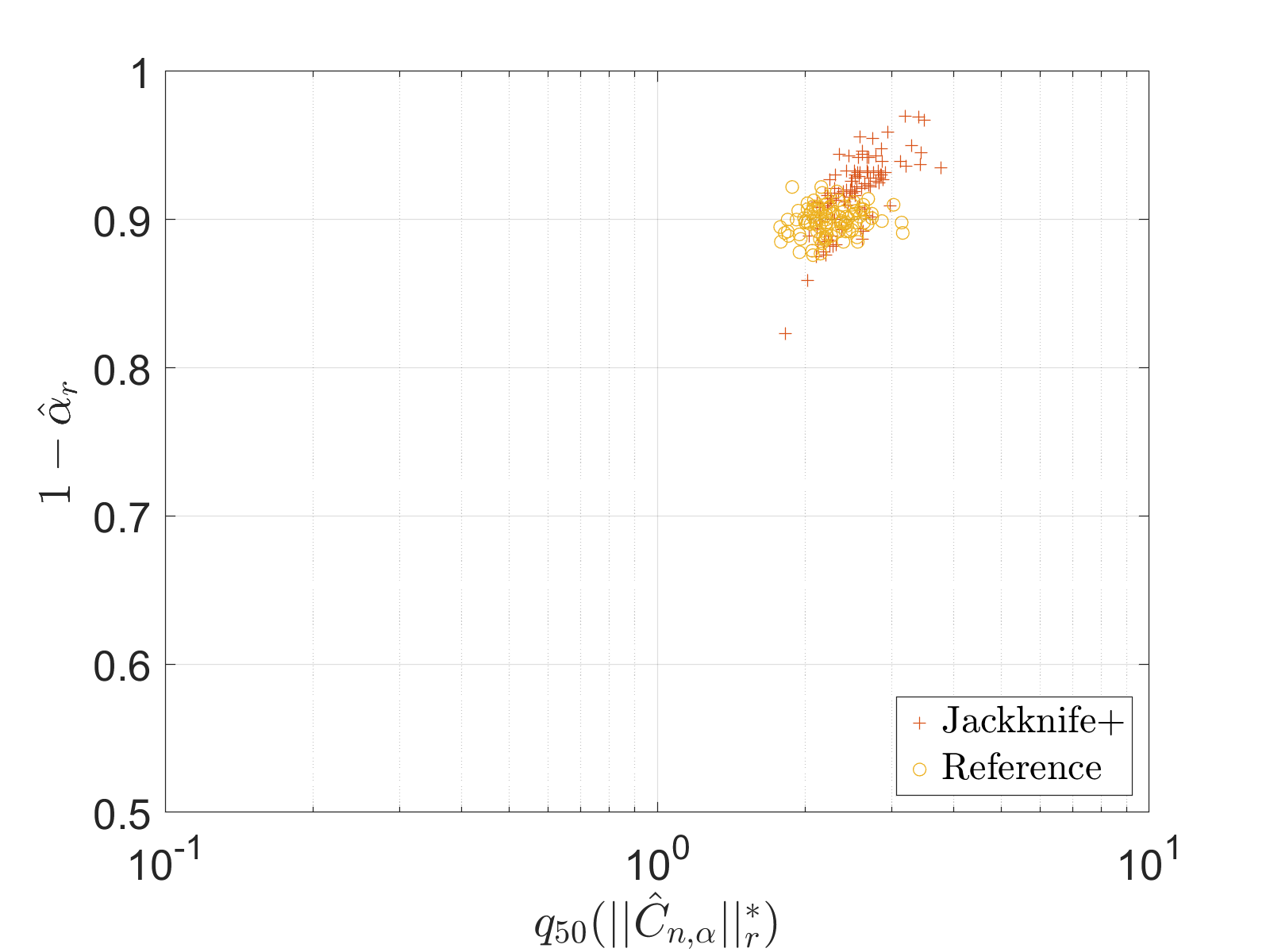} 
        \caption{Jackknife+}
        \label{fig:fpce_ishi_j+}
    \end{subfigure}
    \hfill
    \begin{subfigure}[b]{0.49\textwidth}
        \centering
        \includegraphics[width=\textwidth]{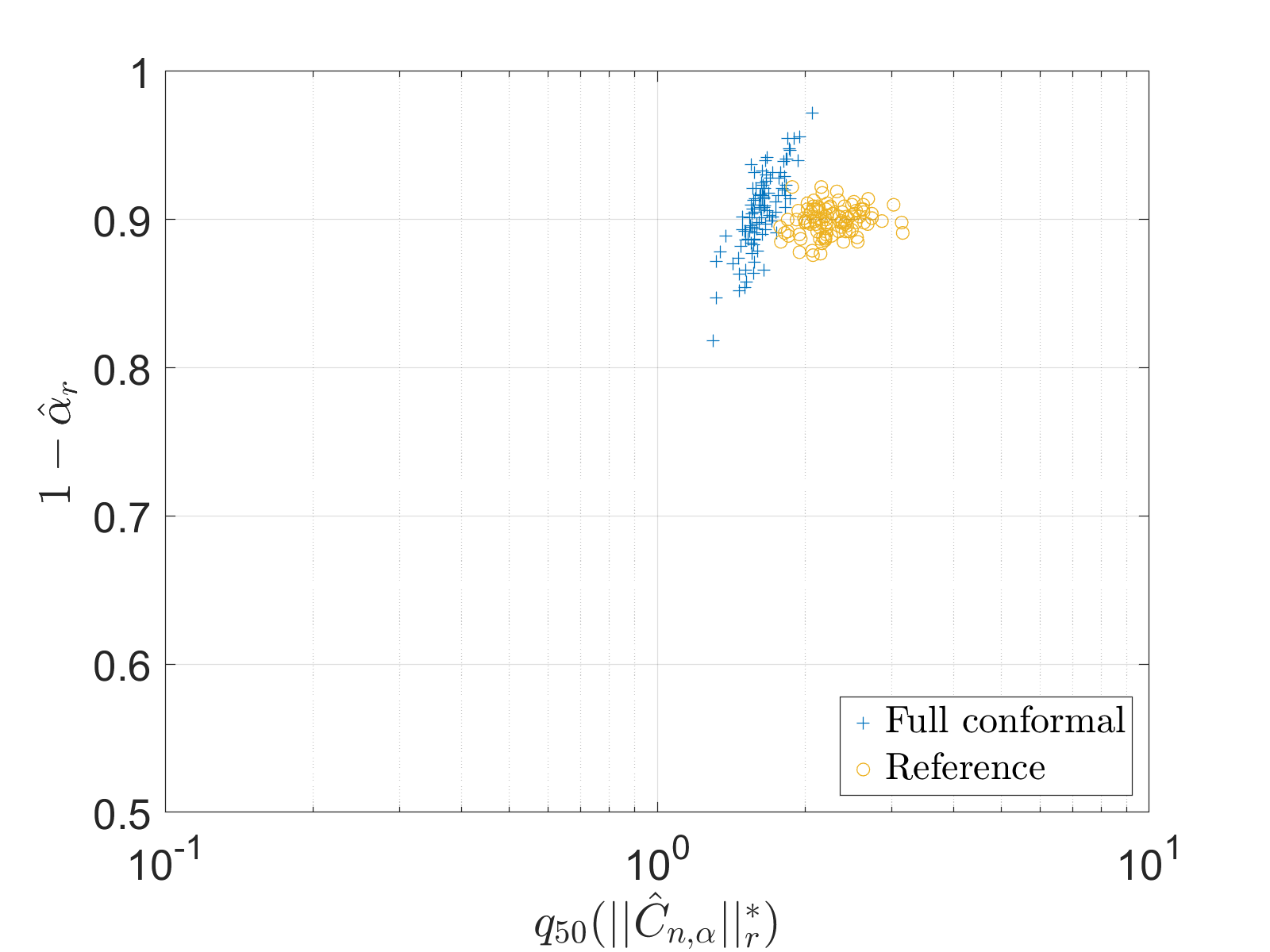}  
        \caption{Full conformal}
        \label{fig:fpce_ishi_fc}
    \end{subfigure}
    \begin{subfigure}[b]{0.49\textwidth} 
        \centering
        \includegraphics[width=\textwidth]{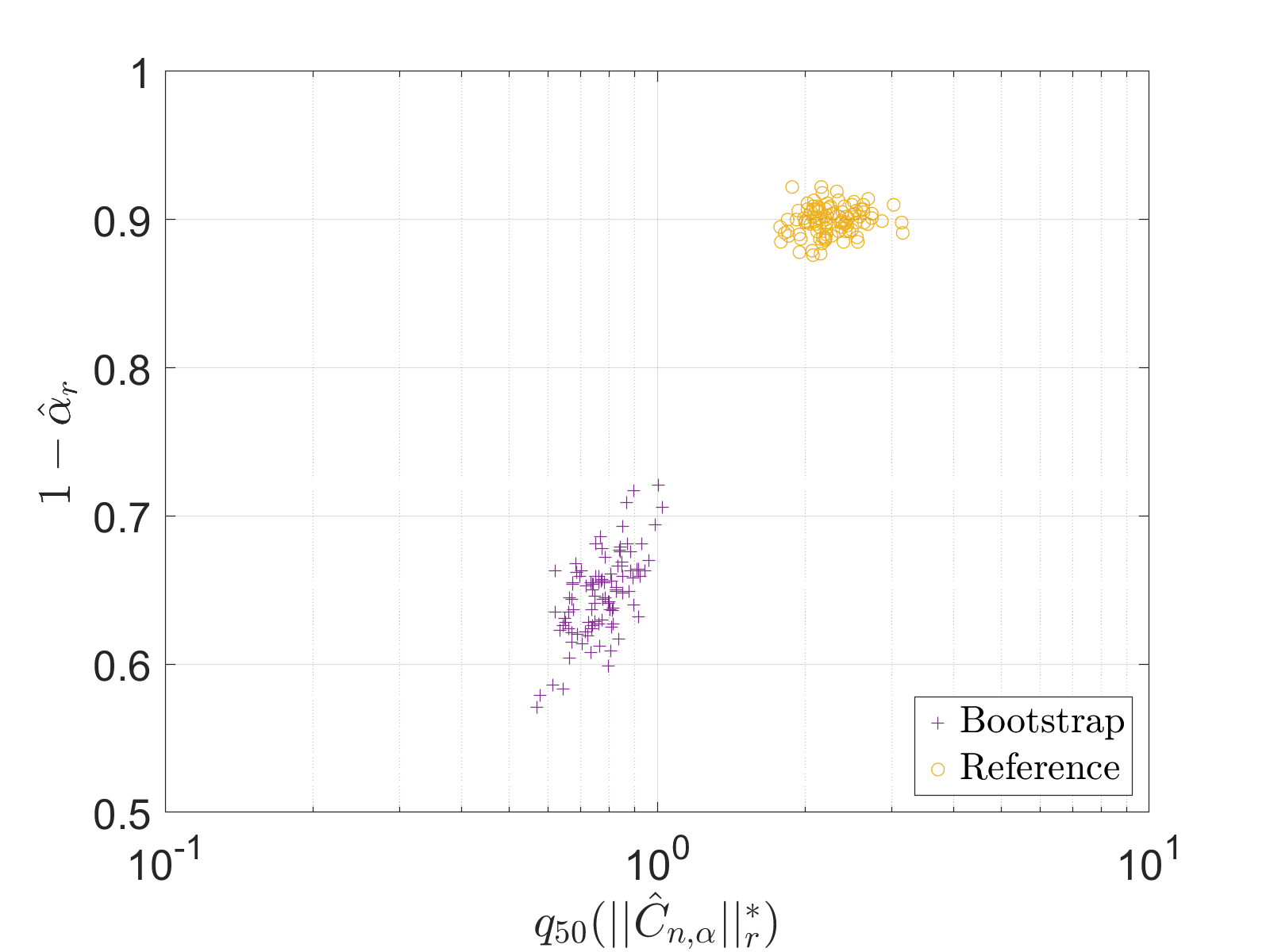}  
        \caption{Bootstrap}
        \label{fig:fpce_ishi_b}
    \end{subfigure}
    \begin{subfigure}[b]{0.49\textwidth} 
        \centering
        \includegraphics[width=\textwidth]{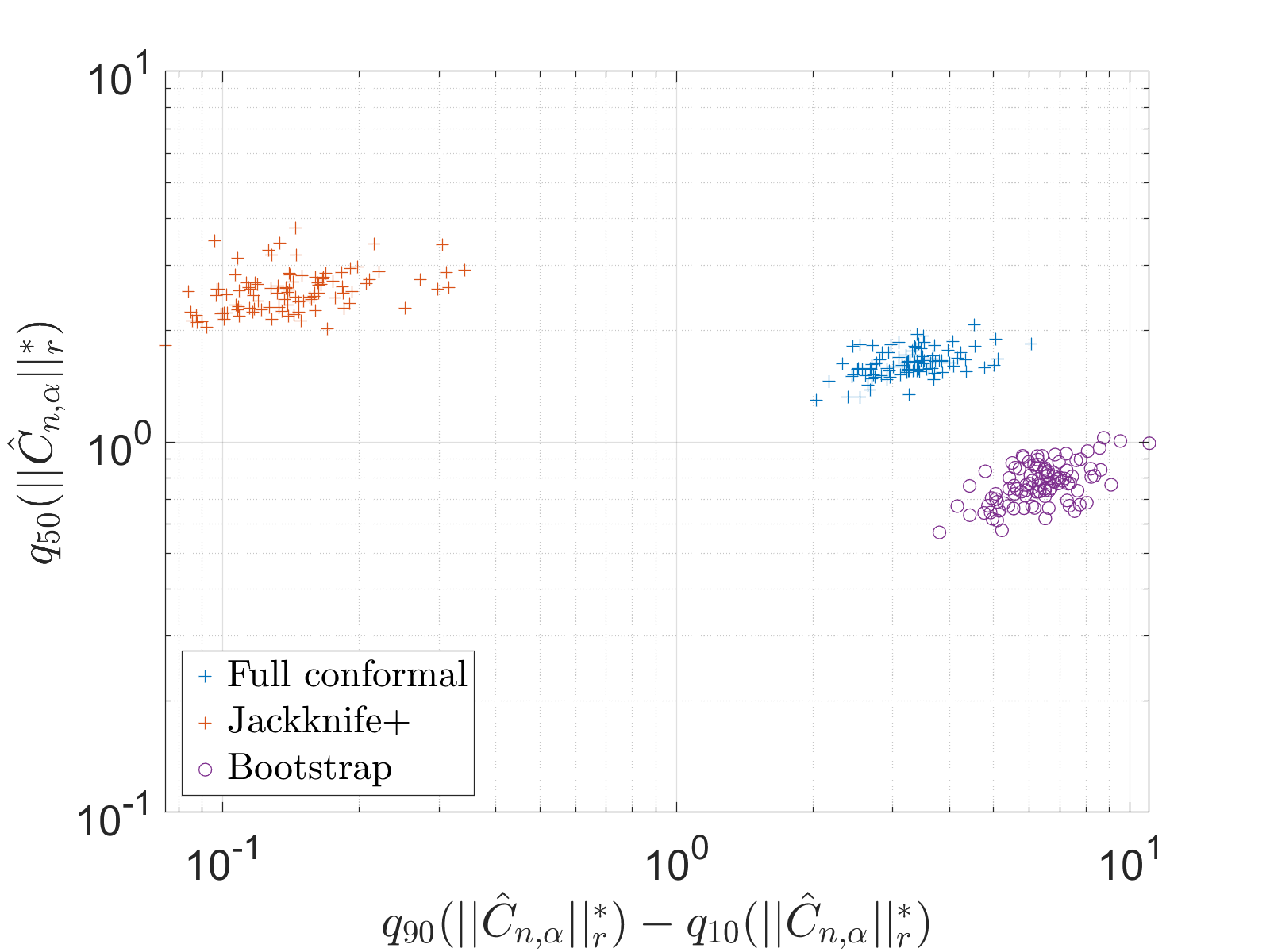}  
        \caption{Spread of prediction intervals widths}
        \label{fig:fpce_ishi_iq}
    \end{subfigure}

    \caption{Comparison of the normalized width of prediction intervals at target coverage level $1-\alpha=0.9$ built based on the PCE surrogate of Ishigami function. Settings of the experiment: $\ned=200$, $\nval=1{,}000$, PCE degree $p=5$, number of regressors $P=56$, $\errval\approx 2\cdot 10^{-1}$.}
    \label{fig:12}
\end{figure}

Furthermore, we present in \Cref{fig:12} the median value of normalized interval widths across replications for bootstrap, Jackknife+, and full conformal procedures, and compare them to the one obtained in the reference setting.   

Across multiple replications, the full conformal \Cref{fig:fpce_ishi_fc} and Jackknife+ (\Cref{fig:fpce_ishi_j+}) yield prediction intervals comparable in size to the reference setting and achieve satisfactory coverage. In contrast, bootstrap provides intervals that are generally smaller than the reference values (\Cref{fig:fpce_ishi_b}), which may explain why the target coverage is rarely achieved. \Cref{fig:fpce_ishi_iq} provides additional insight on the prediction interval widths. On average, Jackknife+ provides slightly larger intervals than full conformal, and significantly larger than bootstrap, as seen from the plot of the y-coordinate $q_{50}(\norm{\cna}^*_r)$. However, the spread of interval widths over the $\nval$ points is more than one order of magnitude smaller when using Jackknife+, compared to the spread observed for full conformal and bootstrap. In other words Jackknife+ provides larger, but more homogeneous (over the validation set) intervals.

\subsubsection{Borehole function}
\label{sec:fpce_results_bore}
The 8-dimensional Borehole function models water flowing through a borehole that is drilled from the ground surface through the two aquifers \citep{Harper1983}. It is typically used to benchmark surrogate models \citep{KersaudySudret2015,Morris1993,An:Owen:2001}. The analytical expression of the flow rate $Q$ is given by
\begin{equation}
    Q(r_w,r,T_u,H_u,T_l,H_l,L,K_w)=\frac{2\pi T_u(H_u-H_l)}{\ln(r/r_w)\left[1+\frac{2LT_u}{ln(r/r_w)r_w^2K_w}+\frac{T_u}{T_l}\right]}.
\end{equation}
The distribution and respective parameters of the random variables considered in the problem are summarized in \Cref{tab:1}. 
\begin{table}[H]
\centering
\small
\begin{tabular}{@{}lll@{}}
\toprule
{Variable} & {Distribution} & {Parameters} \\ \midrule
 $r_w$ & Gaussian & $\mu_{r_w}=0.10$ , $\sigma^2_{r_w}=0.0161812$ \\ 
 $r$ & Lognormal & $\lambda_r=7.71$ , $\xi_r=1.0056$ \\ 
 $T_u$ & Uniform & $T_{u,min}=63{,}070$ , $T_{u,max}=115{,}600$ \\ 
  $H_u$ & Uniform & $H_{u,min}=990$ , $H_{u,max}=1{,}100$ \\ 
 $T_l$ & Uniform & $T_{l,min}=63.1$ , $T_{l,max}=116$ \\ 
 $H_l$ & Uniform & $H_{l,min}=700$ , $H_{l,max}=820$ \\ 
 $L$ & Uniform & $L_{min}=1{,}120$ , $L_{max}=1{,}680$ \\ 
 $K_w$ & Uniform & $K_{w,min}=9{,}885$ , $K_{w,max}=12{,}045$ \\ \bottomrule
\end{tabular}
\caption{Distributions and the associated distribution parameters of the random variables for the Borehole function}
\label{tab:1}
\end{table}

\begin{figure}[H] 
    \centering
        \begin{subfigure}[b]{0.49\textwidth}
        \centering
        \includegraphics[width=\textwidth]{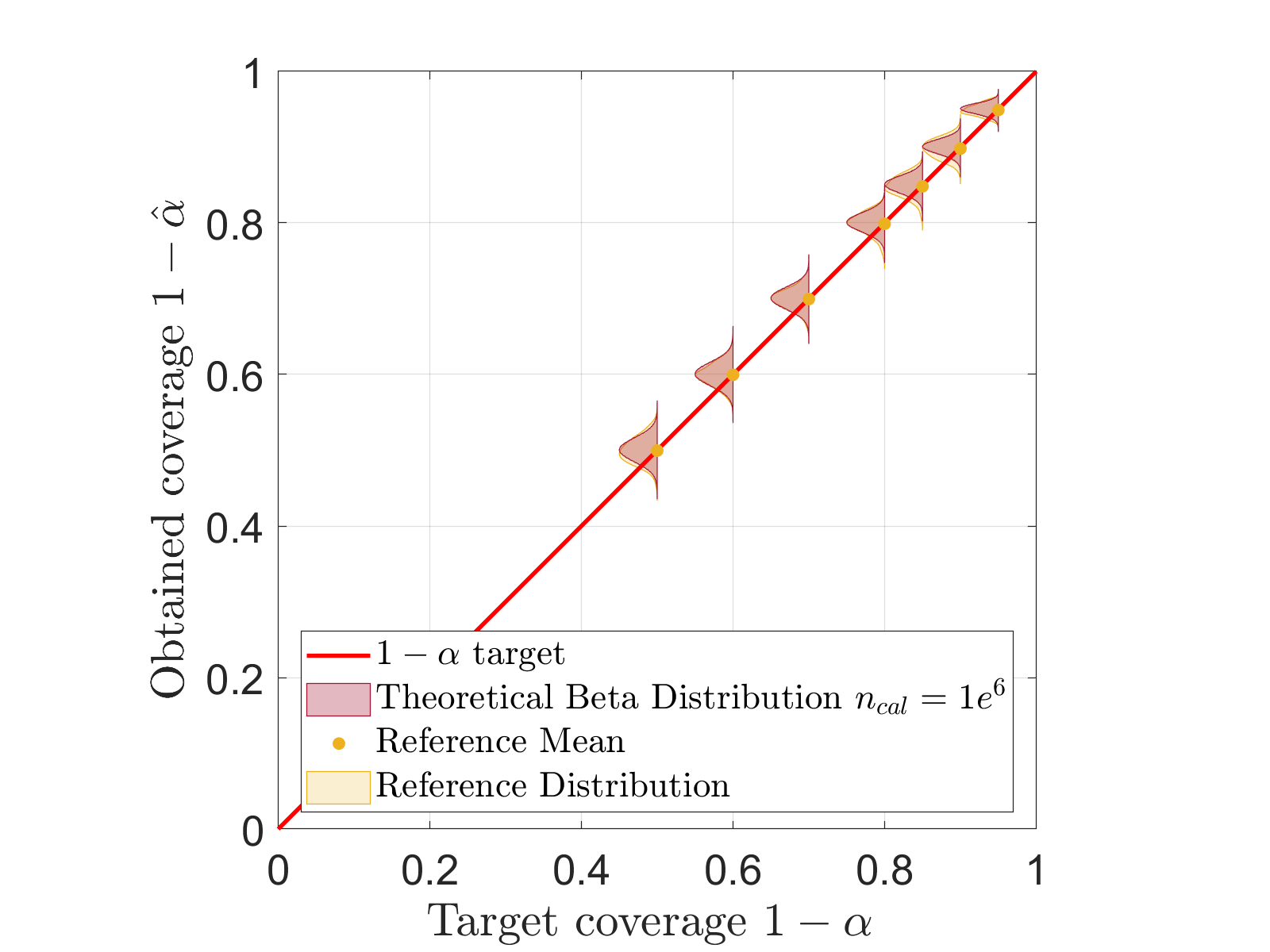}  
        \caption{Reference case}
        \label{fig:fpce_bore_ref_cov}
    \end{subfigure}
    \hfill
    \begin{subfigure}[b]{0.49\textwidth} 
        \centering
        \includegraphics[width=\textwidth]{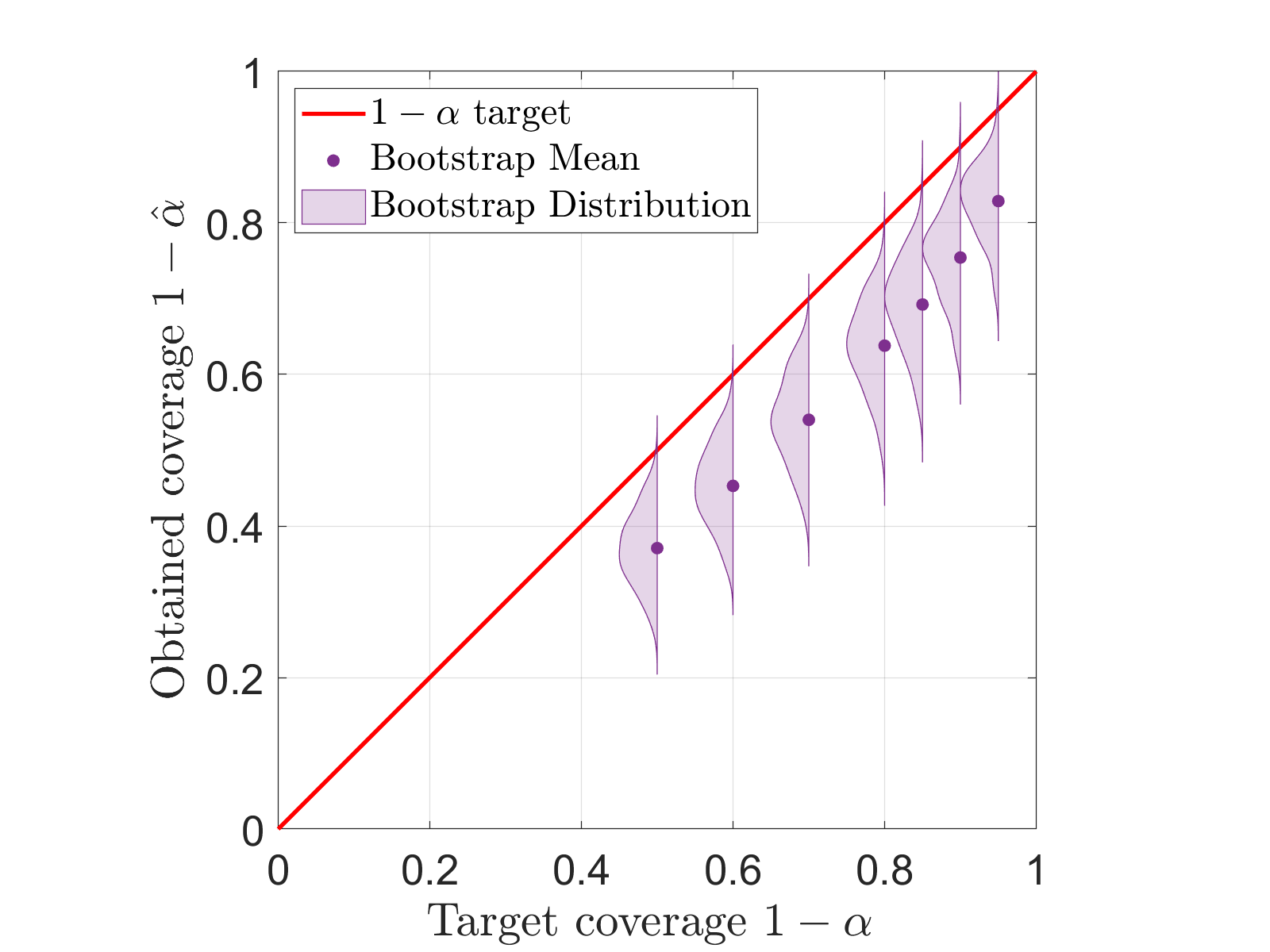}  
        \caption{Bootstrap}
        \label{fig:fpce_bore_b_cov}
    \end{subfigure}
    \begin{subfigure}[b]{0.49\textwidth}
        \centering
        \includegraphics[width=\textwidth]{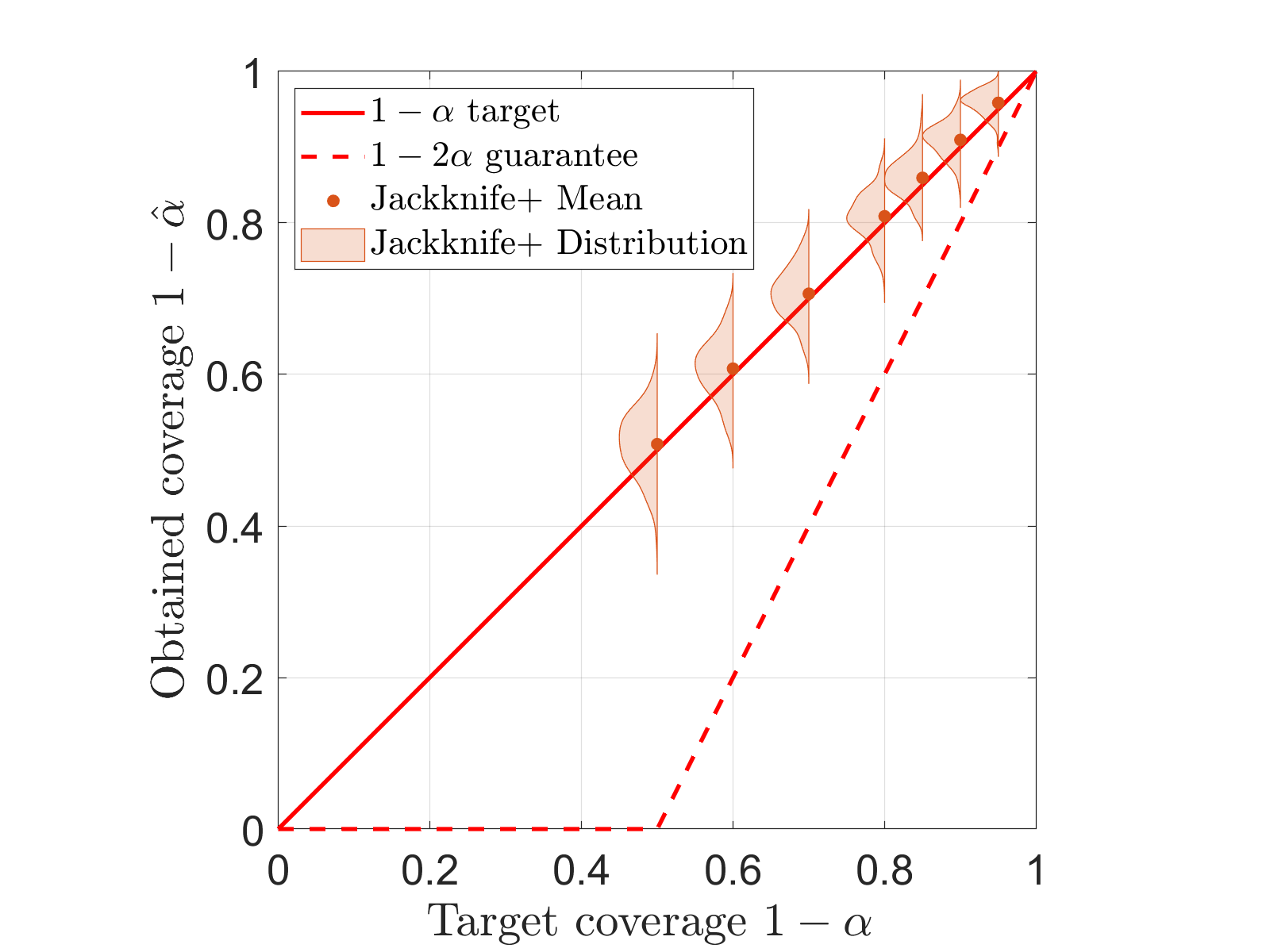} 
        \caption{Jackknife+}
        \label{fig:fpce_bore_j+_cov}
    \end{subfigure}
    \hfill
    \begin{subfigure}[b]{0.49\textwidth}
        \centering
        \includegraphics[width=\textwidth]{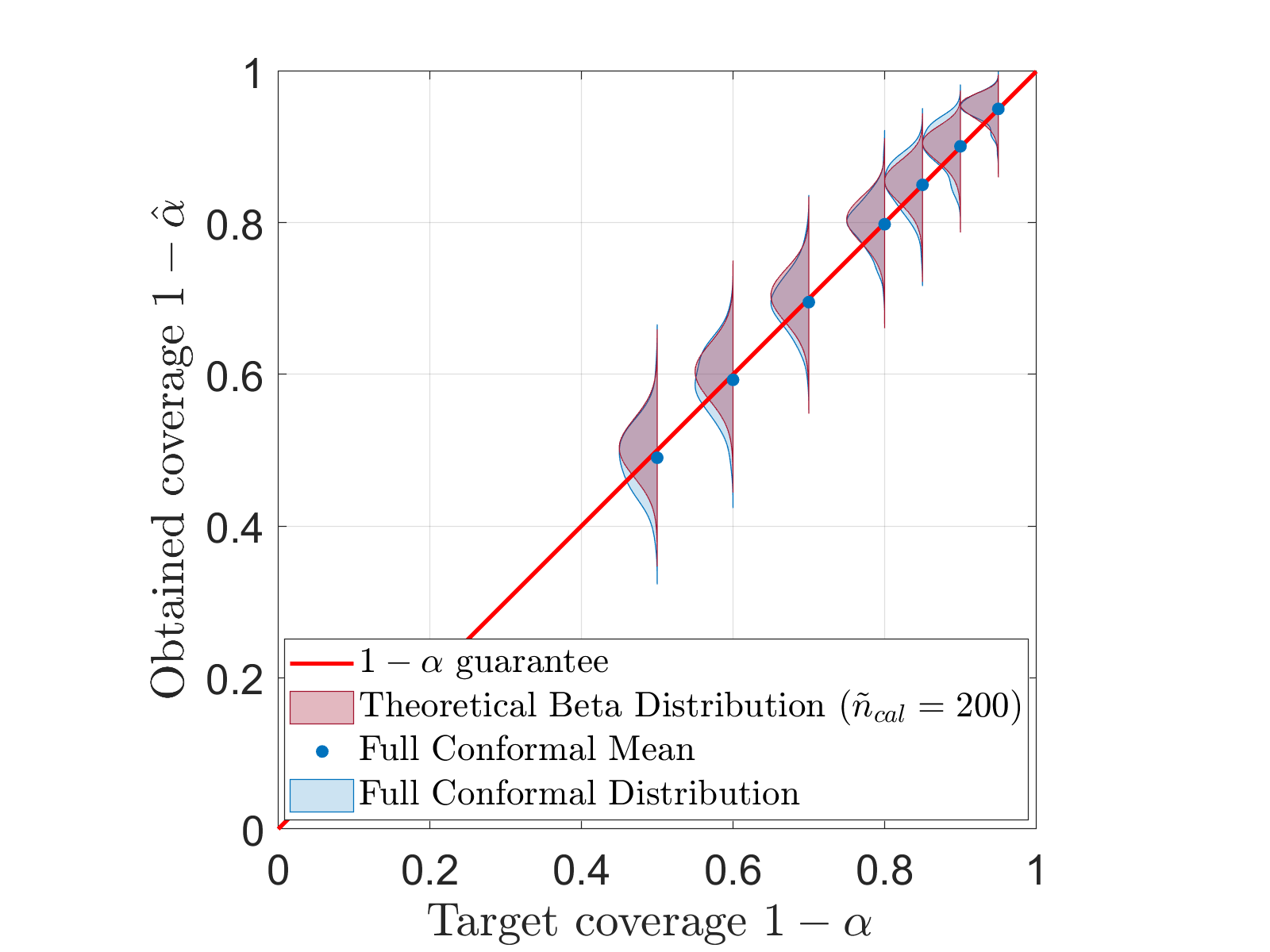}  
        \caption{Full conformal}
       \label{fig:fpce_bore_fc_cov}
    \end{subfigure}
    \caption{Evaluation of prediction intervals built based on the PCE surrogate of the Borehole function. Settings of the experiment: $\ned=200$, $\nval=1{,}000$, PCE degree $p=2$, number of regressors $P=45$, $\errval\approx 1\cdot 10^{-3}$, $\nr=100$.}
    \label{fig:21}
\end{figure}

In this analysis, we chose experimental designs of $\ned=200$ points obtained by Latin Hypercube Sampling, and a PCE degree of $p=2$. The OLS solution with the resulting $P=45$ regressors yields a relative mean-square error $\errval \approx 1\cdot 10^{-3}$ on a validation set of size $\nval=1{,}000$~points. For each replication, prediction intervals are constructed with the different methods described for the $\nval$ points. This process is repeated $\nr=100$ times, and the results for the coverage levels are presented in terms of both distribution and mean value in \Cref{fig:21}.

First, the same conclusion as for the Ishigami function holds for the reference case (\Cref{fig:fpce_bore_ref_cov}). Both conformal prediction procedures reach the theoretical guarantees for coverage, as depicted in \Cref{fig:fpce_bore_fc_cov,fig:fpce_bore_j+_cov}. The distribution of coverage of the full conformal method could be achieved by the split conformal method with an additional calibration set of size $\Tilde{n}_{\textrm{cal}}=200$. Again, bootstrap does not attain the target coverage level (\Cref{fig:fpce_bore_b_cov}). 

\begin{figure}[H] 
    \centering
    \begin{subfigure}[b]{0.49\textwidth}
        \centering
        \includegraphics[width=\textwidth]{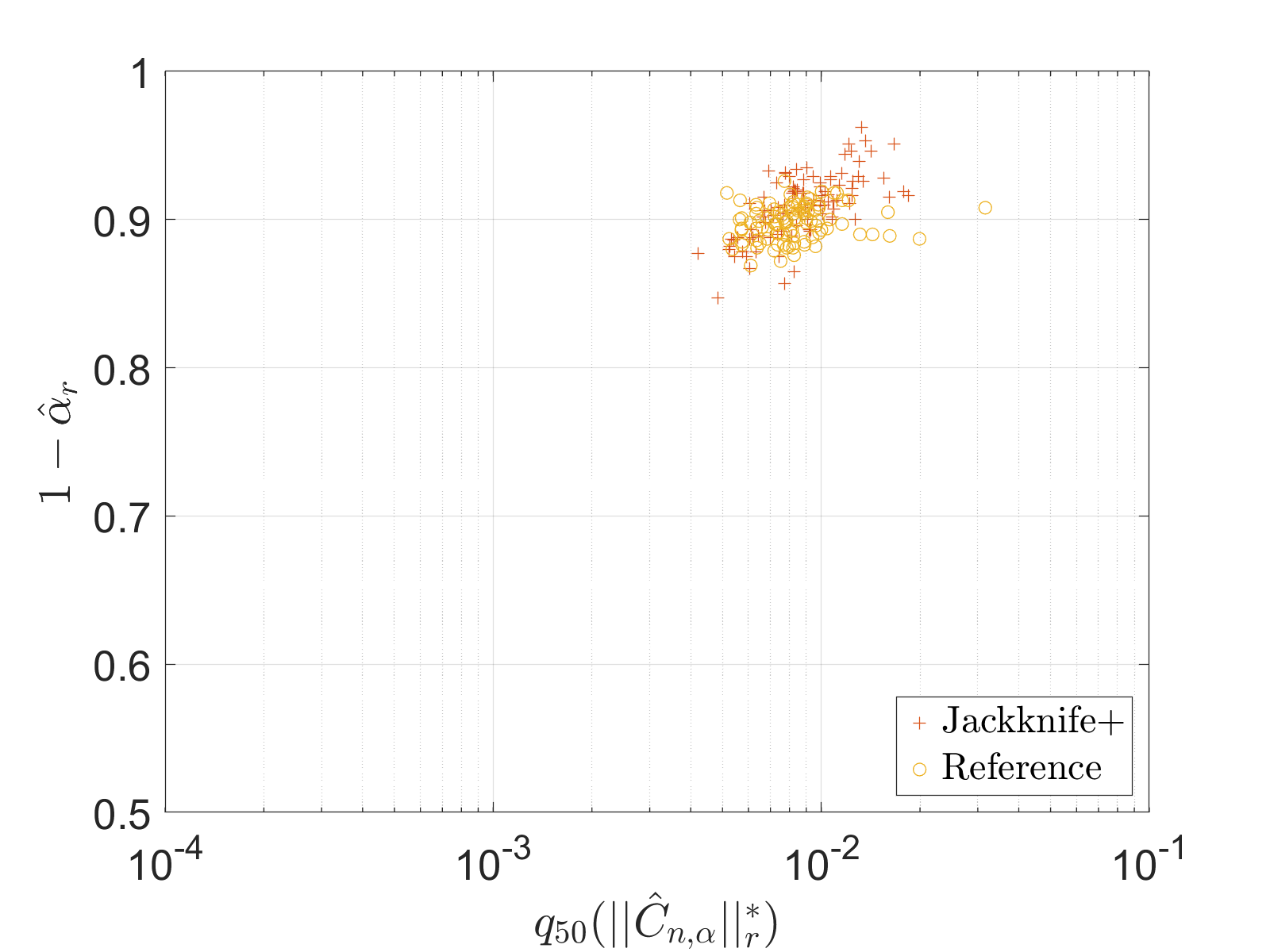} 
        \caption{Jackknife+}
        \label{fig:fpce_bore_j+}
    \end{subfigure}
    \hfill
    \begin{subfigure}[b]{0.49\textwidth}
        \centering
        \includegraphics[width=\textwidth]{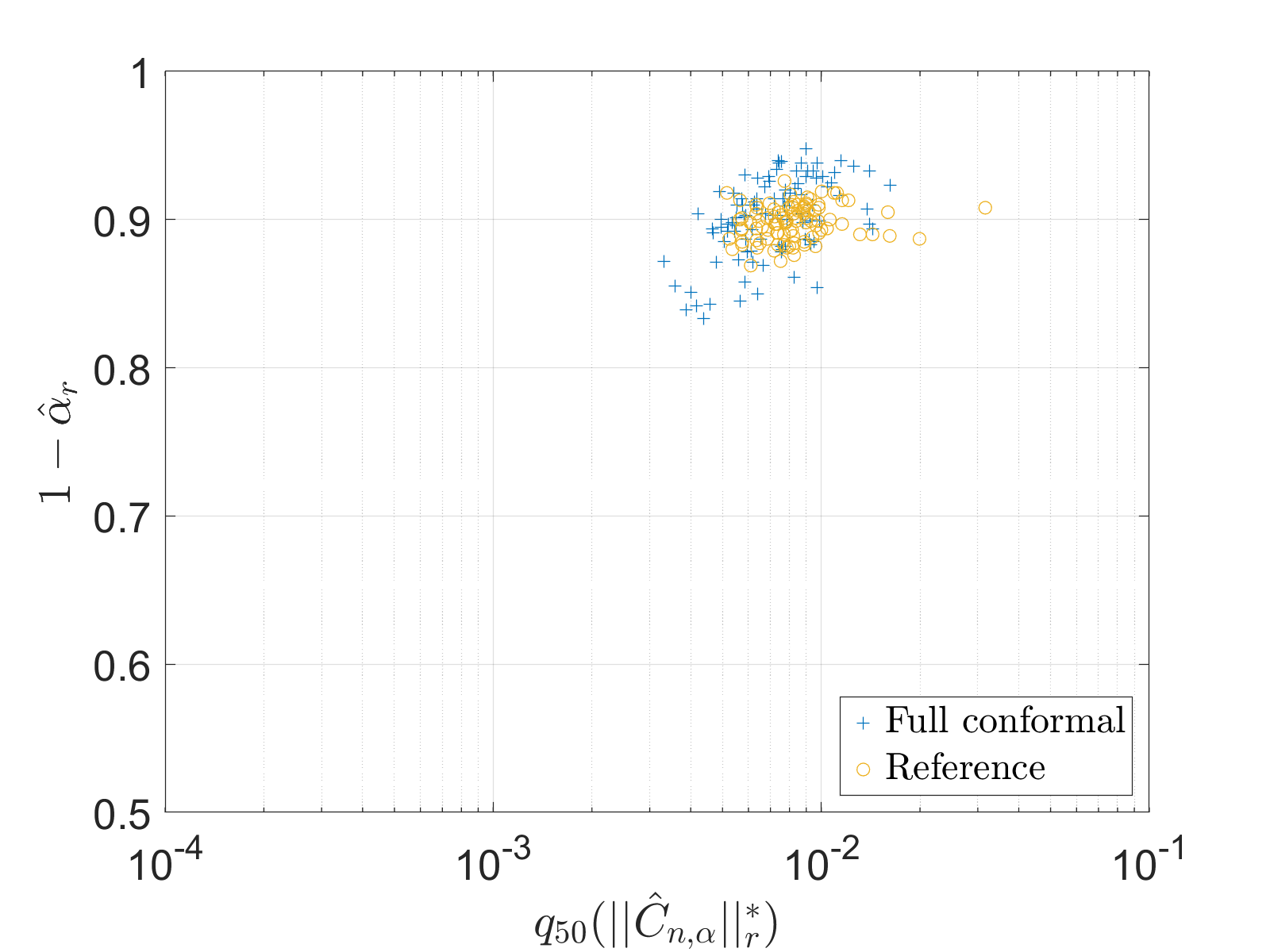}  
        \caption{Full conformal}
        \label{fig:fpce_bore_fc}
    \end{subfigure}
    \begin{subfigure}[b]{0.49\textwidth} 
        \centering
        \includegraphics[width=\textwidth]{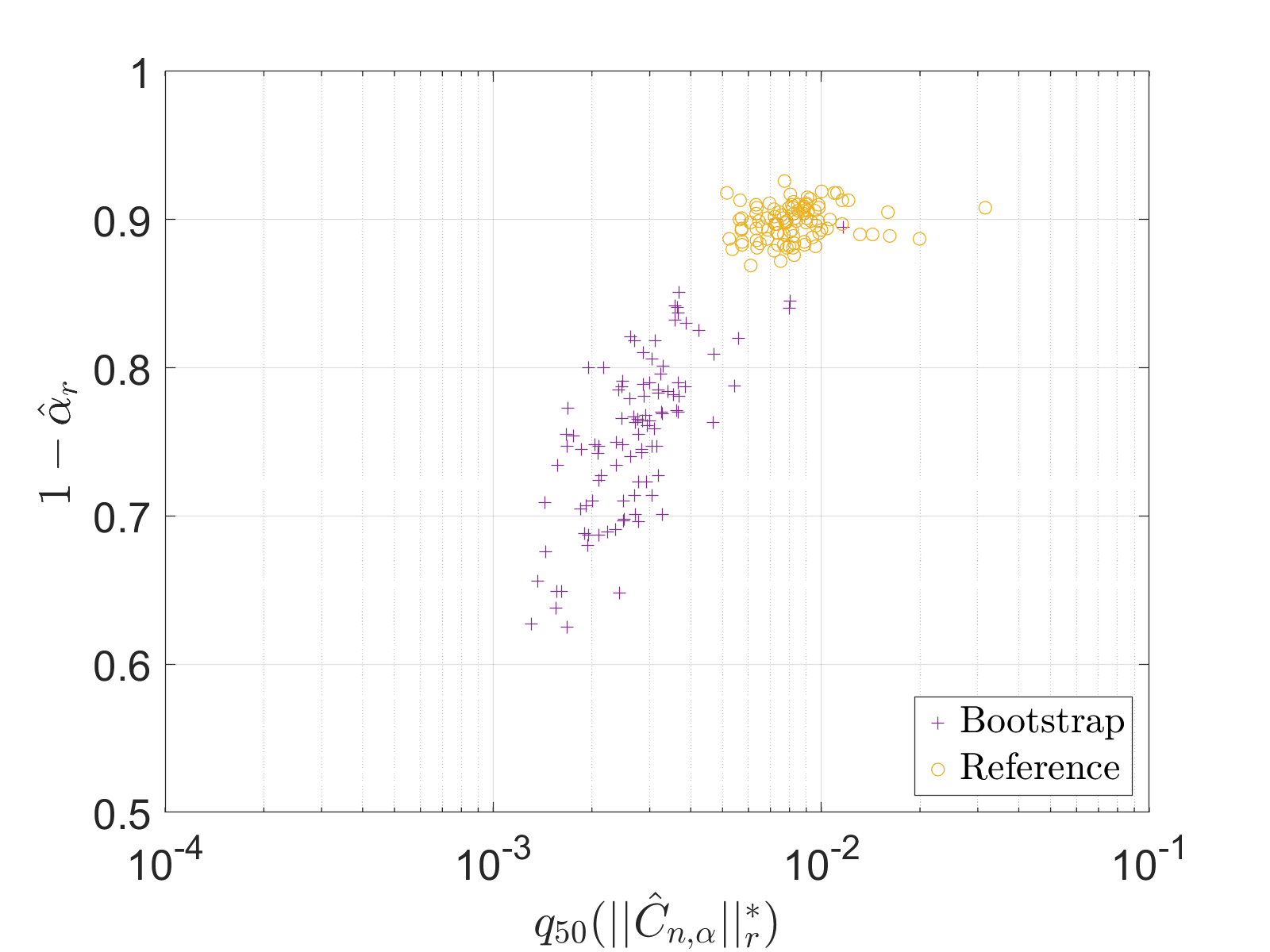}  
        \caption{Bootstrap}
        \label{fig:fpce_bore_b}
    \end{subfigure}
    \begin{subfigure}[b]{0.49\textwidth} 
        \centering
        \includegraphics[width=\textwidth]{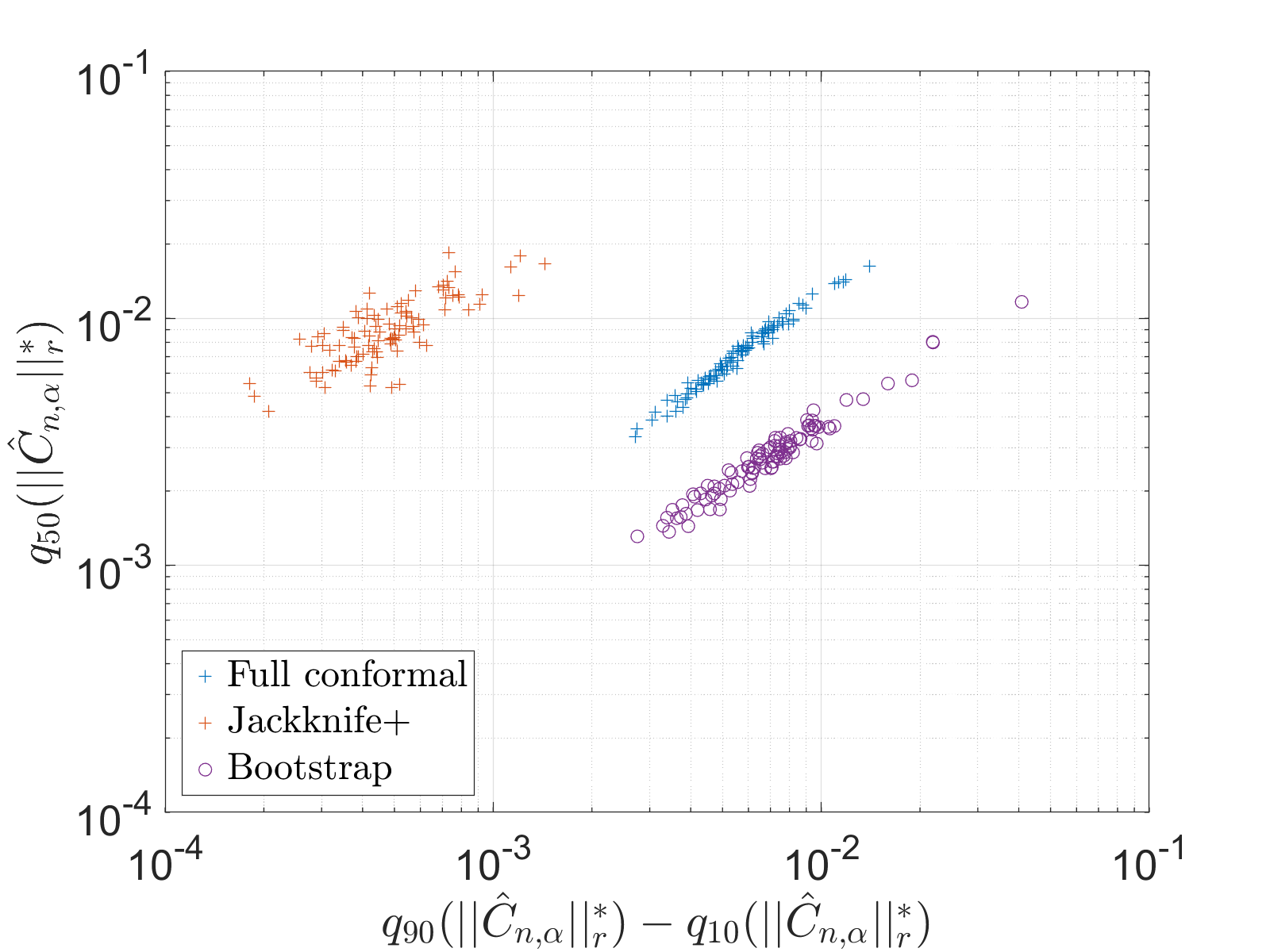}  
        \caption{Spread of prediction interval widths}
        \label{fig:fpce_bore_iq}
    \end{subfigure}
    
    \caption{Comparison of the normalized width of prediction intervals at target coverage level $1-\alpha=0.9$ built based on the PCE surrogate of the Borehole function. Settings of the experiment: $\ned=200$, $\nval=1{,}000$, PCE degree $p=2$,  number of regressors $P=45$, $\errval\approx 1\cdot 10^{-3}$.}
    \label{fig:22}
\end{figure}

The normalized prediction interval widths are compared across different methods and replications at a target coverage level of $1-\alpha=0.9$ and depicted in \Cref{fig:22}. 

Similarly to the previous example, the Jackknife+ yields prediction intervals of quasi-constant width (\Cref{fig:fpce_bore_iq}), close to the ones provided by the reference case, as shown in \Cref{fig:fpce_bore_j+}. In contrast, the full conformal procedure leads to prediction intervals of variable width, with the median width being close to that provided by the reference case (\Cref{fig:fpce_bore_fc}). Bootstrap again demonstrates a wide spread of normalized interval widths (\Cref{fig:fpce_bore_iq}). They are significantly smaller than the reference values, which explains the poor coverage shown in \Cref{fig:fpce_bore_b_cov}.

\newpage
\section{Conformal prediction for sparse PCE}

A closer examination of the solution procedure for the sparse regression problem is required to incorporate conformal predictions with sparse PCE. In this paper, we focus on the commonly used least-angle regression algorithm (LARS) \citep{BlatmanJCP2011}. LARS relies on incrementally increasing the number of regressors. At each iteration, it introduces a new regressor that is most correlated to the current residual. The LARS solver does not provide a single solution but a set of solutions (the so-called \emph{LARS solution path}) with an increasing number of active regressors. Finally, a selection procedure, typically performed through cross-validation, is used to select the best basis that avoids overfitting. Building on the LARS procedure, the Hybrid LARS algorithm \citep{BlatmanJCP2011} performs an additional OLS regression on the selected sparse basis at each LARS iteration. An analytical expression of the associated leave-one-out error is then available and used to select the best sparse basis along the LARS solution path.

\subsection{Why a naive extension from OLS fails}

In the context of conformal prediction, either for the Jackknife+ or the full conformal approach, the essence of the procedures is to fit models $\hat{\cm}_{-i}$ or $\hat{\cm}^y$. One naive idea would be to run Hybrid LARS only once to identify the best sparse solution using all the training data $\mathcal{S}$. The set of $P$ active regressors for this solution is denoted as $\mathcal{A}_P^{LARS}$. The naive solution would then be to use the set $\mathcal{A}_P^{LARS}$ and build the respective models $\hat{\cm}_{-i}$ and $\hat{\cm}^y$ using equations developed for OLS in \Cref{sec:full_conformal_fpce}, that is, \textit{keeping the sparse basis fixed}.

This approach has been tested in a similar setting to the one presented in the previous section, evaluating the coverage of prediction intervals using surrogates with moderate accuracy. 

\begin{figure}[H] 
    \centering
    \begin{subfigure}[b]{0.49\textwidth} 
        \centering
        \includegraphics[width=\textwidth]{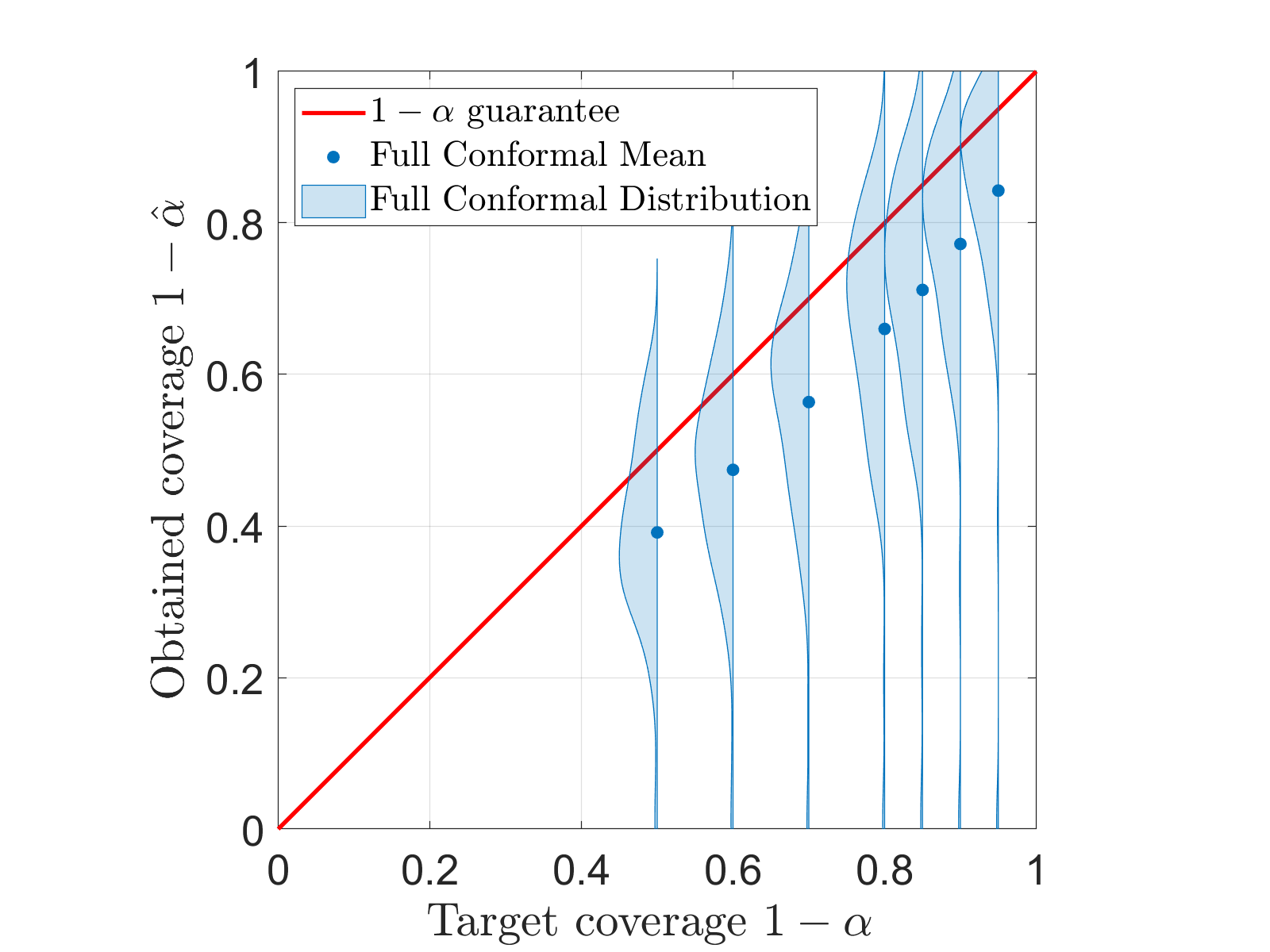}  
        \caption{Full conformal}
        \label{fig:naive1_fc}
    \end{subfigure}
    \hfill
    \begin{subfigure}[b]{0.49\textwidth}
        \centering
        \includegraphics[width=\textwidth]{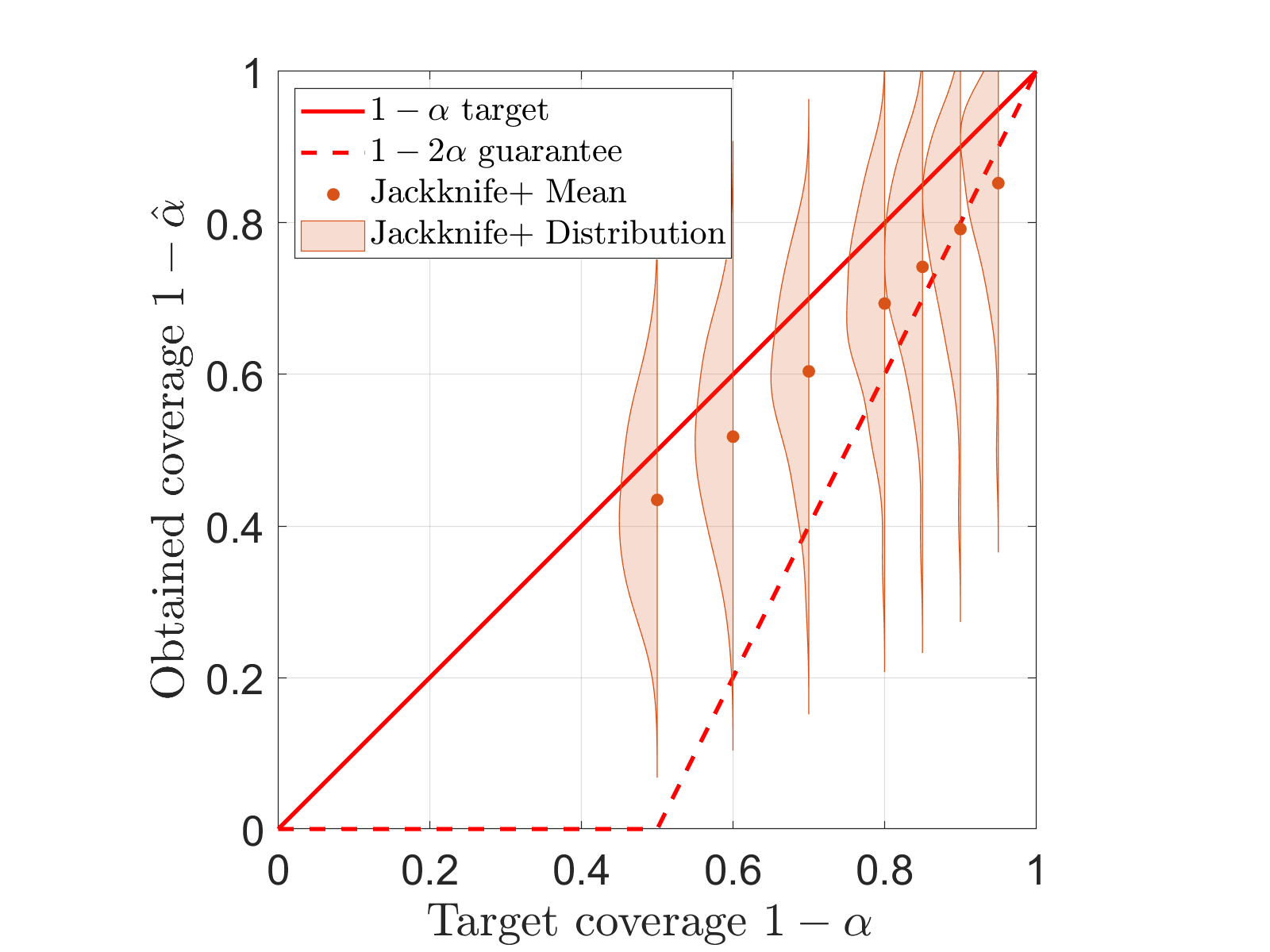}  
        \caption{Jackknife+}
        \label{fig:naive1_j+}
    \end{subfigure}
    \caption{Performance of the naive implementation of conformal LARS predictions for the Ishigami function. Settings of the experiment: $\ned=40$, $\nval=1{,}000$, PCE degree $p=6$, number of regressors $P\approx 23$, $\errval \approx 1\cdot 10^{-1}$, $\nr=100$.}
    \label{fig:naive1}
\end{figure}

\begin{figure}[H] 
    \centering
    \begin{subfigure}[b]{0.49\textwidth} 
        \centering
        \includegraphics[width=\textwidth]{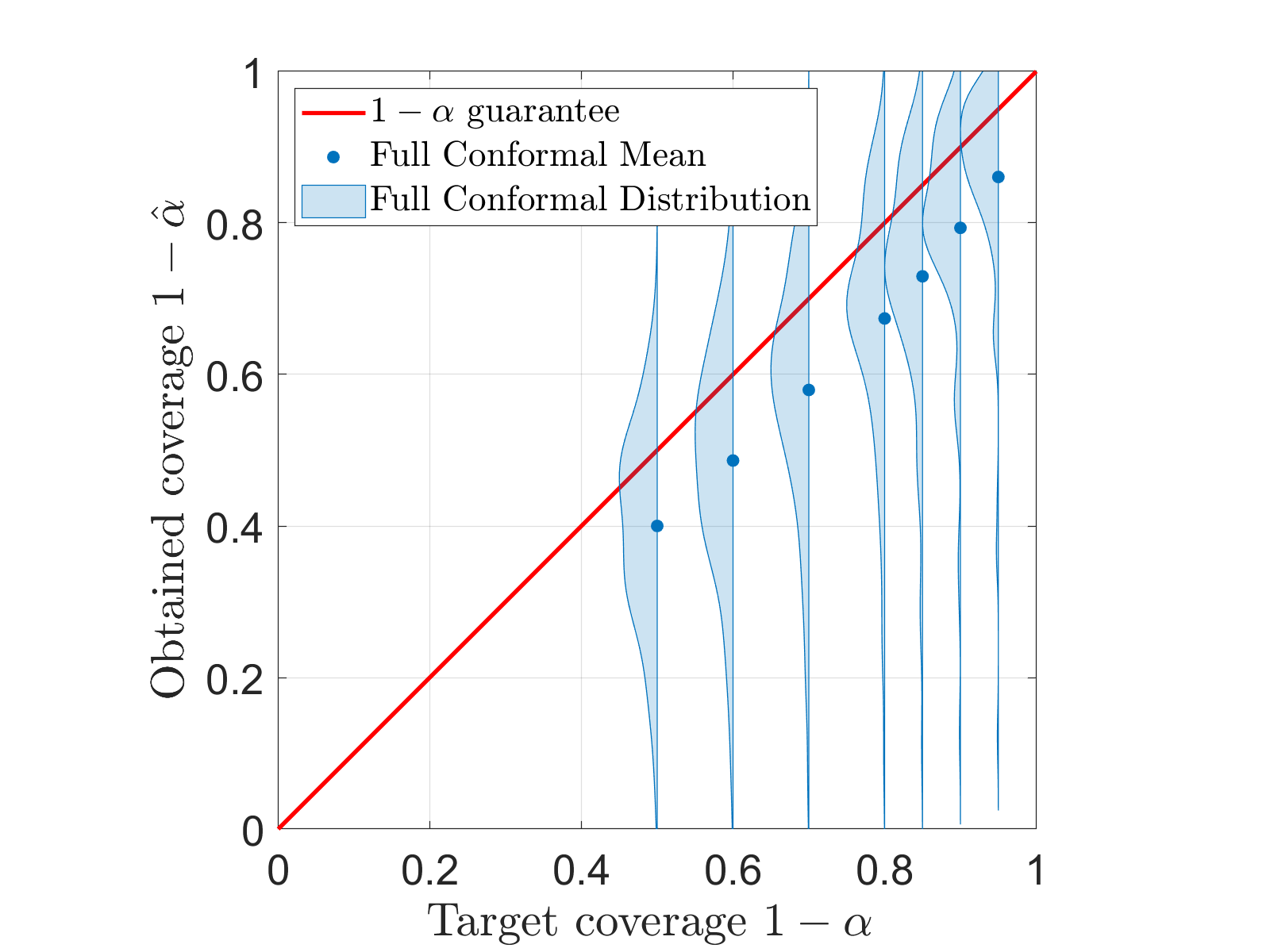}  
        \caption{Full conformal}
        \label{fig:naive2_fc}
    \end{subfigure}
    \hfill
    \begin{subfigure}[b]{0.49\textwidth}
        \centering
        \includegraphics[width=\textwidth]{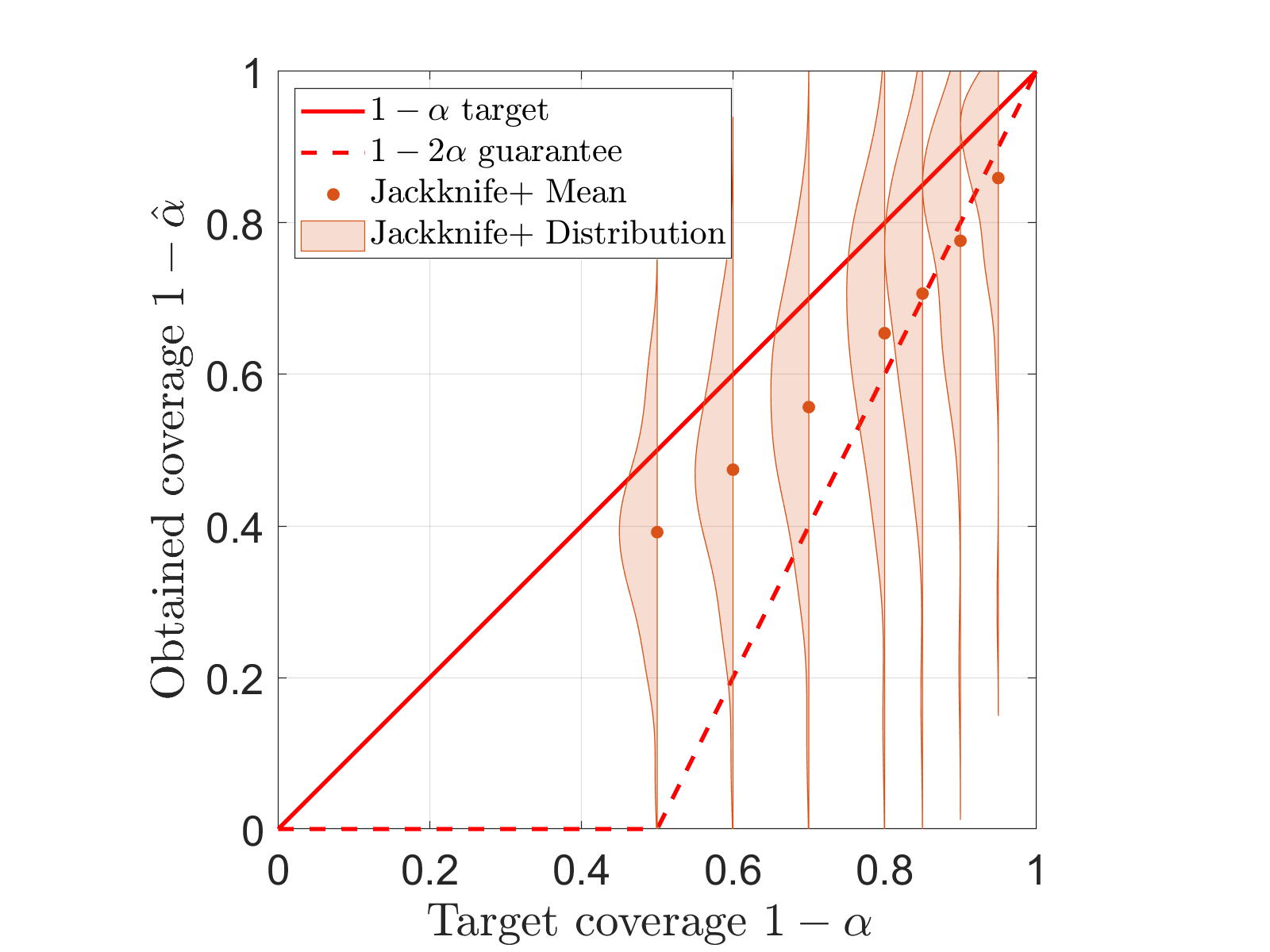}  
        \caption{Jackknife+}
        \label{fig:naive2_j+}
    \end{subfigure}
    \caption{Performance of the naive implementation of conformal LARS predictions for the Borehole function. Settings of the experiment: $\ned=40$, $\nval=1{,}000$, PCE degree $p=2$, number of regressors $P\approx19$, $\errval \approx 1\cdot 10^{-3}$, $\nr=100$.}
    \label{fig:naive2}
\end{figure}

For both test cases, the average observed coverage for full conformal prediction intervals is significantly lower than the theoretical guarantee (\Cref{fig:naive1_fc,fig:naive2_fc}). The same behavior is observed for the Jackknife+ method, where the spread of the distribution of observed coverage is even larger (\Cref{fig:naive1_j+,fig:naive2_j+}). Although the theoretically guaranteed coverage of $1-2\alpha$ is generally achieved on average, the performance of Jackknife+ in comparison to the full PCE setting has considerably decreased. 

To gain insight into this issue, we further investigated the reasons for this lack of performance. As detailed in \Cref{sec:intro_symm}, a central condition for conformal procedures is to have a symmetrical algorithm. For full conformal prediction, the iterative process should fit the model $\hat{\cm}^y$ to the augmented data set $\acc{(\ve{x}_i,y_i),\,i=1\enu n}\cup (\ve{x}_{n+1},y^{\textrm{trial}})$. This is done, however, in two steps in the approach described above. First, the sparse set of regressors $\mathcal{A}_{P}^{LARS}$ is established based on the set $\mathcal{S}$ with Hybrid LARS. Then, the OLS solution is computed with the regressors $\mathcal{A}_{P}^{LARS}$ on the data set $\mathcal{S}\cup (\ve{x}_{n+1},y^{\textrm{trial}})$. As a consequence, if we permute the data points, i.e., define a permutation of the training set: 
\begin{equation}
    \mathcal{S}'\cup (\ve{x}_l,y_l)=\acc{(\ve{x}_1,y_1)\enu (\ve{x}_{l-1},y_{l-1}), (\ve{x}_{l+1},y_{l+1})\enu (\ve{x}_n,y_n),(\ve{x}_{n+1},y^{\textrm{trial}})}\cup(\ve{x}_l,y_l)~,
\end{equation}
the procedure detailed above may not provide the exact same solution. Indeed, there is no guarantee that the identified sets of sparse regressors for $\mathcal{S}$ and $\mathcal{S}'$ are identical. Therefore, the proposed algorithm does not treat the data symmetrically, and the theoretical guarantees offered by the conformal framework do not hold. The issue with the two-step implementation of the Jackknife+ method stems from a similar reason. One could argue that in data-abundant settings, the LARS procedure may always converge to the same set of regressors for all possible permutations of the training data. This is, however, not the case in practical relevant settings, where the problem is rather data-scarce and surrogates of moderate accuracy are considered. 

The previous numerical experiments and their analysis showed that the naive extension of conformal prediction to sparse PCE is not successful. This leads us to extend the framework as shown in the sequel.

\subsection{Adaptation of algorithms to the sparse context}

\subsubsection{Jackknife+ for sparse PCE}

It is straightforward to extend the Jackknife+ conformal prediction method to Hybrid LARS. More precisely, since the Jackknife+ inference procedure relies on the construction of the $n$ leave-one-out surrogate models $\acc{\hat{\cm}_{-i},\, i = 1 \enu n}$, its computational cost remains manageable in the data-scarce regime considered in this work. This makes it feasible to employ the Hybrid LARS algorithm independently for each leave-one-out fit, ensuring a consistent sparse PCE construction across all leave-one-out surrogates. By adopting this strategy, the assumptions required by the Jackknife+ conformal framework are satisfied, and the resulting prediction intervals retain their validity. This adapted Jackknife+ procedure is therefore used throughout the remainder of the paper.

\subsubsection{Full conformal prediction for sparse PCE}

The full conformal approach solves \Cref{eq:4,eq:5}. Each iteration of the search algorithm requires the computation of the surrogate model $\hat{\cm}^y$, which in the sparse regression context requires the computation of the Hybrid LARS. As a result, full conformal inference becomes computationally prohibitive in this setting. Our main assumption is to keep the number of regressors considered equal to the one identified by Hybrid LARS in the original training set, namely $|\mathcal{A}^{LARS}|$. Note, however, that the regressors may be different. Due to the complex dependence of the LARS solution scheme on the training data, a simple bisection search for solving \Cref{eq:4,eq:5} may converge to a local non-meaningful solution or even diverge. Therefore, the problem of finding the bounds of the prediction intervals is reformulated into two minimization problems, namely: 
\begin{equation}
\label{eq:min1}
    \min_{y} \left(y-\hat{\cm}^y(\ve{x}_{n+1})-\hat{q}_{n,\alpha/2}^+\prt{\acc{y_i-\hat{\cm}^y(\ve{x}_i),\,i=1\enu n}} \right)^2,
\end{equation}
\begin{equation}
\label{eq:min2}
    \min_{y} \left( y-\hat{\cm}^y(\ve{x}_{n+1})-\hat{q}_{n,\alpha/2}^-\prt{\acc{y_i-\hat{\cm}^y(\ve{x}_i),\,i=1\enu n}}\right)^2. 
\end{equation}

Since the evaluation of the above objective functions is computationally demanding, we employ the bounded optimization algorithm Brent’s method \citep{Brent1973}, as implemented in the \texttt{fminbnd} function of Matlab.

The full conformal procedure associated with sparse PCE obtained from the LARS algorithm is detailed in \Cref{alg:3}. 

\begin{algorithm}[H]
\small
\caption{Full conformal prediction for sparse PCE}
\label{alg:3}
\begin{algorithmic}
\State \textbf{Step 1:} Train the surrogate model $\hat{\cm}$ on the set $\mathcal{S}$ with Hybrid LARS, and gather the number of active regressors $|\mathcal{A}^{LARS}|$
\State \textbf{Step 2:} Identify the lower and upper bound of the confidence interval $\hat{C}_n^{\alpha}(\ve{x}_{n+1})$ by solving the respective minimization problems: 
\begin{align*}
    \min_{y} \left(y-\hat{\cm}^y(\ve{x}_{n+1})-\hat{q}_{n,\alpha/2}^+\prt{\acc{y_i-\hat{\cm}^y(\ve{x}_i),\,i=1\enu n}} \right)^2\\
   \min_{y} \left( y-\hat{\cm}^y(\ve{x}_{n+1})-\hat{q}_{n,\alpha/2}^-\prt{\acc{y_i-\hat{\cm}^y(\ve{x}_i),\,i=1\enu n}}\right)^2
\end{align*}
Where at each iteration on $y$, the surrogate model  $\hat{\cm}^y$, based on the the set $\mathcal{S}\cup (\ve{x}_{n+1},y)$ is built as follows: 

    \Statex \hspace{\algorithmicindent} \textbf{Step 2a:} Build the set of sparse solutions for the set $\mathcal{S}\cup (\ve{x}_{n+1},y)$ with the LARS solver.
    \Statex \hspace{\algorithmicindent} Select the sparse basis of $|\mathcal{A}^{LARS}|$ regressors: $\Tilde{\mathcal{A}}^{LARS}$.
    \Statex \hspace{\algorithmicindent} \textbf{Step 2b:} Run the OLS solver to find the coefficients for the basis $\Tilde{\mathcal{A}}^{LARS}$

\State \textbf{Step 3:} Return $\hat{C}_{n,\alpha}^{FC}(\ve{x}_{n+1})$
\end{algorithmic}
\end{algorithm}


In practice, this implementation attains coverage close to the theoretical guarantees of conformal prediction; however, it is computationally intensive. Specifically, the method requires repeated training of surrogate models, involving numerous executions of the LARS algorithm with only marginal changes to the training dataset. To mitigate this computational burden, we propose a heuristic simplification. As noted by \citet{Efron2004}, the LARS algorithm can be viewed as an alternative approach to solving the LASSO regression problem. A correspondence can be established between the number of active regressors selected with Hybrid LARS, denoted $|\mathcal{A}^{LARS}|$, and a specific regularization parameter $\lambda$ in the LASSO formulation given in \Cref{eq:19}.

While the solutions obtained via LARS and LASSO may differ in their coefficient estimates, we denote by $\hat{\lambda}$ the value of $\lambda$ that would render the LARS solution optimal for the LASSO objective. As discussed in \citet{Lei2019b}, an optimal solution $\hat{c}_{\alpha}$ to the LASSO problem satisfies the following condition:
\begin{equation}
    -\sum_{i=1}^{\ned}\left(\ve{\phi}(\ve{x}_i)\hat{\ve{c}}_{\alpha}-y_i\right)\ve{\phi}(\ve{x}_i)+\ve{v}=0~, 
\end{equation}
where $\ve{v} \in \mathbb{R}^P$ denotes the dual variable associated with the LASSO optimization problem. This dual variable satisfies the conditions $v_j = \lambda\,\textnormal{sign}(\hat{c}_{\alpha,j})$ if $\hat{c}_{\alpha,j} \neq 0$, and $v_j \in [-\lambda, \lambda]$ if $\hat{c}_{\alpha,j} = 0$. Consequently, the pseudo-regularization parameter $\hat{\lambda}$ can be interpreted as the value of $\lambda$ that would yield the LARS solution $\hat{\ve{c}}_{\alpha}$ as the corresponding LASSO solution. It is thus computed from $\hat{\ve{c}}_{\alpha}$ as follows:
\begin{align}
    \hat{\lambda} &= \textnormal{max}(\abs{\ve{v}_j}),\,j=1\enu P \\
    &= \textnormal{max}\left(\abs{\sum_{i=1}^{\ned}(\ve{\phi}(\ve{x}_i)\hat{\ve{c}}_{\alpha}-y_i)\ve{\phi}(\ve{x}_i))}\right),\,j=1\enu P.
\end{align}

\begin{figure}[H] 
\centering
\includegraphics[width=0.6\textwidth]{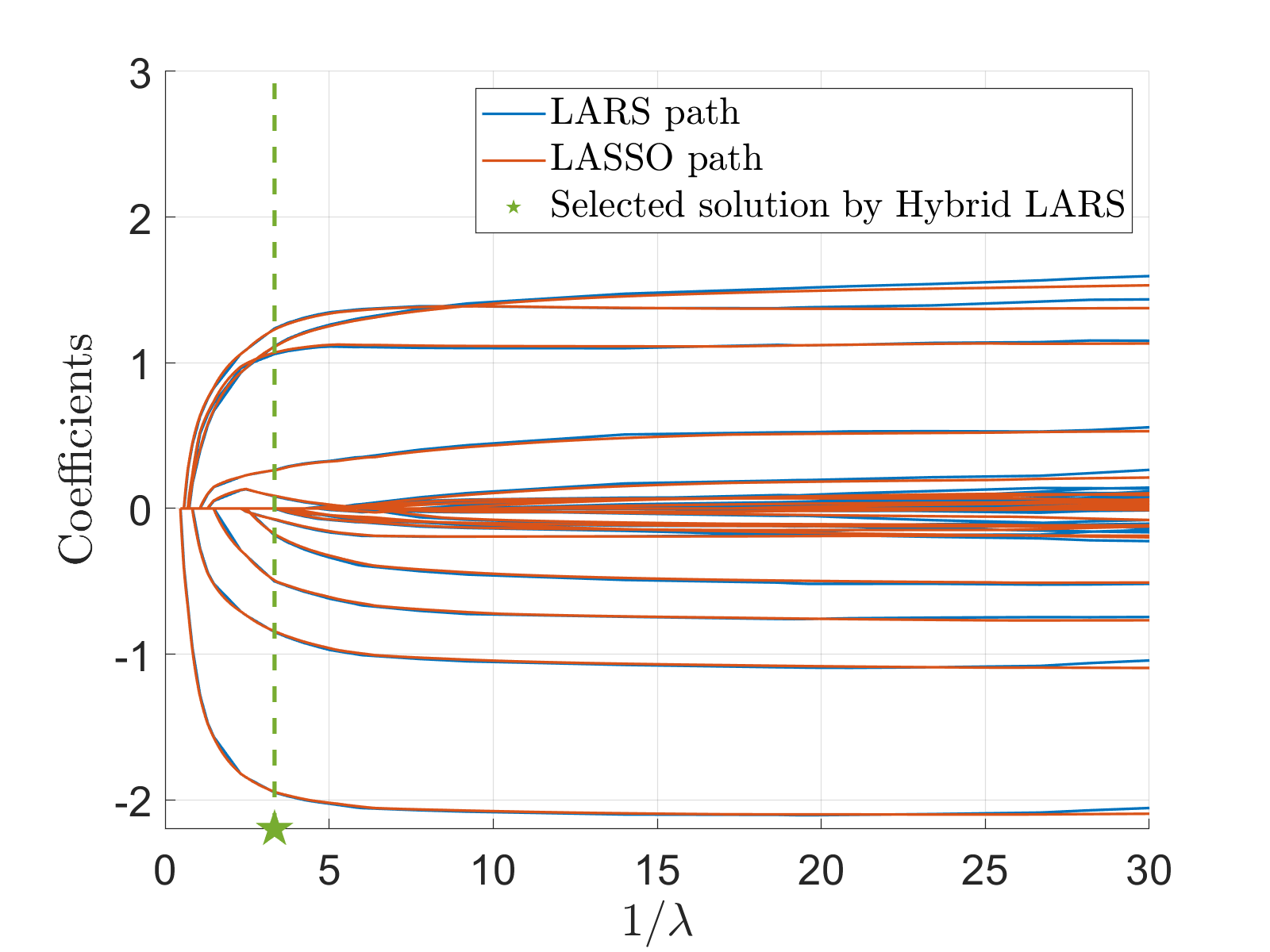}  
\caption{Evolution of LASSO and LARS solution paths as a function of $\lambda$ and $\hat{\lambda}$ respectively for the PCE surrogate of Ishigami function. Settings of the experiment: $\ned=40$, PCE degree $p=8$, total number of regressors $P=165$, $\errval\approx 6\cdot 10^{-2}$.}
\label{fig:LASSO_lars}
\end{figure}

This approximation is exact when using the Lasso-modified version of the LARS algorithm \citep{Efron2004}, since in this case the Lasso and modified LARS solution paths coincide exactly. Although our approach may introduce some approximation error, it has been empirically validated and demonstrates a notable degree of robustness. This is largely attributable to the fact that the selected solutions tend to be highly sparse and significantly deviate from the ordinary least-squares estimate. This behavior is shown in \Cref{fig:LASSO_lars}, which illustrates the evolution of the regression coefficients obtained via LARS and LASSO as functions of the pseudo-regularization parameter $\hat{\lambda}$ and the true regularization parameter $\lambda$, respectively. It is observed that both sets of curves are virtually identical. The figure also highlights the final solution selected by the Hybrid LARS procedure with a green star.

The primary advantage of reformulating the problem within the LASSO framework lies in the availability of an analytical homotopy path for the LASSO solutions over the augmented dataset $\mathcal{S} \cup (\ve{x}_{n+1}, y)$, for any value of $y$. The implementation adopted here strictly follows the procedure outlined by \citet{Lei2019b}, accounting for both positive and negative values of the homotopy parameter $t$. As a result, the training of the surrogate model $\hat{\cm}^y$, especially Steps~2a and 2b of \Cref{alg:3}, is effectively replaced by the evaluation of a piecewise-linear function of $y$, yielding a substantial computational speedup.

Our implementation relies on solving an optimization problem, which contrasts with the approach proposed by \citet{Lei2019b}. The latter is based on solving equations similar to the ones described in \Cref{sec:full_conformal_fpce}. In our experiments, Lei's formulation appeared to suffer from error accumulation, negatively affecting the numerical stability of the results. In comparison, our minimization-based setup produced smoother outcomes with reduced variability (see \Cref{sec:spce_results}). Furthermore, this increased stability is achieved without a significant increase in computational cost, as the underlying objective function mainly relies on a piecewise linear function of $y$.

Since the value of $\hat{\lambda}$ was fixed to the one selected using $\mathcal{S}$, this approach does not strictly preserve the theoretical guarantees of full conformal prediction. Nevertheless, the numerical experiments presented in the following sections consistently achieve coverage levels that are very close to the nominal targets. These results indicate that the practical impact of the approximation of fixing the regularization parameter $\hat{\lambda}$ is limited. Furthermore, the proposed methodology could be modified to fully recover the theoretical guarantees of conformal prediction with a relatively modest additional computational cost. In particular, following the strategy proposed by \citep{Gasparin2024}, we could construct multiple full conformal prediction intervals corresponding to a set of Lasso regularization parameters $\acc{\lambda_m,\,m = 1 \enu M}$, rather than relying on a single value $\hat{\lambda}$, and using our approach for each single $\lambda_m$. The resulting intervals could then be aggregated through a majority-vote mechanism to obtain a single prediction interval that satisfies the theoretical conformal guarantees. Given that the intervals obtained using a single regularization parameter $\hat{\lambda}$ already exhibit satisfactory empirical coverage and closely match the target levels, this extension was not pursued further in the present work.

\subsection{Results}
\label{sec:spce_results}
This section presents the results of both conformal prediction procedures applied to sparse polynomial chaos expansions. The considered Ishigami and Borehole functions are the same as those used in the full PCE analysis (\Cref{sec:fpce_results}). As in the previous case study, the considered surrogate models exhibit moderate accuracy by design, making prediction intervals with formal guarantees particularly valuable in this context. Moreover, to account for statistical variability, the entire procedure is repeated $\nr=100$ times, ensuring a robust assessment of the predictive performance.

\subsubsection{Ishigami function}
\label{sec:spce_results_ishi}
The Ishigami function is first examined. The polynomial degree of the PCE candidate basis is set to $p=6$, leading to an identified sparse basis of approximately $P\approx22$ regressors, 
with slight variations depending on the training dataset. The experimental designs comprise $\ned=40$ points sampled with the LHS method. The sparse PCE surrogate achieves a validation error of $\errval\approx 1\cdot 10^{-1}$ when evaluated on a separate validation set of $\nval=1{,}000$~points. To account for statistical variability, the entire procedure is repeated $\nr=100$~times, ensuring a robust assessment of the predictive performance. 
\begin{figure}[H] 
    \centering
    \begin{subfigure}[b]{0.49\textwidth}
        \centering
        \includegraphics[width=\textwidth]{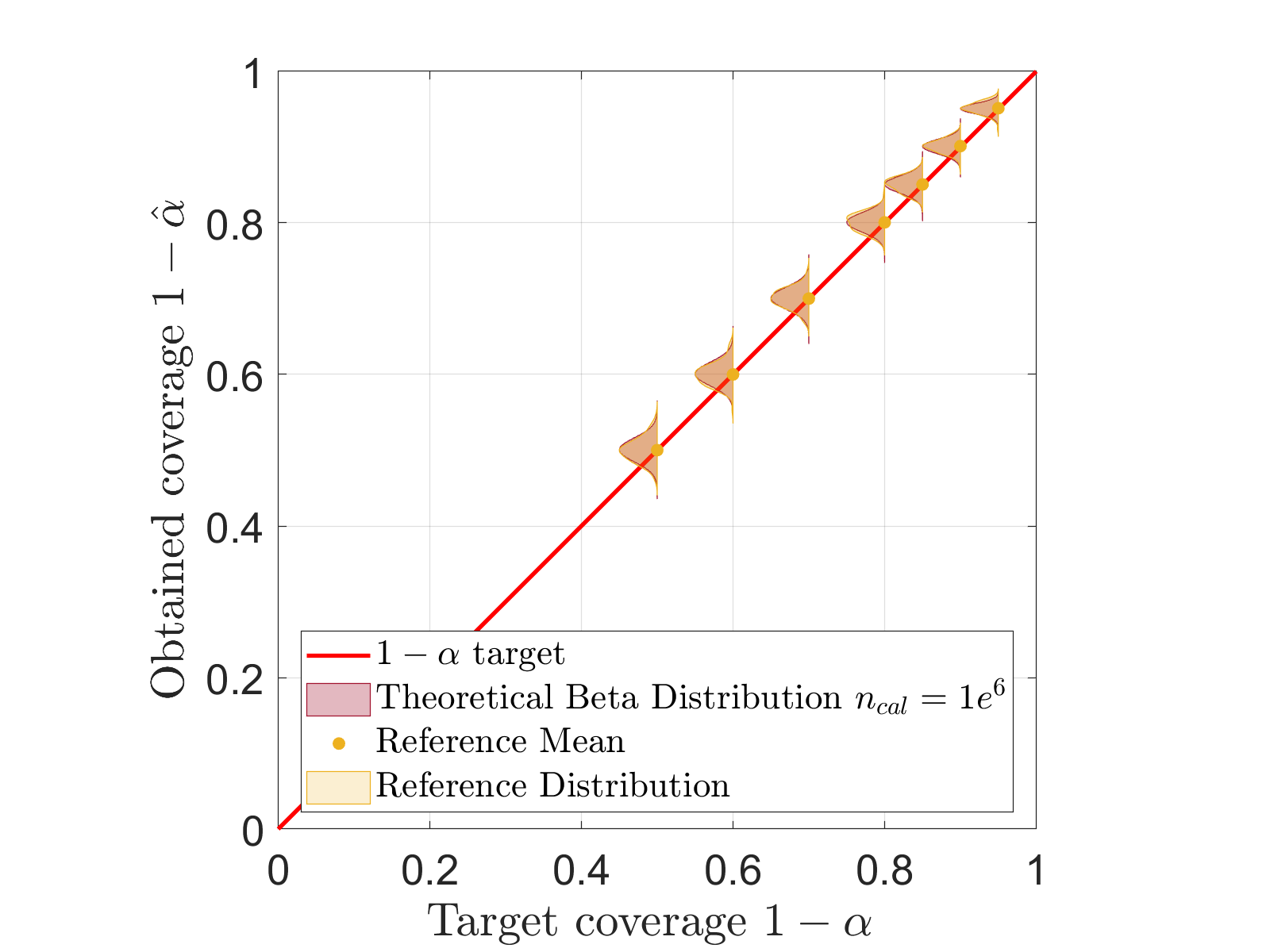}  
        \caption{Reference case}
        \label{fig:spce_ishi_ref}
    \end{subfigure}
    \hfill
    \begin{subfigure}[b]{0.49\textwidth} 
        \centering
        \includegraphics[width=\textwidth]{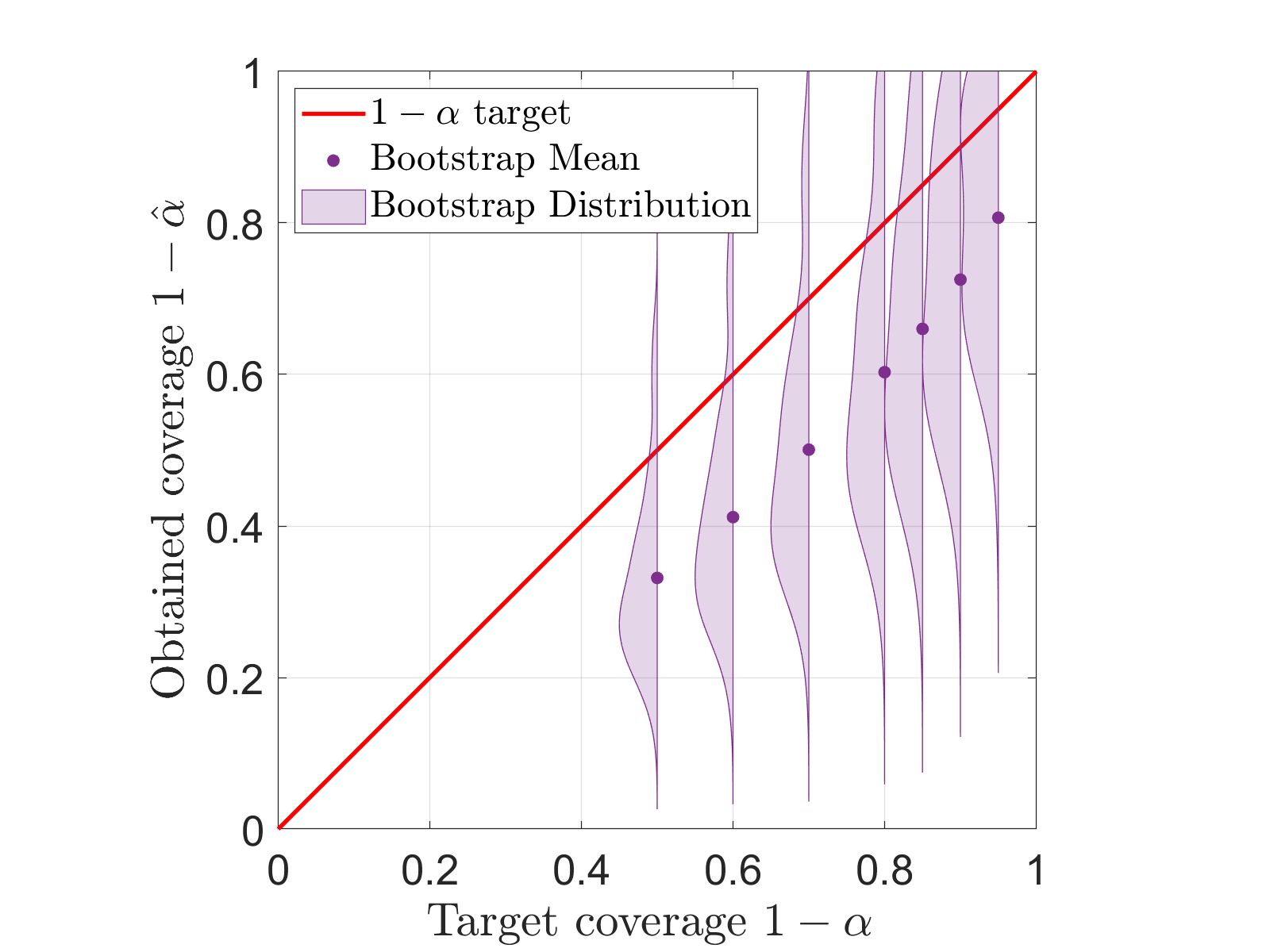}  
        \caption{Bootstrap}
        \label{fig:spce_ishi_boot}
    \end{subfigure}
    \begin{subfigure}[b]{0.49\textwidth}
        \centering
        \includegraphics[width=\textwidth]{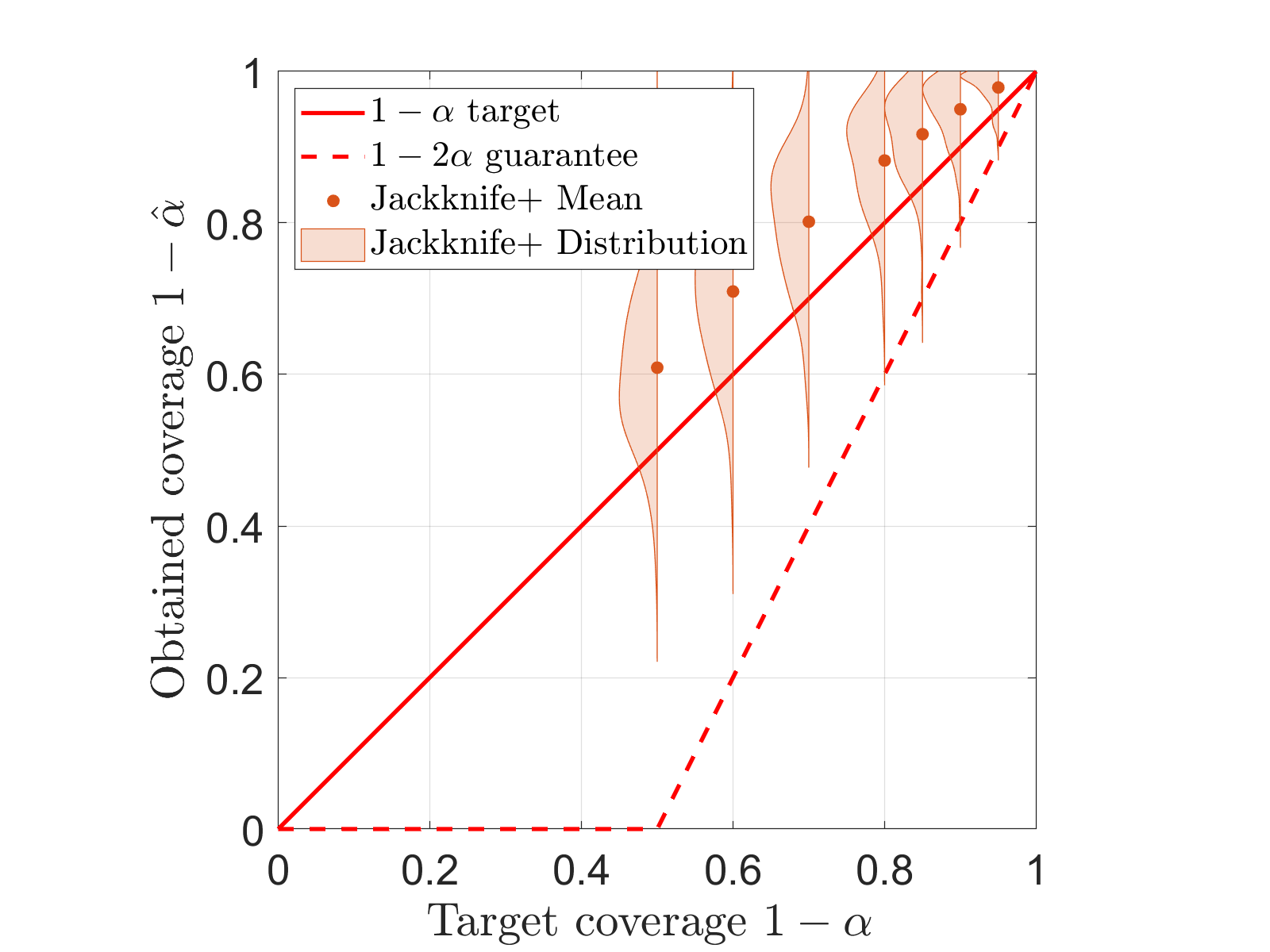} 
        \caption{Jackknife+}
        \label{fig:spce_ishi_j+}
    \end{subfigure}
    \hfill
    \begin{subfigure}[b]{0.49\textwidth}
        \centering
        \includegraphics[width=\textwidth]{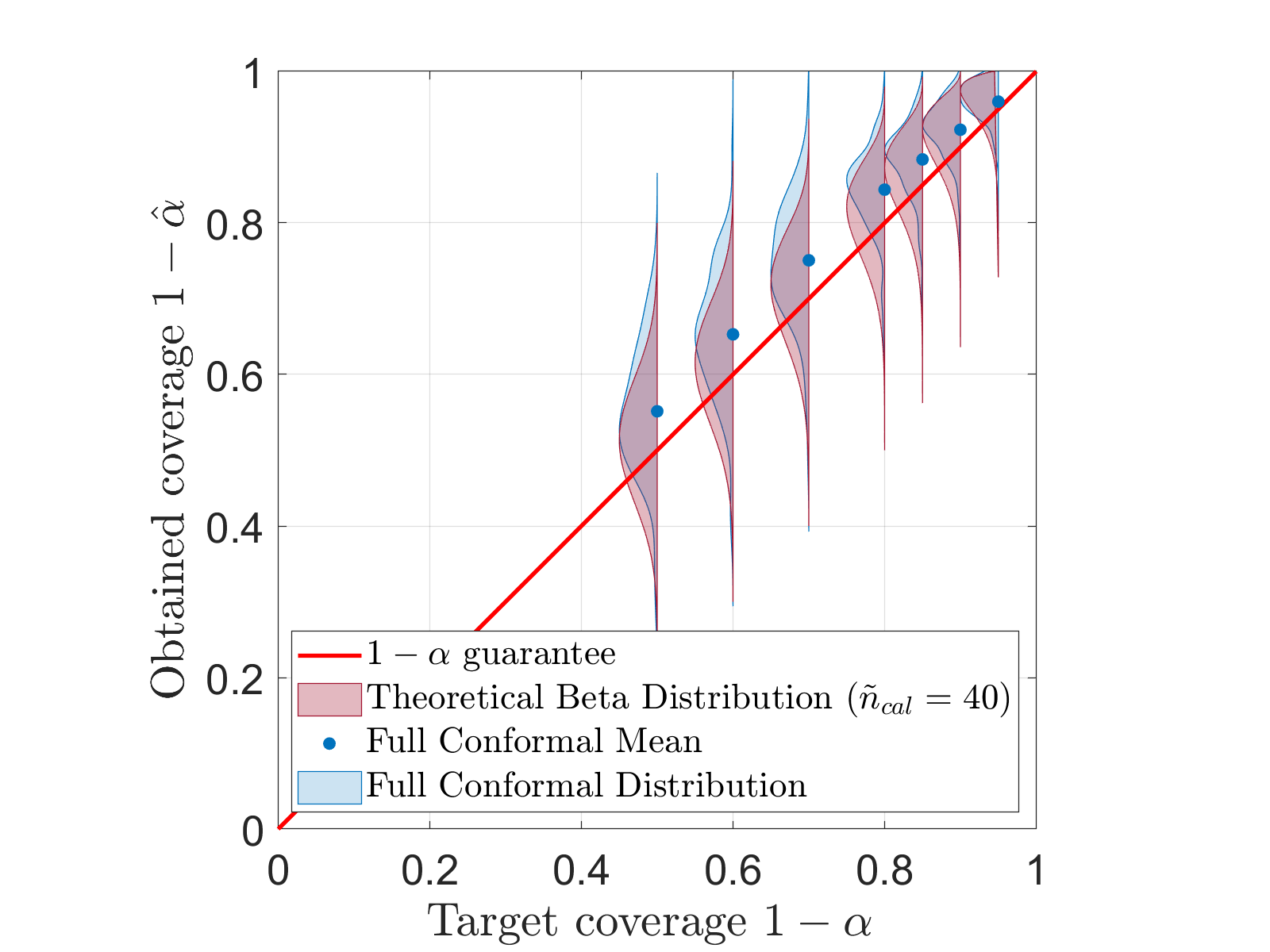}  
        \caption{Full conformal}
        \label{fig:spce_ishi_fc}
    \end{subfigure}
    \caption{Evaluation of prediction intervals built based on the sparse PCE surrogate of the Ishigami function. Settings of the experiment: $\ned=40$, $\nval=1{,}000$, PCE degree $p=6$, number of identified regressors $P\approx22$, $\errval\approx 1\cdot 10^{-1}$, $\nr=100$.}
    \label{fig:spce_ishi}
\end{figure}

Applying the proposed procedure rather than the naive extension from OLS as described in \Cref{alg:3} enables the conformal prediction approaches to achieve now the desired coverage.
First, \Cref{fig:spce_ishi_ref} confirms that the theoretical beta coverage distributions (red) match the observed coverage distributions of the reference split conformal method (yellow). Full conformal prediction reaches the target coverage level, as shown in \Cref{fig:spce_ishi_fc}. The observed coverage of the Jackknife+ procedure remains close to the target level, significantly exceeding the theoretical lower bound of $1-2\alpha$ (\Cref{fig:spce_ishi_j+}). The distribution of coverage of the full conformal method could be achieved by the split conformal method with an additional calibration set of size $\Tilde{n}_{\textrm{cal}}=40$. In contrast, the bootstrap method exhibits suboptimal performance, as its mean coverage falls below the target level and shows considerable variability across different replications (\Cref{fig:spce_ishi_boot}).

\begin{figure}[H] 
    \centering
    \begin{subfigure}[b]{0.49\textwidth}
        \centering
        \includegraphics[width=\textwidth]{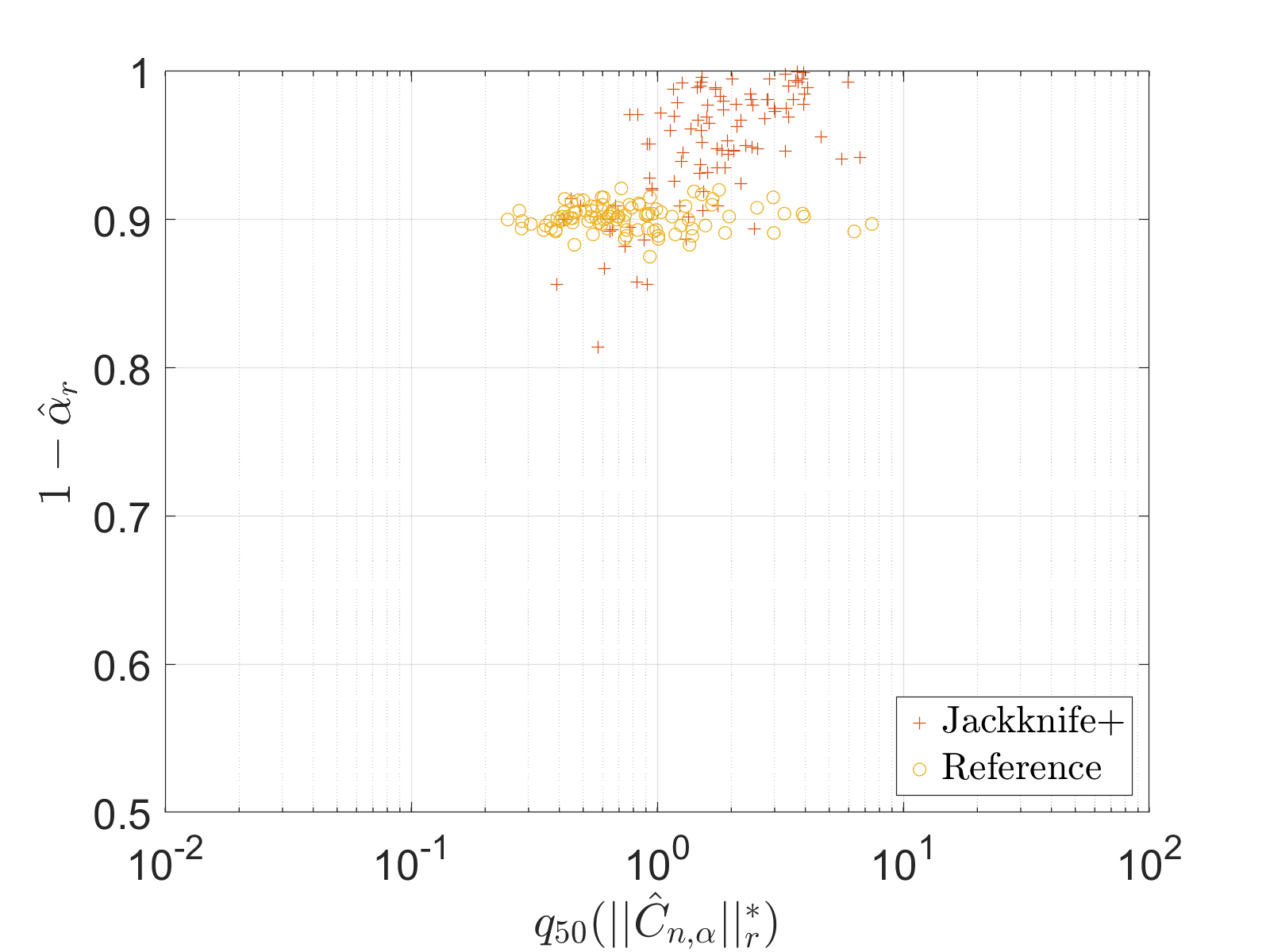} 
        \caption{Jackknife+}
        \label{fig:spce_ishi_w_j+}
    \end{subfigure}
    \hfill
    \begin{subfigure}[b]{0.49\textwidth}
        \centering
        \includegraphics[width=\textwidth]{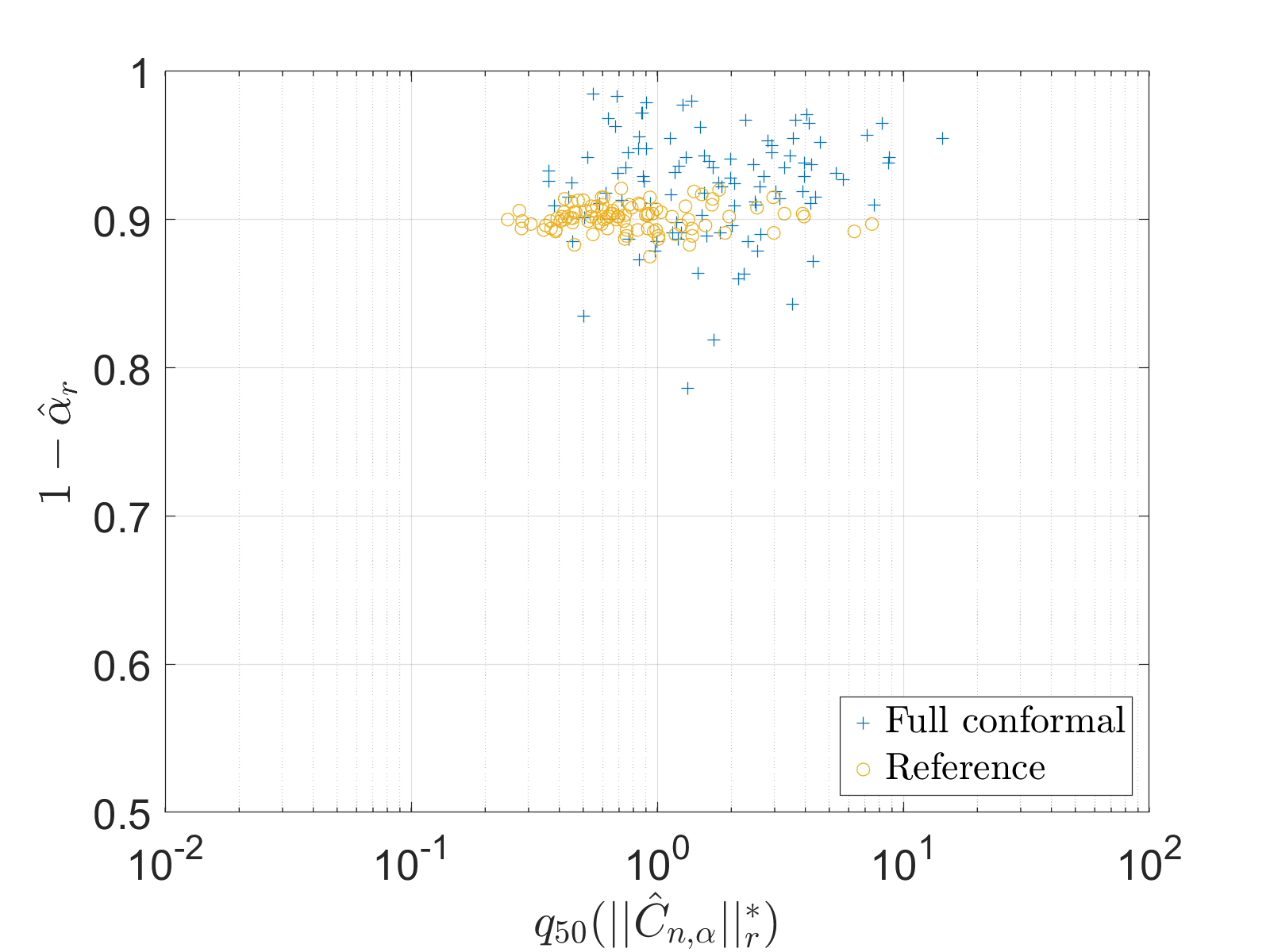}  
        \caption{Full conformal}
        \label{fig:spce_ishi_w_fc}
    \end{subfigure}
    \begin{subfigure}[b]{0.49\textwidth} 
        \centering
        \includegraphics[width=\textwidth]{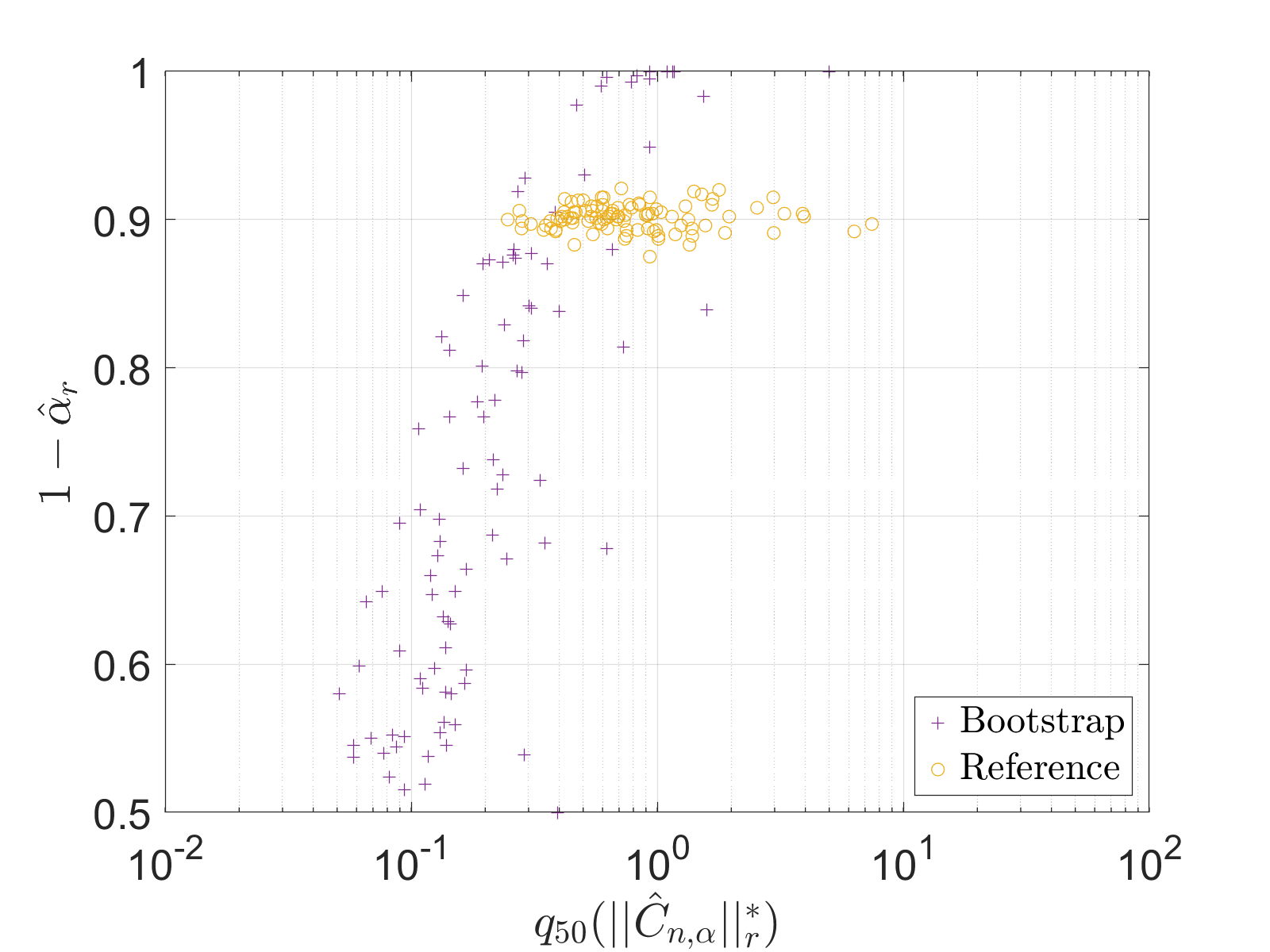} 
        \caption{Bootstrap}
        \label{fig:spce_ishi_w_boot}
    \end{subfigure}
    \begin{subfigure}[b]{0.49\textwidth} 
    \centering
    \includegraphics[width=\textwidth]{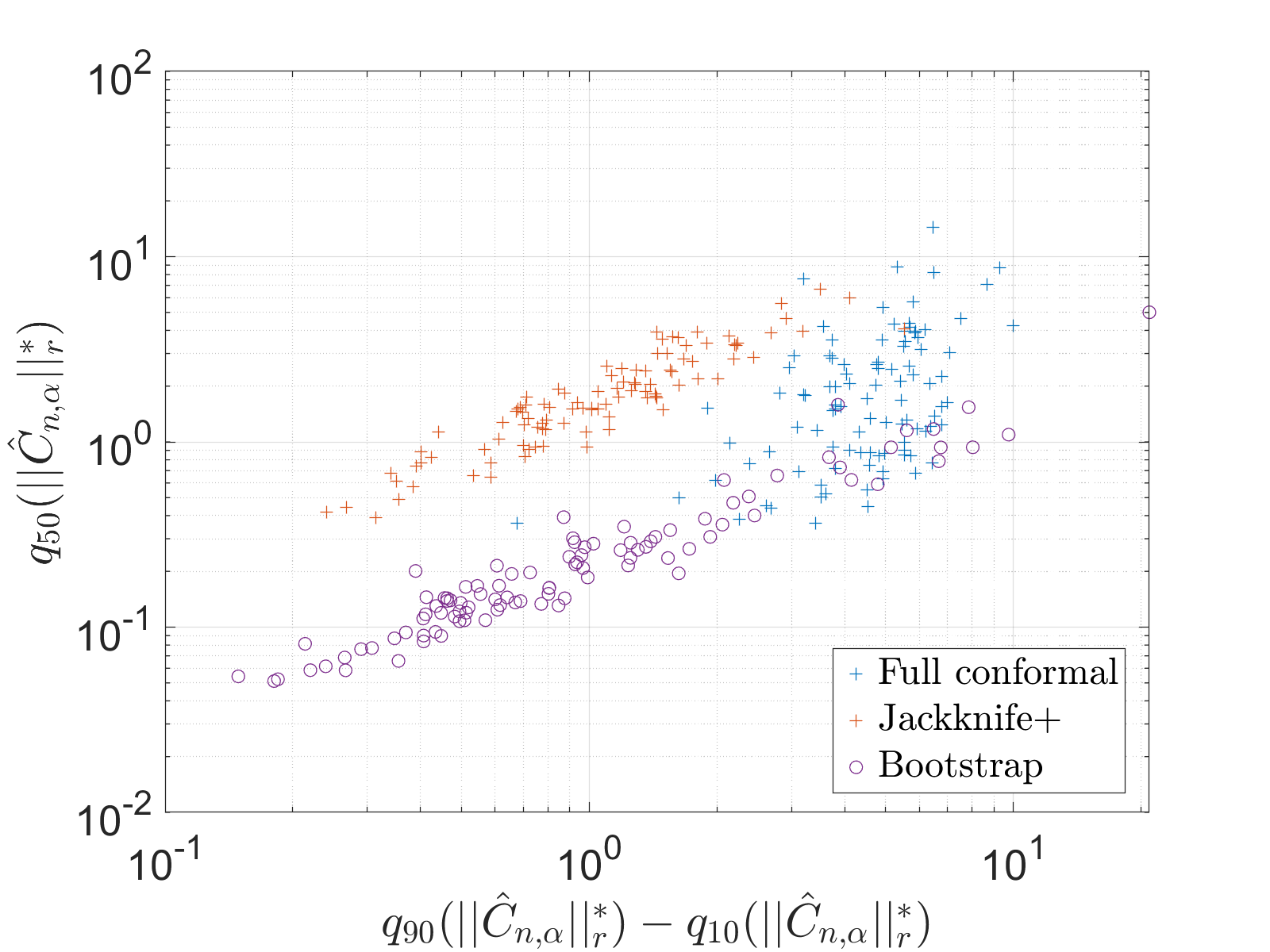}  
    \caption{Spread of prediction interval widths}
    \label{fig:spce_ishi_iq}
    \end{subfigure}

    \caption{Comparison of the normalized interval width of prediction intervals at target coverage level $1-\alpha=0.9$ built based on the sparse PCE surrogate of the Ishigami function. Settings of the experiment: $\ned=40$, $\nval=1{,}000$, PCE degree $p=6$, number of regressors $P\approx22$, $\errval\approx 1\cdot 10^{-1}$. }
    \label{fig:spce_ishi_width}
\end{figure}

A comparable trend emerges when analyzing the distribution of prediction interval widths across methods, as illustrated in \Cref{fig:spce_ishi_width}. 

In this example, the Jackknife+ prediction intervals tend to be somewhat conservative, achieving coverage levels above the target, which consequently leads to intervals slightly wider than those of the reference case (\Cref{fig:spce_ishi_w_j+}). Conversely, the full conformal approach provides coverage closer to the target and prediction interval widths close to the reference, as shown in \Cref{fig:spce_ishi_w_fc}. Finally, the bootstrap method, although occasionally achieving near-complete coverage in certain replications, suffers from excessive variability, leading to prediction intervals with widely fluctuating widths and sometimes very poor coverage (\Cref{fig:spce_ishi_w_boot}). 
\Cref{fig:spce_ishi_iq} provides additional insight on the prediction interval widths. On average, Jackknife+ provides intervals that are of similar width to those from full conformal, and significantly larger than those from bootstrap, as seen from the plot of the y-coordinate $q_{50}(\norm{\cna}^*_r)$. However, the spread of interval widths over the $\nval$ points is one order of magnitude smaller when using Jackknife+, compared to the spread observed for full conformal. In other words Jackknife+ provides more homogeneous (over the validation set) intervals.

\subsubsection{Borehole function}
\label{sec:spce_results_bore}

\begin{figure}[H] 
    \centering
        \begin{subfigure}[b]{0.49\textwidth}
        \centering
        \includegraphics[width=\textwidth]{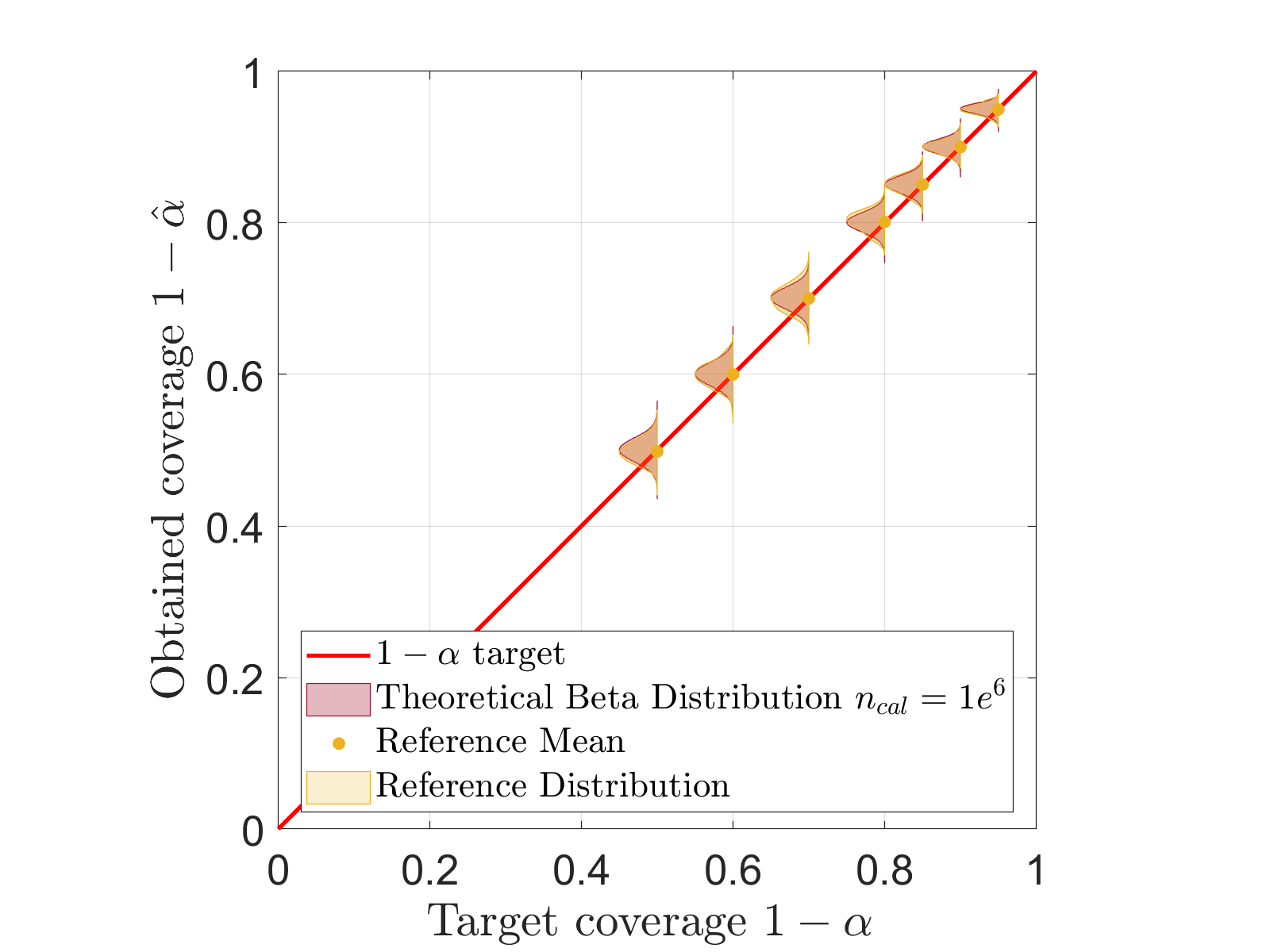}  
        \caption{Reference case}
        \label{fig:spce_bore_ref}
    \end{subfigure}
    \hfill
    \begin{subfigure}[b]{0.49\textwidth} 
        \centering
        \includegraphics[width=\textwidth]{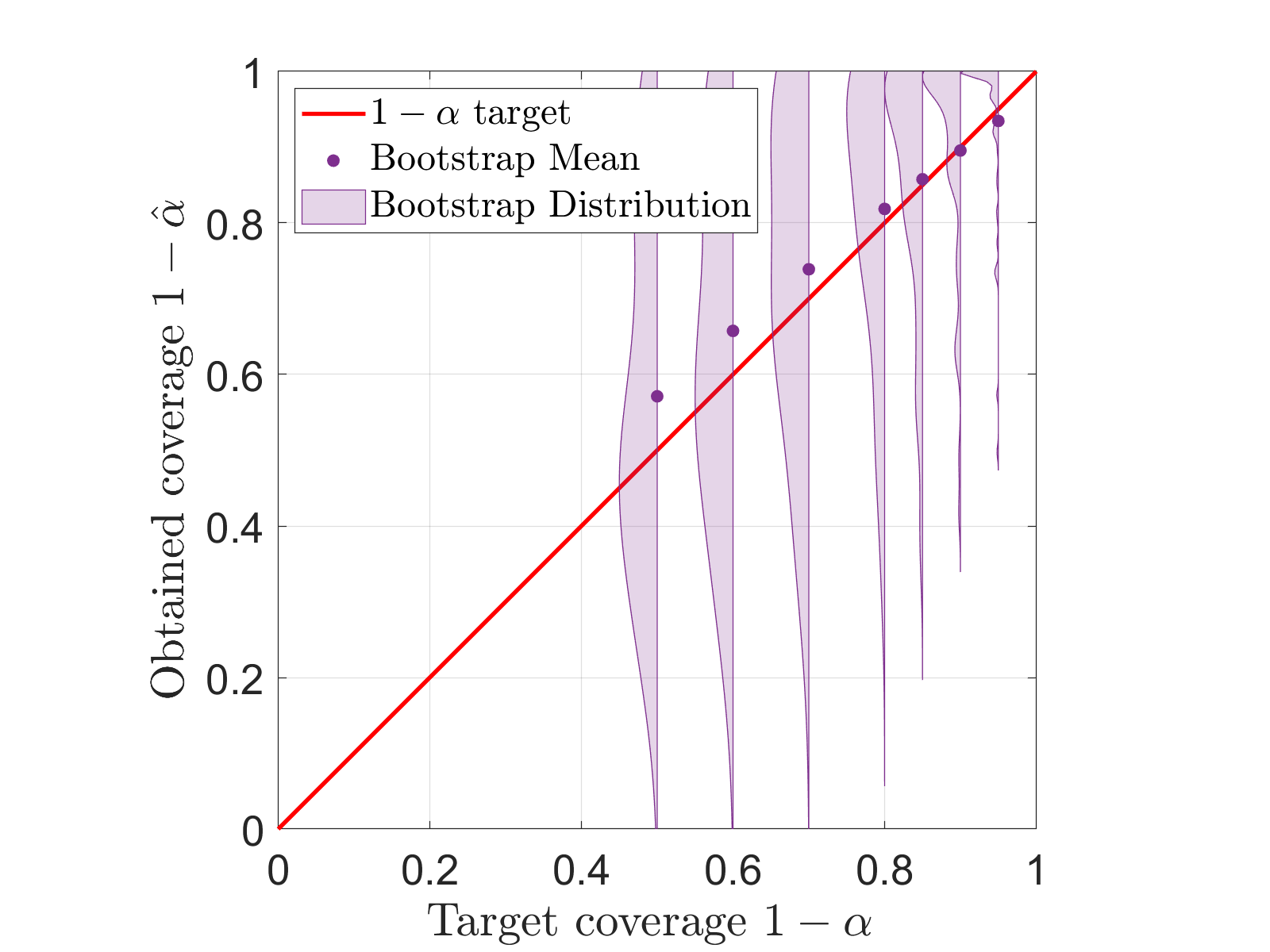}    
        \caption{Bootstrap}
        \label{fig:spce_bore_boot}
    \end{subfigure}
    \begin{subfigure}[b]{0.49\textwidth}
        \centering
        \includegraphics[width=\textwidth]{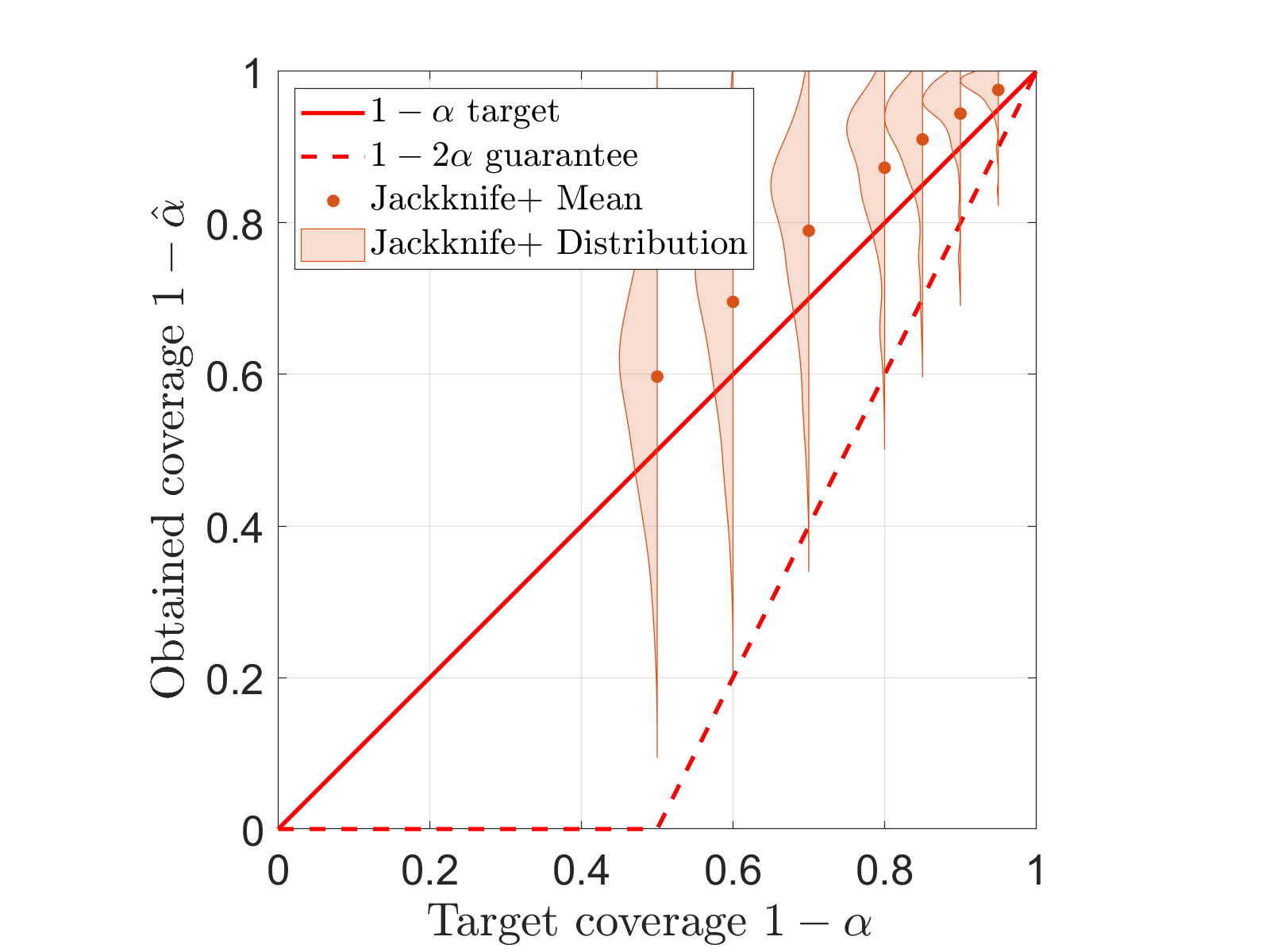}  
        \caption{Jackknife+}
        \label{fig:spce_bore_j+}
    \end{subfigure}
    \hfill
    \begin{subfigure}[b]{0.49\textwidth}
        \centering
        \includegraphics[width=\textwidth]{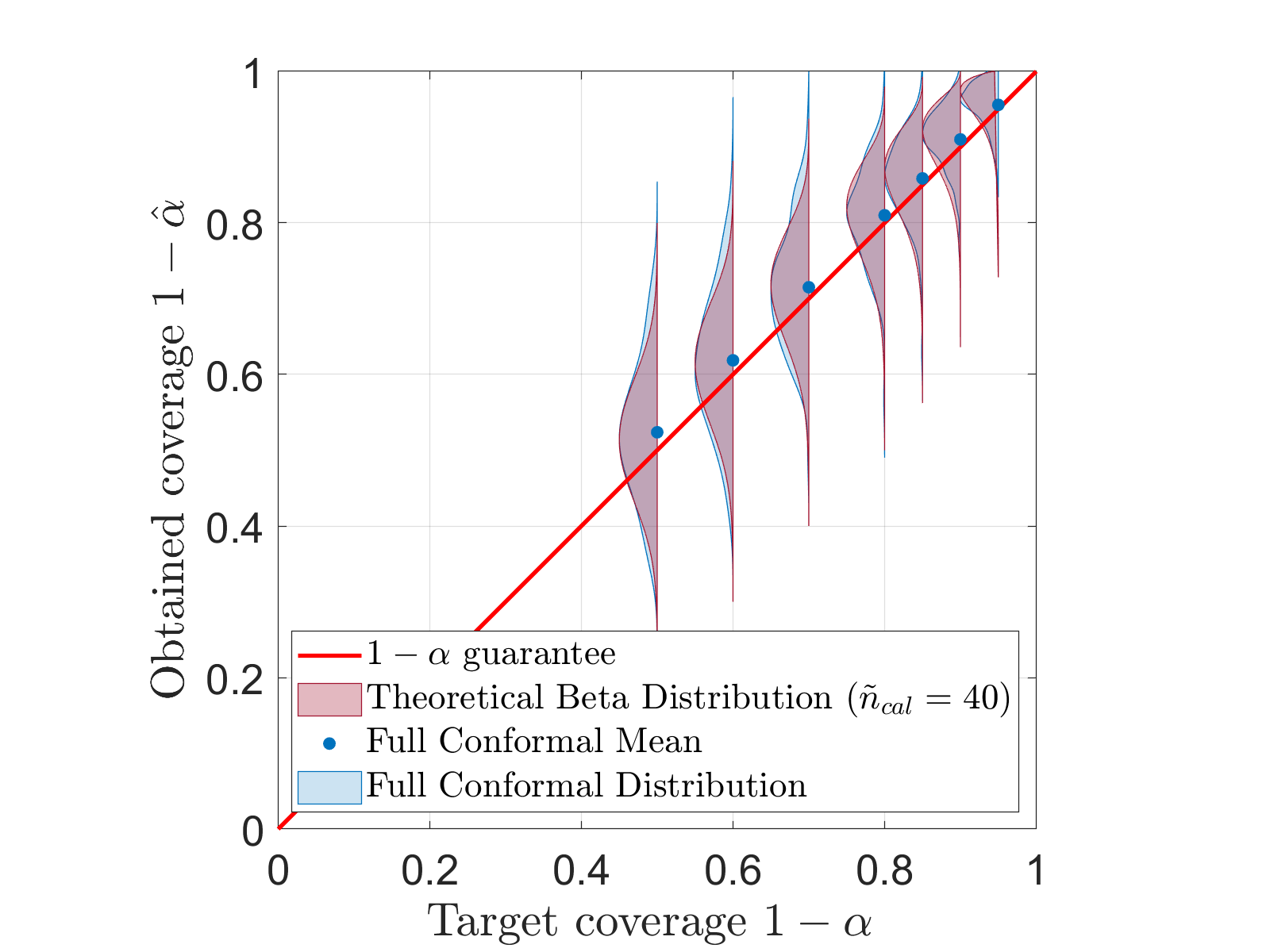} 
        \caption{Full conformal}
        \label{fig:spce_bore_fc}
    \end{subfigure}
    \caption{Evaluation of prediction intervals built based on the sparse PCE surrogate of the Borehole function. Settings of the experiment: $\ned=40$, $\nval=1{,}000$, PCE degree $p=2$, number of regressors $P\approx27$, $\errval\approx 1\cdot 10^{-3}$, $\nr=100$.}
    \label{fig:spce_bore}
\end{figure}

The adapted procedure for sparse PCE is also applied to the Borehole function, where the sparsity-of-effects principle becomes even more pronounced. In this case, experimental designs consisting of $\ned=40$ points are generated by Latin Hypercube Sampling. A sparse PCE surrogate of degree $p=2$ is then constructed, achieving a moderate predictive performance with a validation error $\errval \approx 10^{-3}$. The experiment is again repeated $\nr=100$~times in order to account for statistical uncertainty in the training set. The obtained coverages are evaluated for all methods and presented in \Cref{fig:spce_bore}.

Similar to the observation in the previous case, both the Jackknife+ and full conformal methods demonstrate consistency with their theoretical coverage guarantees, reinforcing their reliability (\Cref{fig:spce_bore_fc,fig:spce_bore_j+}). The distribution of coverage of the full conformal method could be achieved by the split conformal method with an additional calibration set of size $\Tilde{n}_{\textrm{cal}}=40$. In contrast, the bootstrap-based prediction intervals exhibit significant variability in their achieved coverage across different replications, thereby exhibiting poor robustness (\Cref{fig:spce_bore_boot}).

\begin{figure}[H] 
    \centering
    \begin{subfigure}[b]{0.49\textwidth}
        \centering
        \includegraphics[width=\textwidth]{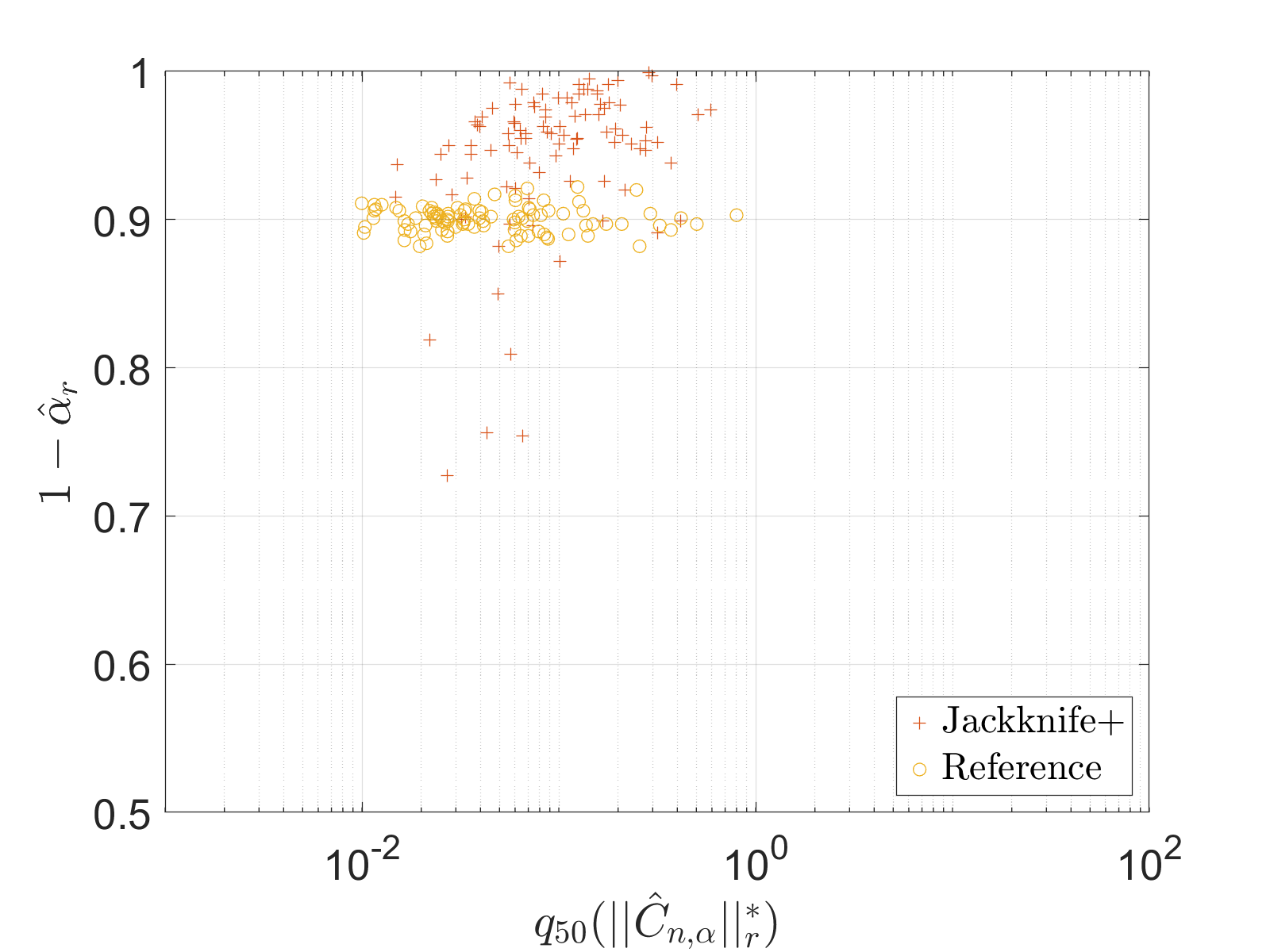} 
        \caption{Jackknife+}
        \label{fig:spce_bore_w_j+}
    \end{subfigure}
    \hfill
    \begin{subfigure}[b]{0.49\textwidth}
        \centering
        \includegraphics[width=\textwidth]{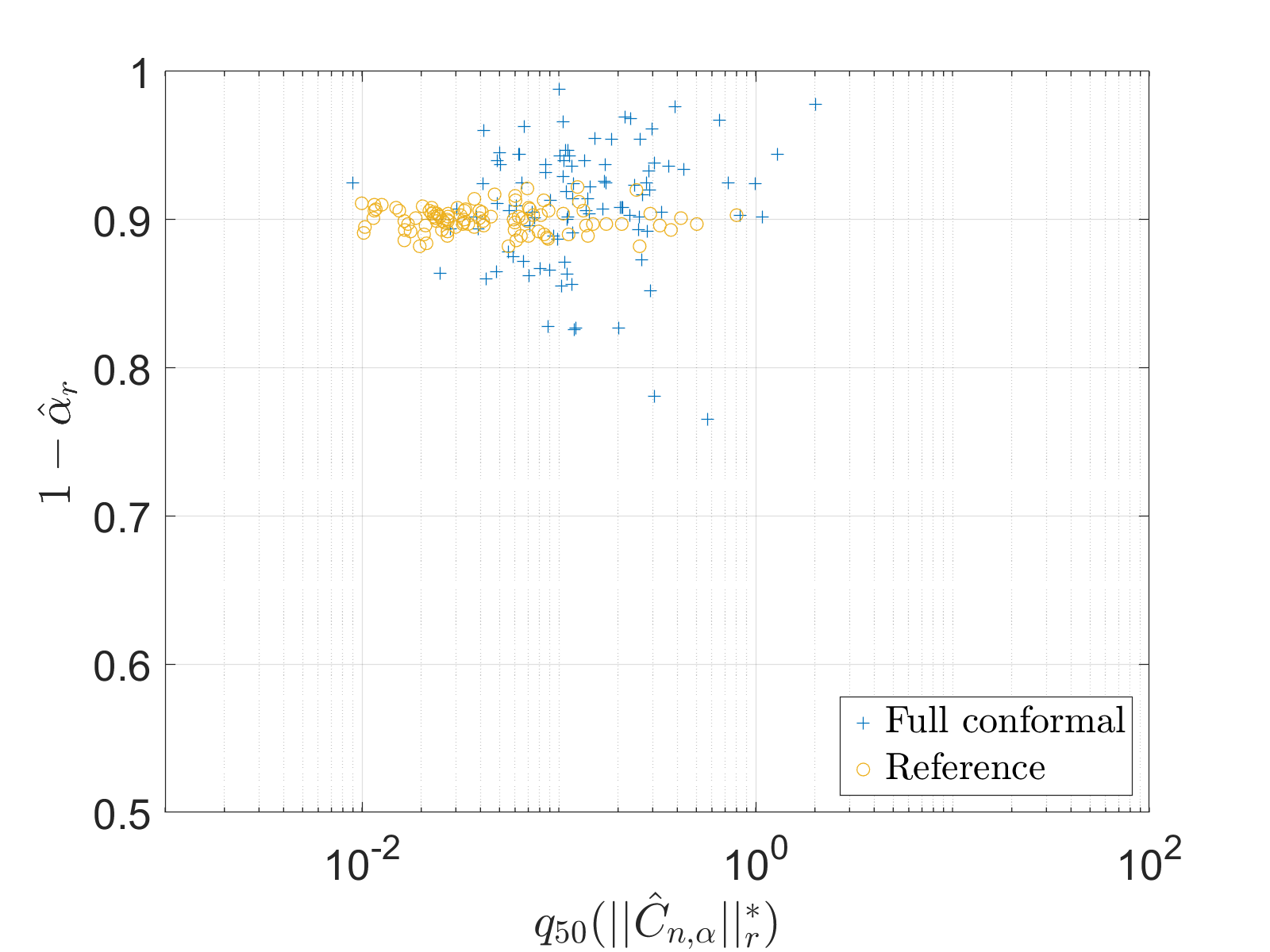}  
        \caption{Full conformal}
        \label{fig:spce_bore_w_fc}
    \end{subfigure}
    \begin{subfigure}[b]{0.49\textwidth} 
        \centering
        \includegraphics[width=\textwidth]{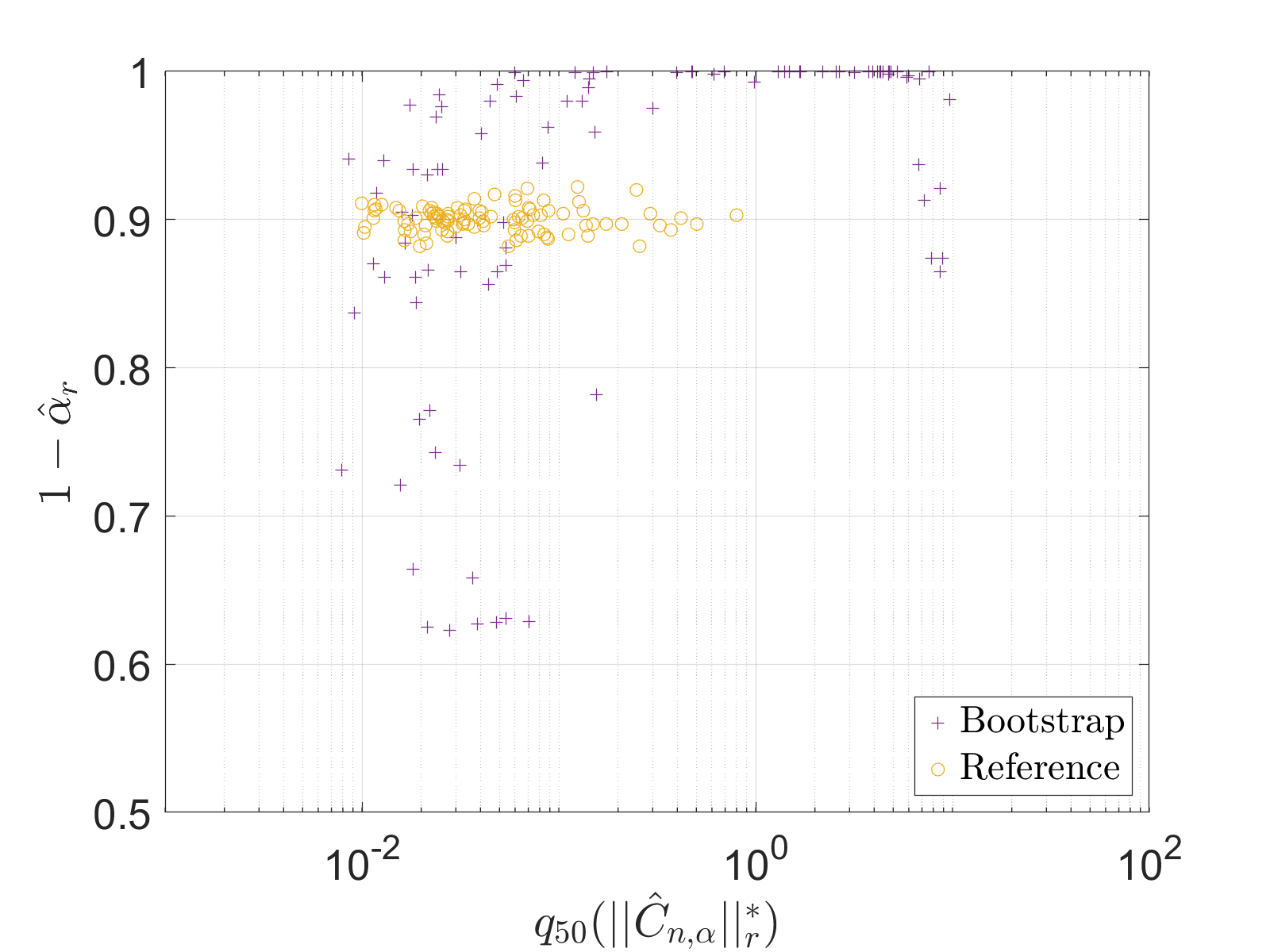} 
        \caption{Bootstrap}
        \label{fig:spce_bore_w_boot}
    \end{subfigure}
    \begin{subfigure}[b]{0.49\textwidth} 
    \centering
    \includegraphics[width=\textwidth]{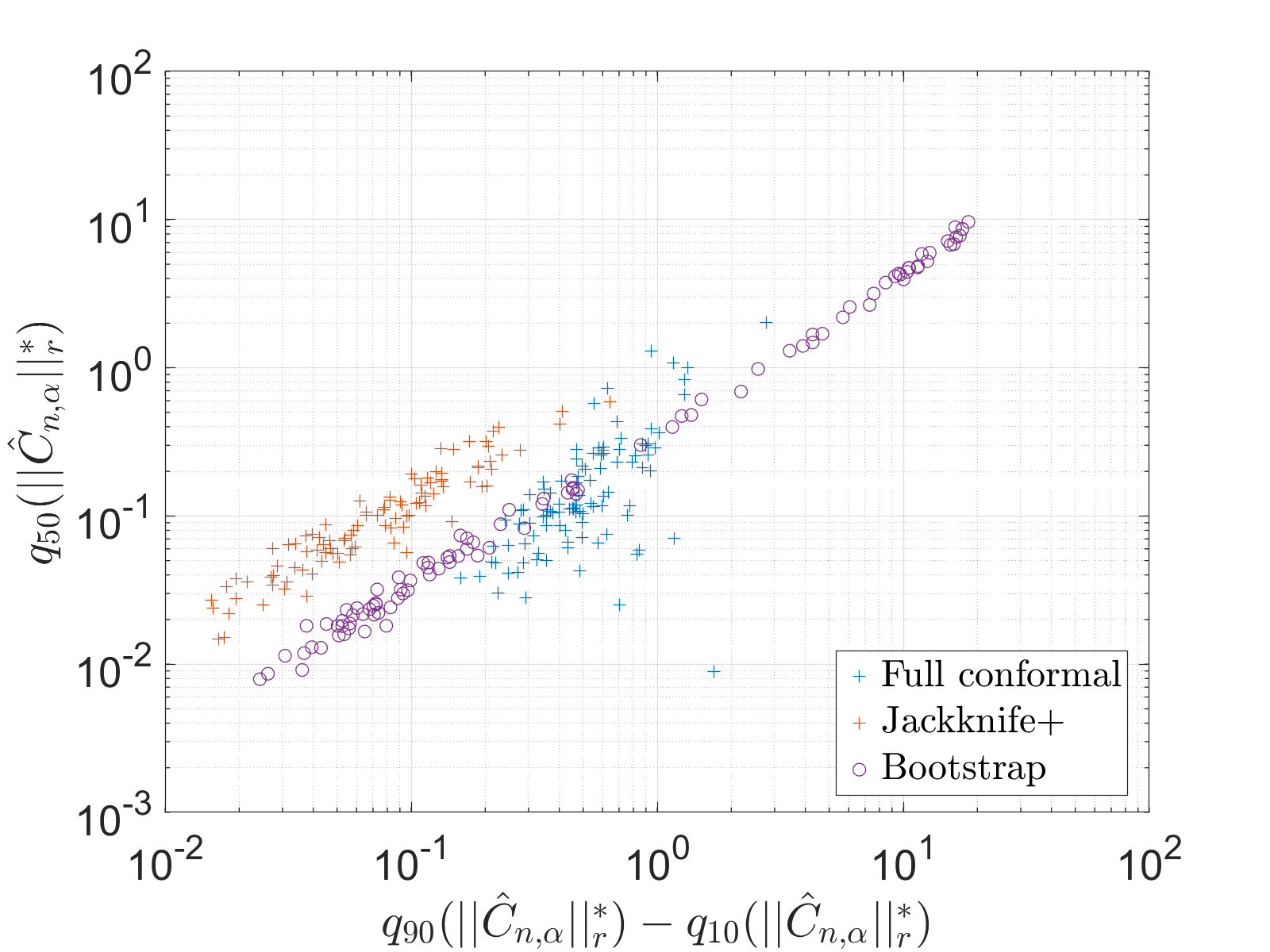}  
    \caption{Spread of prediction interval widths}
    \label{fig:spce_bore_iq}
    \end{subfigure}
    
    \caption{Comparison of the normalized interval width of prediction intervals built based on the sparse PCE surrogate of the Borehole function. Settings of the experiment: $\ned=40$, $\nval=1{,}000$, PCE degree $p=2$,  number of regressors $P\approx27$, $\errval\approx 1\cdot 10^{-3}$. }
    \label{fig:spce_bore_width}
\end{figure}

A similar pattern is observed when examining the distribution of prediction interval widths across methods for the Borehole function, as shown in \Cref{fig:spce_bore_width}. 

The full conformal and Jackknife+ prediction intervals achieve the target coverage with a median interval width close to that of the reference case (\Cref{fig:spce_bore_w_fc,fig:spce_bore_w_j+}). The bootstrap method remains highly inconsistent, occasionally attaining near-total coverage in some replications but displaying substantial variability in median interval widths, further highlighting its lack of reliability (\Cref{fig:spce_bore_w_boot}). Finally, the full conformal procedure provides a larger spread (i.e, spatial heterogeneity) of prediction interval widths than the Jackknife+, as depicted by the difference in their respective interquantile ranges of prediction interval widths (\Cref{fig:spce_bore_iq}).

\section{Discussion}

This section presents a discussion of the implemented methods, beginning with an analysis of their computational efficiency, followed by an examination of the local heterogeneity of the created conformal prediction intervals.

\subsection{Computational requirements}

The various examples have demonstrated the validity and reliability of the proposed conformal prediction methods for both full and sparse PCE. Both approaches successfully achieve their theoretical coverage guarantees, yet differences emerge in their adaptivity and computational efficiency. Notably, the full conformal method generates prediction intervals that are more heterogeneous (that is, with more scattered widths across a validation set) compared to the Jackknife+ procedure. The Jackknife+ approach tends to produce prediction interval widths with a relatively narrow spread across the input space, especially in the full PCE case, where interval widths remain nearly constant over the validation set. A similar behavior is observed in the sparse PCE setting, even though the LARS algorithm induces additional variability in the construction of these intervals. In contrast, the full conformal method allows for more local variability in prediction interval widths by design, as observed in the studied examples.  

However, the enhanced flexibility of the full conformal approach comes at a higher computational cost since generating prediction intervals requires solving \cref{eq:min1,eq:min2}. Nevertheless, the proposed approach, which leverages the homotopy path of the LASSO algorithm, renders the full conformal procedure computationally feasible even for sparse PCEs. To provide a more tangible comparison, execution times for generating prediction intervals using both methods are reported in \Cref{tab:2}. These values correspond to the time required to compute prediction intervals for $\nval=1{,}000$ validation points, based on the same experimental designs and with a target confidence level of $1-\alpha=0.9$, across all cases discussed in \Cref{sec:fpce_results,sec:spce_results}. All calculations have been carried out with a laptop equipped with an $\textnormal{Intel}^\textrm{R}~\textnormal{Core}^\textrm{TM}~\textnormal{Ultra}~7~155\textnormal{H}~1.40~\textnormal{GHz}$ processor.

\begin{table}[H]
\centering
\small
\caption{Execution times of prediction interval generation using an $\textnormal{Intel}^\textrm{R}~\textnormal{Core}^\textrm{TM}~\textnormal{Ultra}~7~155\textnormal{H}~1.40~\textnormal{GHz}$~(serial execution)}
\begin{tabular}{llll}
\hline
{PCE type} & {Model} & {Execution time Jackknife+ $[s]$} & {Execution time full conformal $[s]$} \\ \hline

Full & Ishigami & $0.53$ & $0.63$ \\ 
Full & Borehole & $0.47$ & $0.59$ \\ 
Sparse & Ishigami & $0.50$ & $5.58$ \\ 
Sparse & Borehole & $1.09$ & $6.72$ \\ \hline
\end{tabular}
\label{tab:2}
\end{table}

For full PCE, both conformal prediction methods exhibit comparable computational complexity. The computational cost of the full conformal approach increases significantly in the case of sparse PCE because it requires solving an additional optimization problem. This raises an important question: to what extent do the prediction intervals generated by this more expensive method exhibit higher local variability of width?

\subsection{Adaptivity of prediction intervals}

To address this question, we assess the variability of the prediction interval widths, and in particular their ability to widen in regions where the surrogate model is less accurate and to narrow in regions of higher accuracy. We call this feature \emph{adaptivity}. This property of prediction intervals is particularly useful in active-learning contexts, as it enables the identification of critical regions where it is necessary to improve the surrogate model \citep{MarelliSS2018}. Typically, the prediction intervals should be smaller at validation points close to training data points and wider as we move further away from them. This is exactly the motivation behind credible intervals in Gaussian Process regression.

As adaptivity is desired, the prediction interval width should be positively correlated with the true model error in the same point. For PCE, bootstrap resampling has been shown to produce prediction intervals exhibiting such a correlation with the prediction error, as shown by \citet{MarelliSS2018} and successfully deployed for active learning problems. As shown above, the prediction intervals built by bootstrap resampling may not achieve the desired coverage level, but they serve as a fairly good proxy of the local prediction error. Therefore, we investigate the adaptivity of the prediction intervals generated by the Jackknife+ and the full conformal procedures. 

\begin{figure}[H] 
    \centering
    \begin{subfigure}[b]{0.49\textwidth}
       \centering
        \includegraphics[width=\textwidth]{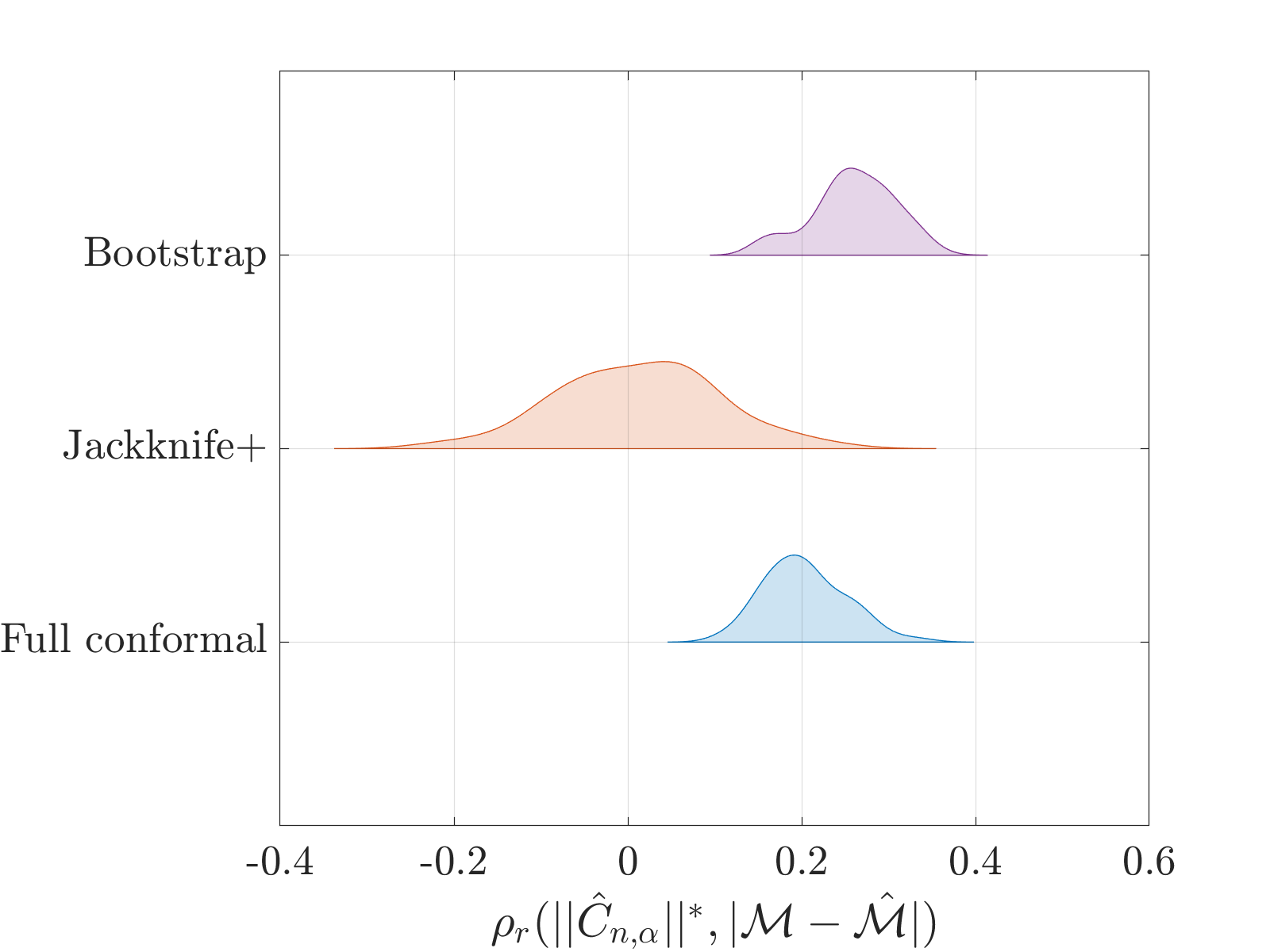}  
        \caption{Ishigami}
    \end{subfigure}
    \hfill
    \begin{subfigure}[b]{0.49\textwidth}
        \centering
        \includegraphics[width=\textwidth]{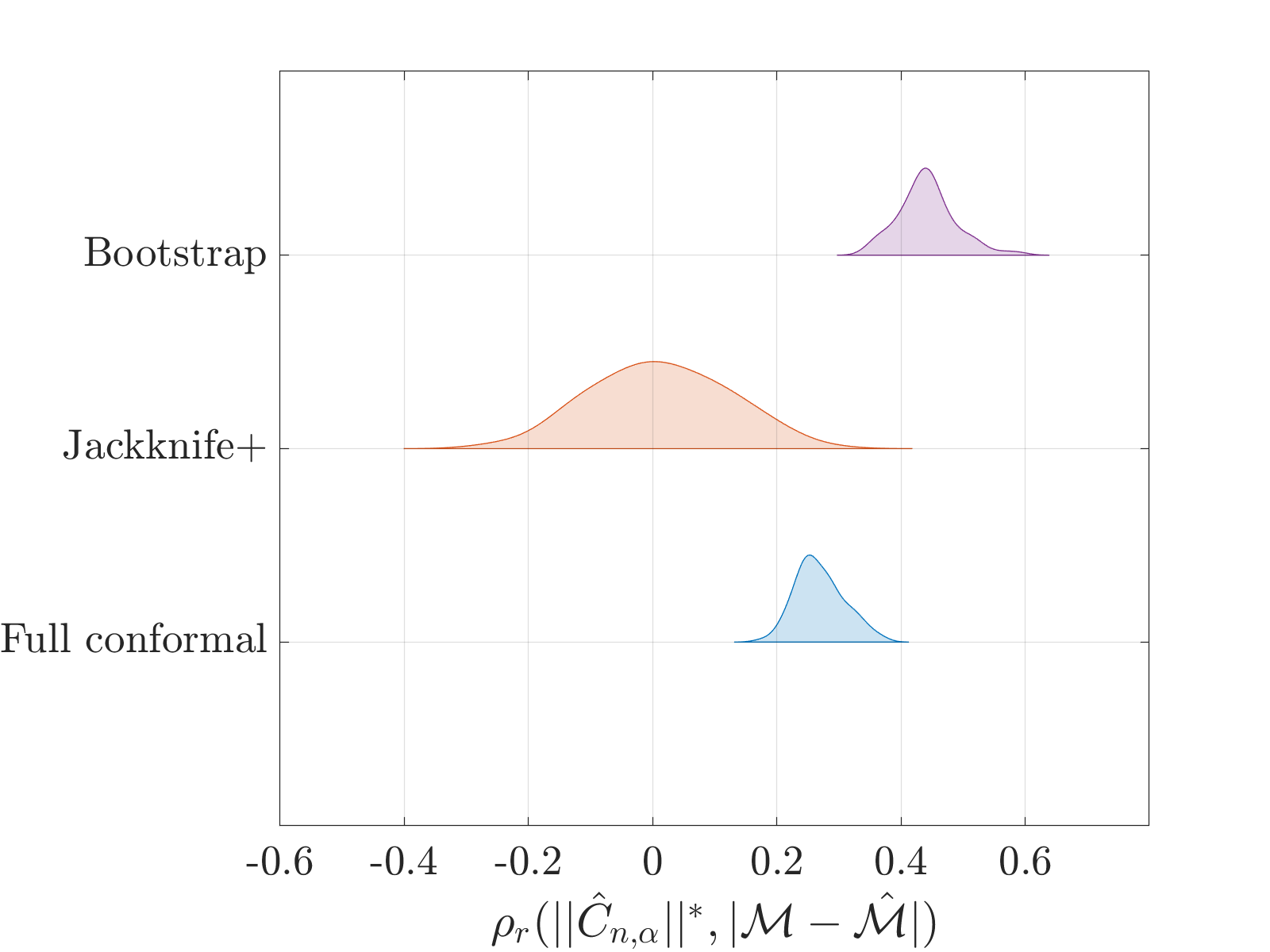}  
        \caption{Borehole}
    \end{subfigure}
    \caption{Correlation between the normalized interval widths of prediction intervals at target coverage level $1-\alpha=0.9$ built based on the full PCE surrogate and the absolute prediction error for the Ishigami (a) and the Borehole (b) functions (settings of the experiment: see \Cref{sec:fpce_results_ishi}~(a) and \Cref{sec:fpce_results_bore}~(b))}
    \label{fig:fpce_correlation}
\end{figure}

\begin{figure}[H] 
    \centering
    \begin{subfigure}[b]{0.49\textwidth}
       \centering
        \includegraphics[width=\textwidth]{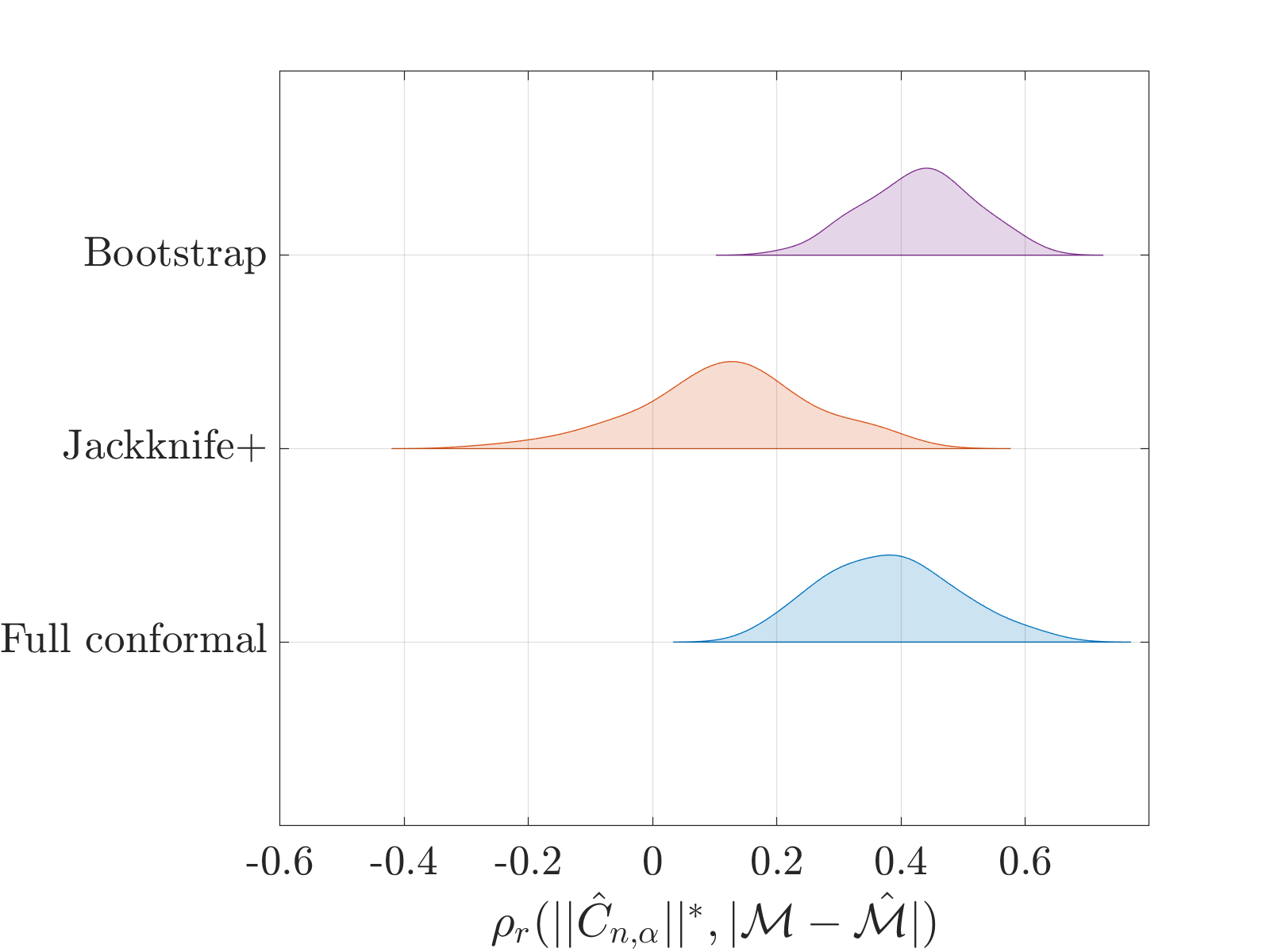}  
        \caption{Ishigami}
    \end{subfigure}
    \hfill
    \begin{subfigure}[b]{0.49\textwidth}
        \centering
        \includegraphics[width=\textwidth]{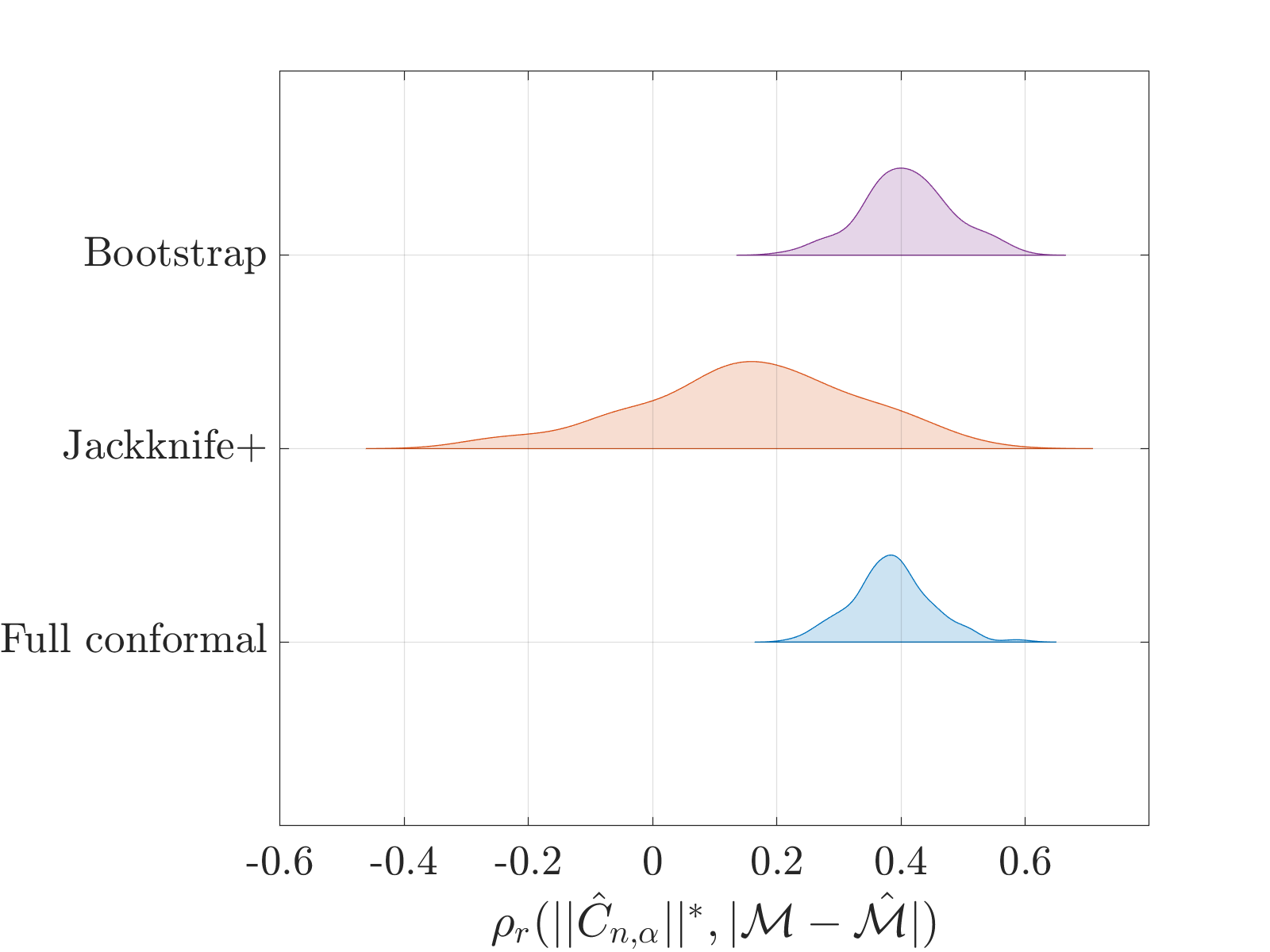}  
        \caption{Borehole}
    \end{subfigure}
    \caption{Correlation between the normalized interval widths of prediction intervals at target coverage level $1-\alpha=0.9$ built based on the sparse PCE surrogate and the absolute prediction error for the Ishigami (a) and the Borehole (b) functions (settings of the experiments: see \Cref{sec:spce_results_ishi}~(a) and \Cref{sec:spce_results_bore}~(b))}
    \label{fig:spce_correlation}
\end{figure}

To this end, we study the association between the normalized interval widths and the absolute prediction residuals for the studied examples, both for full PCE (\Cref{sec:fpce_results}) and sparse PCE (\Cref{sec:spce_results}). Since no linear relationship between the normalized interval widths and the absolute prediction errors is assumed, we compute Spearman’s rank correlation coefficient. Precisely, for each replication $r$, we compute the correlation coefficient between the normalized interval widths at target coverage level $1-\alpha$ and the absolute model error over the validation set, as:
\begin{equation}
    \rho_r = \rho^{\textrm{Spearman}}\left(\acc{ \norm{\hat{C}_{n,\alpha}(\ve{x}_{\textrm{val},i})}^{*}_r,\, i=1\enu \nval};\acc{\abs{\cm(\ve{x}_{\textrm{val},i})-\hat{\cm}_r(\ve{x}_{\textrm{val},i})},\, i=1\enu \nval}\right). 
\end{equation}
For compactness, this quantity is denoted by $\rho_r\left(\norm{\hat{C}_{n,\alpha}}^{*},\,|\cm-\hat{\cm}|\right)$ throughout the remainder of the paper. We represent the distribution of these correlation coefficients among $\nr=100$ replications to account for statistical variability. The results for full PCE are presented in \Cref{fig:fpce_correlation}. 
Analogously, the results for sparse PCE are presented in \Cref{fig:spce_correlation}.

For both the full and sparse PCE surrogates, and for the Ishigami and Borehole functions, very similar conclusions hold. First, the width of the prediction intervals generated by the Jackknife+ method are essentially \emph{not correlated} to the prediction error. In fact, the correlation is close to zero, and oscillates between positive and negative values across replications. Therefore, the Jackknife+ prediction intervals can serve as an input-domain-wise prediction error proxy, with guaranteed coverage properties, but cannot not be used when local adaptivity is desired.  

As expected from existing literature, the correlation coefficient is positive for the prediction intervals from the bootstrap method. For the full conformal approach, the correlation coefficient between the prediction interval width and the absolute error is positive, yet with smaller values than those of the bootstrap method (especially for the full PCE setting, as shown in \Cref{fig:fpce_correlation}). Nevertheless, the replications do not present negative correlations, underlining the robustness of the conclusions. Therefore, as shown in \Cref{sec:fpce_results,sec:spce_results}, the full conformal approach provides intervals with a more variable width than the Jackknife+ method, which is positively correlated with the local prediction error. Although correlation coefficients can be relatively small, they are in the same order of magnitude as those from the bootstrap method, which has been demonstrated to be sufficient for active learning purposes \citep{MarelliSS2018}. As a conclusion, the full conformal prediction intervals perform similarly to the ones provided by bootstrap in terms of local adaptivity and are more robust as they offer additional coverage guarantees.

\newpage 
\section{Conclusion}
In this work, we addressed the challenge of quantifying \emph{local prediction uncertainty} in polynomial chaos expansions (PCEs), a crucial aspect of surrogate modeling that remains insufficiently explored in the literature. While bootstrap resampling provides an estimate of prediction uncertainty, it lacks rigorous statistical guarantees, particularly for small training datasets. As an alternative, we developed \emph{conformalized} full and sparse PCEs, which is the first attempt to integrate conformal prediction and PCEs, to our knowledge. 

We developed two conformal prediction methods, the full conformal approach and the Jackknife+, into both full and sparse PCE frameworks. For full PCEs, we introduced computational simplifications and shortcuts to enhance efficiency without compromising statistical validity.

In the case of sparse PCEs, specific adaptations were introduced to address the non-symmetric nature of the regression procedure. For the Jackknife+ approach, these adaptations are straightforward and preserve the theoretical coverage guarantees of the conformal prediction framework. For the full conformal method, we proposed additional computational simplifications; while this modification does not strictly retain the theoretical guarantees, the numerical experiments on benchmark functions demonstrate that the resulting prediction intervals achieve coverage levels that are empirically very close to the nominal targets. Overall, these developments lead to well-calibrated prediction intervals for sparse PCEs, with coverage properties that consistently improve upon those obtained using traditional bootstrap resampling.

The results demonstrate that conformal prediction significantly enhances the reliability of uncertainty quantification in PCEs. The Jackknife+ method tends to be slightly conservative, whereas the full conformal approach provides tighter intervals with better adherence to target coverage. Both methods outperform bootstrap-based approaches in terms of coverage, offering greater robustness across different problem settings and replications. 

The choice between these methods ultimately depends on the specific requirements of the target application. For risk-critical models, where obtaining a reliable prediction interval at an unseen point is essential, especially in regions with limited knowledge, the full conformal procedure should be the preferred approach. Its ability to adapt to local uncertainty makes it particularly well-suited for scenarios demanding precise, yet narrow intervals. However, this method comes with increased computational costs, which may pose a challenge for large-scale applications.

Conversely, when the primary objective is a broader, input-space-wide assessment of prediction interval widths, the Jackknife+ method offers a practical alternative. It provides robust coverage guarantees while maintaining computational efficiency, making it a suitable choice for applications where a balance between accuracy and computational feasibility is required.

While this study demonstrates the effectiveness of conformal prediction in enhancing PCE surrogate models, several potential research directions remain open for exploration. One promising direction is extending these methods to more complex surrogate models, such as multi-fidelity PCE frameworks, where uncertainty quantification remains a significant challenge. Additionally, these techniques could be integrated into adaptive experimental design strategies, further improving model reliability in data-scarce environments.

\section*{Acknowledgments}
This project is part of the Optimization, Reliability And Calibration using Emulators of Stochastic computational models (ORACLES) project and has received funding from the Swiss National Science Foundation under Grant Agreement No. 10004826.


\bibliographystyle{chicago}
\bibliography{thebibliography}

\begin{thebibliography}{62}
\expandafter\ifx\csname natexlab\endcsname\relax\def\natexlab#1{#1}\fi
\providecommand{\url}[1]{\texttt{#1}}
\providecommand{\href}[2]{#2}
\providecommand{\path}[1]{#1}
\providecommand{\DOIprefix}{doi:}
\providecommand{\ArXivprefix}{arXiv:}
\providecommand{\URLprefix}{URL: }
\providecommand{\Pubmedprefix}{pmid:}
\providecommand{\doi}[1]{\href{http://dx.doi.org/#1}{\path{#1}}}
\providecommand{\Pubmed}[1]{\href{pmid:#1}{\path{#1}}}
\providecommand{\bibinfo}[2]{#2}
\ifx\xfnm\relax \def\xfnm[#1]{\unskip,\space#1}\fi
\bibitem[{An and Owen(2001)}]{An:Owen:2001}
\bibinfo{author}{An, J.}, \bibinfo{author}{Owen, A.}, \bibinfo{year}{2001}.
\newblock \bibinfo{title}{Quasi-regression}.
\newblock \bibinfo{journal}{Journal of Complexity} \bibinfo{volume}{17},
  \bibinfo{pages}{588--607}.
\bibitem[{Angelopoulos and Bates(2023)}]{Angelopoulos2023}
\bibinfo{author}{Angelopoulos, A.}, \bibinfo{author}{Bates, S.},
  \bibinfo{year}{2023}.
\newblock \bibinfo{title}{Conformal prediction: a gentle introduction}.
\newblock \bibinfo{journal}{Foundations and Trends in Machine Learning}
  \bibinfo{volume}{16}, \bibinfo{pages}{494--591}.
\bibitem[{Barber et~al.(2021)Barber, Candès, Ramdas and
  Tibshirani}]{Barber2021}
\bibinfo{author}{Barber, R.}, \bibinfo{author}{Candès, E.},
  \bibinfo{author}{Ramdas, A.}, \bibinfo{author}{Tibshirani, R.},
  \bibinfo{year}{2021}.
\newblock \bibinfo{title}{Predictive inference with the {J}ackknife+}.
\newblock \bibinfo{journal}{The Annals of Statistics} \bibinfo{volume}{49},
  \bibinfo{pages}{486--507}.
\bibitem[{Berveiller et~al.(2006)Berveiller, Sudret and
  Lemaire}]{Berveiller2006}
\bibinfo{author}{Berveiller, M.}, \bibinfo{author}{Sudret, B.},
  \bibinfo{author}{Lemaire, M.}, \bibinfo{year}{2006}.
\newblock \bibinfo{title}{Stochastic finite elements: a non intrusive approach
  by regression}.
\newblock \bibinfo{journal}{European Journal of Computational Mechanics}
  \bibinfo{volume}{15}, \bibinfo{pages}{81--92}.
\bibitem[{Blatman(2009)}]{BlatmanThesis}
\bibinfo{author}{Blatman, G.}, \bibinfo{year}{2009}.
\newblock \bibinfo{title}{Adaptive sparse polynomial chaos expansions for
  uncertainty propagation and sensitivity analysis}.
\newblock Ph.D. thesis. Universit\'e Blaise Pascal, Clermont-Ferrand.
\bibitem[{Blatman and Sudret(2010)}]{Blatman2010a}
\bibinfo{author}{Blatman, G.}, \bibinfo{author}{Sudret, B.},
  \bibinfo{year}{2010}.
\newblock \bibinfo{title}{Efficient computation of global sensitivity indices
  using sparse polynomial chaos expansions}.
\newblock \bibinfo{journal}{Reliability Engineering {\&} System Safety}
  \bibinfo{volume}{95}, \bibinfo{pages}{1216--1229}.
\bibitem[{Blatman and Sudret(2011)}]{BlatmanJCP2011}
\bibinfo{author}{Blatman, G.}, \bibinfo{author}{Sudret, B.},
  \bibinfo{year}{2011}.
\newblock \bibinfo{title}{Adaptive sparse polynomial chaos expansion based on
  {L}east {A}ngle {R}egression}.
\newblock \bibinfo{journal}{Journal of Computational Physics}
  \bibinfo{volume}{230}, \bibinfo{pages}{2345--2367}.
\bibitem[{Brent(1973)}]{Brent1973}
\bibinfo{author}{Brent, R.P.}, \bibinfo{year}{1973}.
\newblock \bibinfo{title}{Algorithms for Minimization without Derivatives}.
\newblock \bibinfo{publisher}{Prentice Hall Inc}.
\bibitem[{Candès and Wakin(2008)}]{Candes_sparsity}
\bibinfo{author}{Candès, E.}, \bibinfo{author}{Wakin, M.},
  \bibinfo{year}{2008}.
\newblock \bibinfo{title}{An introduction to compressive sampling: {A}
  sensing/sampling paradigm that goes against the common knowledge in data
  acquisition}.
\newblock \bibinfo{journal}{IEEE Signal Processing Magazine}
  \bibinfo{volume}{25}, \bibinfo{pages}{21--30}.
\bibitem[{Dai and Milenkovic(2009)}]{Dai2009}
\bibinfo{author}{Dai, W.}, \bibinfo{author}{Milenkovic, O.},
  \bibinfo{year}{2009}.
\newblock \bibinfo{title}{Subspace pursuit for compressive sensing signal
  reconstruction}.
\newblock \bibinfo{journal}{IEEE Transactions on Information Theory}
  \bibinfo{volume}{55}, \bibinfo{pages}{2230--2249}.
\bibitem[{Deb et~al.(2001)Deb, Babu\u{s}ka and Oden}]{Babuska01}
\bibinfo{author}{Deb, M.}, \bibinfo{author}{Babu\u{s}ka, I.},
  \bibinfo{author}{Oden, J.}, \bibinfo{year}{2001}.
\newblock \bibinfo{title}{Solution of stochastic partial differential equations
  using {Galerkin} finite element techniques}.
\newblock \bibinfo{journal}{Computer Methods in Applied Mechanics and
  Engineering} \bibinfo{volume}{190}, \bibinfo{pages}{6359--6372}.
\bibitem[{Demay and Iooss(2021)}]{Demay2021}
\bibinfo{author}{Demay, C.}, \bibinfo{author}{Iooss, B.}, \bibinfo{year}{2021}.
\newblock \bibinfo{title}{Model selection based on validation criteria for
  {G}aussian process regression: An application with highlights on the
  predictive variance}.
\newblock \bibinfo{journal}{Quality and Reliability Engineering International}
  \bibinfo{volume}{38}, \bibinfo{pages}{1482--1500}.
\bibitem[{Diaz et~al.(2018)Diaz, Doostan and Hampton}]{Diaz2018}
\bibinfo{author}{Diaz, P.}, \bibinfo{author}{Doostan, A.},
  \bibinfo{author}{Hampton, J.}, \bibinfo{year}{2018}.
\newblock \bibinfo{title}{Sparse polynomial chaos expansions via compressed
  sensing and {D}-optimal design}.
\newblock \bibinfo{journal}{Computer Methods in Applied Mechanics and
  Engineering} \bibinfo{volume}{336}, \bibinfo{pages}{640--666}.
\bibitem[{Doostan and Owhadi(2011)}]{Doostan2011}
\bibinfo{author}{Doostan, A.}, \bibinfo{author}{Owhadi, H.},
  \bibinfo{year}{2011}.
\newblock \bibinfo{title}{A non-adapted sparse approximation of {PDEs} with
  stochastic inputs}.
\newblock \bibinfo{journal}{Journal of Computational Physics}
  \bibinfo{volume}{230}, \bibinfo{pages}{3015--3034}.
\bibitem[{Efron(1979)}]{Efron1979}
\bibinfo{author}{Efron, B.}, \bibinfo{year}{1979}.
\newblock \bibinfo{title}{{Bootstrap methods: another look at the Jackknife}}.
\newblock \bibinfo{journal}{Annals of Statistics} \bibinfo{volume}{7},
  \bibinfo{pages}{1--26}.
\bibitem[{Efron(1983)}]{Efron1983}
\bibinfo{author}{Efron, B.}, \bibinfo{year}{1983}.
\newblock \bibinfo{title}{Estimating the error rate of a prediction rule:
  Improvement on cross-validation}.
\newblock \bibinfo{journal}{Journal of the American Statistical Association}
  \bibinfo{volume}{78}, \bibinfo{pages}{316--331}.
\bibitem[{Efron et~al.(2004)Efron, Hastie, Johnstone and
  Tibshirani}]{Efron2004}
\bibinfo{author}{Efron, B.}, \bibinfo{author}{Hastie, T.},
  \bibinfo{author}{Johnstone, I.}, \bibinfo{author}{Tibshirani, R.},
  \bibinfo{year}{2004}.
\newblock \bibinfo{title}{Least angle regression}.
\newblock \bibinfo{journal}{Annals of Statistics} \bibinfo{volume}{32},
  \bibinfo{pages}{407--499}.
\bibitem[{Gammerman et~al.(1998)Gammerman, Vovk and Vapnik}]{Vovk1998}
\bibinfo{author}{Gammerman, A.}, \bibinfo{author}{Vovk, V.},
  \bibinfo{author}{Vapnik, V.}, \bibinfo{year}{1998}.
\newblock \bibinfo{title}{Learning by transduction}, in:
  \bibinfo{booktitle}{UAI'98: Proceedings of the Fourteenth conference on
  Uncertainty in artificial intelligence Madison, USA}, pp. \bibinfo{pages}{148
  -- 155}.
\bibitem[{Gasparin and Ramdas(2024)}]{Gasparin2024}
\bibinfo{author}{Gasparin, M.}, \bibinfo{author}{Ramdas, A.},
  \bibinfo{year}{2024}.
\newblock \bibinfo{title}{Merging uncertainty sets via majority vote}
  \bibinfo{note}{ArXiv:2401.09379}.
\bibitem[{Ghanem and Spanos(1991)}]{Ghanem1991}
\bibinfo{author}{Ghanem, R.G.}, \bibinfo{author}{Spanos, P.D.},
  \bibinfo{year}{1991}.
\newblock \bibinfo{title}{Stochastic finite elements: {A} spectral approach}.
\newblock \bibinfo{publisher}{Springer New York}.
\bibitem[{Harper and Gupta(1983)}]{Harper1983}
\bibinfo{author}{Harper, W.V.}, \bibinfo{author}{Gupta, S.K.},
  \bibinfo{year}{1983}.
\newblock \bibinfo{title}{{Sensitivity/uncertainty analysis of a borehole
  scenario comparing Latin hypercube sampling and deterministic sensitivity
  approaches}}.
\newblock \bibinfo{type}{Technical Report} \bibinfo{number}{No. BMI/ONWI-516}.
  Battelle Memorial Institute - Office of Nuclear Waste Isolation.
  \bibinfo{address}{Columbus, OH (USA)}.
\bibitem[{Ishigami and Homma(1990)}]{ishigami_function}
\bibinfo{author}{Ishigami, T.}, \bibinfo{author}{Homma, T.},
  \bibinfo{year}{1990}.
\newblock \bibinfo{title}{An importance quantification technique in uncertainty
  analysis for computer models}, in: \bibinfo{booktitle}{Proceedings of ISUMA,
  First International Symposium on Uncertainty Modelling and Analysis},
  \bibinfo{organization}{University of Maryland}. pp.
  \bibinfo{pages}{398--403}.
\bibitem[{Jaber et~al.(2025)Jaber, Blot, Brunel, Chabridon, Remy, Iooss, Lucor,
  Mougeot and Leite}]{Jaber2025}
\bibinfo{author}{Jaber, E.}, \bibinfo{author}{Blot, V.},
  \bibinfo{author}{Brunel, N.}, \bibinfo{author}{Chabridon, V.},
  \bibinfo{author}{Remy, E.}, \bibinfo{author}{Iooss, B.},
  \bibinfo{author}{Lucor, D.}, \bibinfo{author}{Mougeot, M.},
  \bibinfo{author}{Leite, A.}, \bibinfo{year}{2025}.
\newblock \bibinfo{title}{Conformal approach to {G}aussian process surrogate
  evaluation with marginal coverage guarantees}.
\newblock \bibinfo{journal}{Journal of Machine Learning for Modeling and
  Computing} \bibinfo{volume}{6}, \bibinfo{pages}{37--68}.
\bibitem[{Jakeman et~al.(2015)Jakeman, Eldred and Sargsyan}]{Jakeman2015}
\bibinfo{author}{Jakeman, J.}, \bibinfo{author}{Eldred, M.},
  \bibinfo{author}{Sargsyan, K.}, \bibinfo{year}{2015}.
\newblock \bibinfo{title}{Enhancing {$\ell_1$}-minimization estimates of
  polynomial chaos expansions using basis selection}.
\newblock \bibinfo{journal}{Journal of Computational Physics}
  \bibinfo{volume}{289}, \bibinfo{pages}{18--34}.
\bibitem[{Ji et~al.(2008)Ji, Xue and Carin}]{Ji2008}
\bibinfo{author}{Ji, S.}, \bibinfo{author}{Xue, Y.}, \bibinfo{author}{Carin,
  L.}, \bibinfo{year}{2008}.
\newblock \bibinfo{title}{Bayesian compressive sensing}.
\newblock \bibinfo{journal}{IEEE Transactions on Signal Processing}
  \bibinfo{volume}{56}, \bibinfo{pages}{2346--2356}.
\bibitem[{Kersaudy et~al.(2015)Kersaudy, Sudret, Varsier, Picon and
  Wiart}]{KersaudySudret2015}
\bibinfo{author}{Kersaudy, P.}, \bibinfo{author}{Sudret, B.},
  \bibinfo{author}{Varsier, N.}, \bibinfo{author}{Picon, O.},
  \bibinfo{author}{Wiart, J.}, \bibinfo{year}{2015}.
\newblock \bibinfo{title}{A new surrogate modeling technique combining
  {K}riging and polynomial chaos expansions -- {A}pplication to uncertainty
  analysis in computational dosimetry}.
\newblock \bibinfo{journal}{Journal of Computational Physics}
  \bibinfo{volume}{286}, \bibinfo{pages}{103--117}.
\bibitem[{{Le Ma{\^\i}tre} and Knio(2010)}]{LeMaitreBook2010}
\bibinfo{author}{{Le Ma{\^\i}tre}, O.P.}, \bibinfo{author}{Knio, O.M.},
  \bibinfo{year}{2010}.
\newblock \bibinfo{title}{Spectral Methods for Uncertainty Quantification With
  Applications to Computational Fluid Dynamics}.
\newblock Scientific Computation, \bibinfo{publisher}{Springer},
  \bibinfo{address}{Dordrecht, Netherlands}.
\bibitem[{Lei(2019)}]{Lei2019b}
\bibinfo{author}{Lei, J.}, \bibinfo{year}{2019}.
\newblock \bibinfo{title}{Fast exact conformalization of the lasso using
  piecewise linear homotopy}.
\newblock \bibinfo{journal}{Biometrika} \bibinfo{volume}{16},
  \bibinfo{pages}{749–764}.
\bibitem[{Lei et~al.(2018)Lei, G'Sell, Rinaldo, Tibshirani and
  Wasserman}]{Lei2018}
\bibinfo{author}{Lei, J.}, \bibinfo{author}{G'Sell, M.},
  \bibinfo{author}{Rinaldo, A.}, \bibinfo{author}{Tibshirani, R.},
  \bibinfo{author}{Wasserman, L.}, \bibinfo{year}{2018}.
\newblock \bibinfo{title}{Distribution-free predictive inference for
  regression}.
\newblock \bibinfo{journal}{Journal of the American Statistical Association}
  \bibinfo{volume}{113}, \bibinfo{pages}{1094--1111}.
\bibitem[{Lei and Wasserman(2013)}]{Lei2013}
\bibinfo{author}{Lei, J.}, \bibinfo{author}{Wasserman, L.},
  \bibinfo{year}{2013}.
\newblock \bibinfo{title}{Distribution-free prediction bands for non-parametric
  regression}.
\newblock \bibinfo{journal}{Journal of the Royal Statistical Society Series B:
  Statistical Methodology} \bibinfo{volume}{76}, \bibinfo{pages}{71--96}.
\bibitem[{Liang and Barber(2025)}]{Liang2025}
\bibinfo{author}{Liang, R.}, \bibinfo{author}{Barber, R.},
  \bibinfo{year}{2025}.
\newblock \bibinfo{title}{Algorithmic stability implies training-conditional
  coverage for distribution-free prediction methods}.
\newblock \bibinfo{journal}{The Annals of Statistics} \bibinfo{volume}{53},
  \bibinfo{pages}{1457--1482}.
\bibitem[{Linusson et~al.(2014)Linusson, Johansson and
  L{\"o}fstr{\"o}m}]{Linusson2014_signed_residual}
\bibinfo{author}{Linusson, H.}, \bibinfo{author}{Johansson, U.},
  \bibinfo{author}{L{\"o}fstr{\"o}m, T.}, \bibinfo{year}{2014}.
\newblock \bibinfo{title}{Signed-error conformal regression}, in:
  \bibinfo{booktitle}{Advances in Knowledge Discovery and Data Mining}, pp.
  \bibinfo{pages}{224--236}.
\bibitem[{L\"uthen et~al.(2021)L\"uthen, Marelli and
  Sudret}]{LuethenSIAMJUQ2021}
\bibinfo{author}{L\"uthen, N.}, \bibinfo{author}{Marelli, S.},
  \bibinfo{author}{Sudret, B.}, \bibinfo{year}{2021}.
\newblock \bibinfo{title}{Sparse polynomial chaos expansions: {Literature}
  survey and benchmark}.
\newblock \bibinfo{journal}{SIAM/ASA Journal on Uncertainty Quantification}
  \bibinfo{volume}{9}, \bibinfo{pages}{593--649}.
\bibitem[{L\"uthen et~al.(2022)L\"uthen, Marelli and Sudret}]{LuethenIJUQ2022}
\bibinfo{author}{L\"uthen, N.}, \bibinfo{author}{Marelli, S.},
  \bibinfo{author}{Sudret, B.}, \bibinfo{year}{2022}.
\newblock \bibinfo{title}{Automatic selection of basis-adaptive sparse
  polynomial chaos expansions for engineering applications}.
\newblock \bibinfo{journal}{International Journal for Uncertainty
  Quantification} \bibinfo{volume}{12}, \bibinfo{pages}{49--74}.
\bibitem[{L\"uthen et~al.(2023)L\"uthen, Marelli and Sudret}]{LuethenCMAME2023}
\bibinfo{author}{L\"uthen, N.}, \bibinfo{author}{Marelli, S.},
  \bibinfo{author}{Sudret, B.}, \bibinfo{year}{2023}.
\newblock \bibinfo{title}{A spectral surrogate model for stochastic simulators
  computed from trajectory samples}.
\newblock \bibinfo{journal}{Computer Methods in Applied Mechanics and
  Engineering} \bibinfo{volume}{406}, \bibinfo{pages}{1--29}.
\bibitem[{Marelli et~al.(2021)Marelli, L\"uthen and Sudret}]{Marelli2021}
\bibinfo{author}{Marelli, S.}, \bibinfo{author}{L\"uthen, N.},
  \bibinfo{author}{Sudret, B.}, \bibinfo{year}{2021}.
\newblock \bibinfo{title}{{UQLab user manual -- Polynomial chaos expansions}}.
\newblock \bibinfo{type}{Technical Report}. Chair of Risk, Safety and
  Uncertainty Quantification, ETH Zurich, Switzerland.
\newblock \bibinfo{note}{Report \# UQLab-V1.4-104}.
\bibitem[{Marelli and Sudret(2014)}]{MarelliICVRAM2014}
\bibinfo{author}{Marelli, S.}, \bibinfo{author}{Sudret, B.},
  \bibinfo{year}{2014}.
\newblock \bibinfo{title}{{UQLab}: A framework for uncertainty quantification
  in {Matlab}}, in: \bibinfo{booktitle}{Vulnerability, Uncertainty, and Risk
  (Proc. 2nd Int. Conf. on Vulnerability, Risk Analysis and Management
  {(ICVRAM2014)}, Liverpool, United Kingdom)}, \bibinfo{publisher}{American
  Society of Civil Engineers}. pp. \bibinfo{pages}{2554--2563}.
\bibitem[{Marelli and Sudret(2018)}]{MarelliSS2018}
\bibinfo{author}{Marelli, S.}, \bibinfo{author}{Sudret, B.},
  \bibinfo{year}{2018}.
\newblock \bibinfo{title}{An active-learning algorithm that combines sparse
  polynomial chaos expansions and bootstrap for structural reliability
  analysis}.
\newblock \bibinfo{journal}{Structural Safety} \bibinfo{volume}{75},
  \bibinfo{pages}{67--74}.
\bibitem[{Montgomery(2004)}]{Montgomery:2004}
\bibinfo{author}{Montgomery, D.}, \bibinfo{year}{2004}.
\newblock \bibinfo{title}{Design and analysis of experiments}.
\newblock \bibinfo{publisher}{John Wiley and Sons, New York}.
\bibitem[{Morris et~al.(1993)Morris, Mitchell and Ylvisaker}]{Morris1993}
\bibinfo{author}{Morris, M.D.}, \bibinfo{author}{Mitchell, T.J.},
  \bibinfo{author}{Ylvisaker, D.}, \bibinfo{year}{1993}.
\newblock \bibinfo{title}{Bayesian design and analysis of computer experiments:
  use of derivatives in surface prediction}.
\newblock \bibinfo{journal}{Technometrics} \bibinfo{volume}{35},
  \bibinfo{pages}{243--255}.
\bibitem[{Nouretdinov et~al.(2001)Nouretdinov, Melluish and Vovk}]{Vovk1999b}
\bibinfo{author}{Nouretdinov, I.}, \bibinfo{author}{Melluish, T.},
  \bibinfo{author}{Vovk, V.}, \bibinfo{year}{2001}.
\newblock \bibinfo{title}{Ridge regression confidence machine}, in:
  \bibinfo{booktitle}{ICML '01: Proceedings of the Eighteenth International
  Conference on Machine Learning, Williamstown, USA}, pp. \bibinfo{pages}{385
  -- 392}.
\bibitem[{Olive(2007)}]{Olive_empirical_pi}
\bibinfo{author}{Olive, D.}, \bibinfo{year}{2007}.
\newblock \bibinfo{title}{Prediction intervals for regression models}.
\newblock \bibinfo{journal}{Computational Statistics and Data Analysis}
  \bibinfo{volume}{51}, \bibinfo{pages}{3115--3122}.
\bibitem[{Papadopoulos(2024)}]{Papadopoulos2024}
\bibinfo{author}{Papadopoulos, H.}, \bibinfo{year}{2024}.
\newblock \bibinfo{title}{Guaranteed coverage prediction intervals with
  {G}aussian process regression}.
\newblock \bibinfo{journal}{IEEE Transactions on Pattern Analysis and Machine
  Intelligence} \bibinfo{volume}{46}, \bibinfo{pages}{9072--9083}.
\bibitem[{Papadopoulos et~al.(2002)Papadopoulos, Proedrou, Vovk and
  Gammerman}]{Papadopoulos2002}
\bibinfo{author}{Papadopoulos, H.}, \bibinfo{author}{Proedrou, K.},
  \bibinfo{author}{Vovk, V.}, \bibinfo{author}{Gammerman, A.},
  \bibinfo{year}{2002}.
\newblock \bibinfo{title}{Inductive confidence machines for regression}, in:
  \bibinfo{editor}{Elomaa, T.}, \bibinfo{editor}{Mannila, H.},
  \bibinfo{editor}{Toivonen, H.} (Eds.), \bibinfo{booktitle}{Machine Learning:
  13th European Conference on Machine Learning, Helsinki, Finland}, pp.
  \bibinfo{pages}{345--356}.
\bibitem[{Santner et~al.(2003)Santner, Williams and Notz}]{Santner2003}
\bibinfo{author}{Santner, T.J.}, \bibinfo{author}{Williams, B.J.},
  \bibinfo{author}{Notz, W.I.}, \bibinfo{year}{2003}.
\newblock \bibinfo{title}{The {D}esign and {A}nalysis of {C}omputer
  {E}xperiments}.
\newblock \bibinfo{publisher}{Springer, New York}.
\bibitem[{Sargsyan et~al.(2014)Sargsyan, Safta, Najm, Debusschere, Ricciuto and
  Thornton}]{Sargsyan2014}
\bibinfo{author}{Sargsyan, K.}, \bibinfo{author}{Safta, C.},
  \bibinfo{author}{Najm, H.}, \bibinfo{author}{Debusschere, B.},
  \bibinfo{author}{Ricciuto, D.}, \bibinfo{author}{Thornton, P.},
  \bibinfo{year}{2014}.
\newblock \bibinfo{title}{Dimensionality reduction for complex models via
  {B}ayesian compressive sensing}.
\newblock \bibinfo{journal}{International Journal for Uncertainty
  Quantification} \bibinfo{volume}{4}, \bibinfo{pages}{63--93}.
\bibitem[{Saunders et~al.(1999)Saunders, Gammerman and Vovk}]{Vovk1999}
\bibinfo{author}{Saunders, C.}, \bibinfo{author}{Gammerman, A.},
  \bibinfo{author}{Vovk, V.}, \bibinfo{year}{1999}.
\newblock \bibinfo{title}{Transduction with confidence and credibility}, in:
  \bibinfo{booktitle}{IJCAI'99: Proceedings of the 16th International Joint
  Conference on Artificial Intelligence, Stockholm, Sweden}, pp.
  \bibinfo{pages}{722 -- 726}.
\bibitem[{Schmoyer(1992)}]{Schmoyer_pi}
\bibinfo{author}{Schmoyer, R.}, \bibinfo{year}{1992}.
\newblock \bibinfo{title}{Asymptotically valid prediction intervals for linear
  models}.
\newblock \bibinfo{journal}{Technometrics} \bibinfo{volume}{34},
  \bibinfo{pages}{399--408}.
\bibitem[{Sherman and Morrison(1950)}]{Sherman1950}
\bibinfo{author}{Sherman, J.}, \bibinfo{author}{Morrison, W.J.},
  \bibinfo{year}{1950}.
\newblock \bibinfo{title}{Adjustments of an inverse matrix corresponding to a
  change in one element of a given matrix}.
\newblock \bibinfo{journal}{The Annals of Mathematical Statistics}
  \bibinfo{volume}{21}, \bibinfo{pages}{124--127}.
\bibitem[{Smith(2024)}]{SmithUQBook2014}
\bibinfo{author}{Smith, R.C.}, \bibinfo{year}{2024}.
\newblock \bibinfo{title}{Uncertainty {Q}uantification: {T}heory,
  {I}mplementation, and {A}pplications, {S}econd {E}dition}.
\newblock \bibinfo{publisher}{Society for Industrial and Applied Mathematics}.
\bibitem[{Soize and Ghanem(2004)}]{Soize2004}
\bibinfo{author}{Soize, C.}, \bibinfo{author}{Ghanem, R.},
  \bibinfo{year}{2004}.
\newblock \bibinfo{title}{Physical systems with random uncertainties: chaos
  representations with arbitrary probability measure}.
\newblock \bibinfo{journal}{SIAM Journal on Scientific Computing}
  \bibinfo{volume}{26}, \bibinfo{pages}{395--410}.
\bibitem[{Stine(1985)}]{Stine1985}
\bibinfo{author}{Stine, R.}, \bibinfo{year}{1985}.
\newblock \bibinfo{title}{Bootstrap prediction intervals for regression}.
\newblock \bibinfo{journal}{Journal of the American Statistical Association}
  \bibinfo{volume}{80}, \bibinfo{pages}{1026--1031}.
\bibitem[{Tipping(2001)}]{Tipping2001}
\bibinfo{author}{Tipping, M.E.}, \bibinfo{year}{2001}.
\newblock \bibinfo{title}{Sparse {B}ayesian learning and the relevance vector
  machine}.
\newblock \bibinfo{journal}{Journal of Machine Learning Research}
  \bibinfo{volume}{1}, \bibinfo{pages}{211--244}.
\bibitem[{Torre et~al.(2019)Torre, Marelli, Embrechts and Sudret}]{Torre2019a}
\bibinfo{author}{Torre, E.}, \bibinfo{author}{Marelli, S.},
  \bibinfo{author}{Embrechts, P.}, \bibinfo{author}{Sudret, B.},
  \bibinfo{year}{2019}.
\newblock \bibinfo{title}{Data-driven polynomial chaos expansion for machine
  learning regression}.
\newblock \bibinfo{journal}{Journal of Computational Physics}
  \bibinfo{volume}{388}, \bibinfo{pages}{601--623}.
\bibitem[{Tropp and Gilbert(2007)}]{Tropp2007}
\bibinfo{author}{Tropp, J.A.}, \bibinfo{author}{Gilbert, A.C.},
  \bibinfo{year}{2007}.
\newblock \bibinfo{title}{Signal recovery from random measurements via
  orthogonal matching pursuit}.
\newblock \bibinfo{journal}{IEEE Transactions on Information Theory}
  \bibinfo{volume}{53}, \bibinfo{pages}{4655--4666}.
\bibitem[{Tsilifis et~al.(2020)Tsilifis, Papaioannou, Straub and
  Nobile}]{Tsilifis2020}
\bibinfo{author}{Tsilifis, P.}, \bibinfo{author}{Papaioannou, I.},
  \bibinfo{author}{Straub, D.}, \bibinfo{author}{Nobile, F.},
  \bibinfo{year}{2020}.
\newblock \bibinfo{title}{Sparse polynomial chaos expansions using variational
  relevance vector machines}.
\newblock \bibinfo{journal}{Journal of Computational Physics}
  \bibinfo{volume}{416}.
\bibitem[{Vovk(2012)}]{Vovk2012}
\bibinfo{author}{Vovk, V.}, \bibinfo{year}{2012}.
\newblock \bibinfo{title}{Conditional validity of inductive conformal
  predictors}, in: \bibinfo{booktitle}{Proceedings of the Asian Conference on
  Machine Learning, Taoyuan, Taiwan}, pp. \bibinfo{pages}{475--490}.
\bibitem[{Vovk et~al.(2005)Vovk, Gammerman and Shafer}]{Vovk2005}
\bibinfo{author}{Vovk, V.}, \bibinfo{author}{Gammerman, A.},
  \bibinfo{author}{Shafer, G.}, \bibinfo{year}{2005}.
\newblock \bibinfo{title}{Algorithmic Learning in a Random World}.
\newblock \bibinfo{publisher}{Springer}.
\bibitem[{Vovk et~al.(2018)Vovk, Nouretdinov, Manokhin and
  Gammerman}]{Vovk2018}
\bibinfo{author}{Vovk, V.}, \bibinfo{author}{Nouretdinov, I.},
  \bibinfo{author}{Manokhin, V.}, \bibinfo{author}{Gammerman, A.},
  \bibinfo{year}{2018}.
\newblock \bibinfo{title}{Cross-conformal predictive distributions}, in:
  \bibinfo{booktitle}{Proceedings of the Seventh Workshop on Conformal and
  Probabilistic Prediction and Applications, Maastricht, The Netherlands}, pp.
  \bibinfo{pages}{37--51}.
\bibitem[{Xiu(2010)}]{XiuBook2010}
\bibinfo{author}{Xiu, D.}, \bibinfo{year}{2010}.
\newblock \bibinfo{title}{Numerical Methods for Stochastic Computations: A
  Spectral Method Approach}.
\newblock \bibinfo{publisher}{Princeton University Press},
  \bibinfo{address}{Princeton, New Jersey, USA}.
\bibitem[{Xiu and Karniadakis(2002)}]{Xiu2002}
\bibinfo{author}{Xiu, D.}, \bibinfo{author}{Karniadakis, G.E.},
  \bibinfo{year}{2002}.
\newblock \bibinfo{title}{{The Wiener-Askey polynomial chaos for stochastic
  differential equations}}.
\newblock \bibinfo{journal}{SIAM Journal on Scientific Computing}
  \bibinfo{volume}{24}, \bibinfo{pages}{619--644}.
\bibitem[{Xu et~al.(2011)Xu, Caramanis and Mannor}]{Xu2011}
\bibinfo{author}{Xu, H.}, \bibinfo{author}{Caramanis, C.},
  \bibinfo{author}{Mannor, S.}, \bibinfo{year}{2011}.
\newblock \bibinfo{title}{Sparse algorithms are not stable: A no-free-lunch
  theorem}.
\newblock \bibinfo{journal}{IEEE transactions on pattern analysis and machine
  intelligence} \bibinfo{volume}{34}, \bibinfo{pages}{187--193}.

\end{thebibliography}

\end{document}